\documentclass[aps,prd,eqsecnum,11pt,showpacs,preprintnumbers,nofootinbib]{revtex4}
\usepackage{amssymb,amsmath,bm,color}
\usepackage{graphicx}
\usepackage{graphics}

\def\nbox#1#2{\vcenter{\hrule \hbox{\vrule height#2in
\kern#1in \vrule} \hrule}}
\def\sq{\,\raise.5pt\hbox{$\nbox{.09}{.09}$}}
\def\sqb{\,\raise.5pt\hbox{$\overline{\nbox{.09}{.09}}$}}

\newcommand{\bea}{\begin{eqnarray}}
\newcommand{\eea}{\end{eqnarray}}
\newcommand{\be}{\begin{equation}}
\newcommand{\ee}{\end{equation}}

\newcommand{\slK}{\mbox{\,\slash \hspace{-0.62em}$K$}}
\newcommand{\slV}{\mbox{\,\slash \hspace{-0.52em}$V$}}
\newcommand{\slQ}{\mbox{\,\slash \hspace{-0.6em}$Q$}}

\newcommand{\slP}{\mbox{\,\slash \hspace{-0.62em}$P$}}

\renewcommand{\k}{{\bf k}}
\newcommand{\q}{{\bf q}}
\newcommand{\p}{{\bf p}}

\renewcommand{\v}{{\bf v}}

\newcommand{\rt}[1]{{}}
\begin{document}

\allowdisplaybreaks
\vspace{-.5cm}
\preprint{{\it Phys. Rev. D} {\bf 81}, 025014 (2010);\qquad {\rm LA-UR 09-0419}}

\title{Systematics of High Temperature Perturbation Theory: \\
The Two-Loop Electron Self-Energy in QED}

\author{Emil Mottola}
\affiliation{Theoretical Division \\
Los Alamos National Laboratory \\
Los Alamos, NM 87545 USA}
\email{emil@lanl.gov}

\author{Zsolt Sz{\'e}p}
\affiliation{Statistical and Biological Physics Research Group\\
of the Hungarian Academy of Sciences,\\ H-1117 Budapest, Hungary}
\email{szepzs@achilles.elte.hu}

\begin{abstract}
\vskip .2cm

In order to investigate the systematics of the loop expansion in high
temperature gauge theories beyond the leading order hard thermal loop (HTL)
approximation, we calculate the two-loop electron proper self-energy $\Sigma$ 
in high temperature QED. The two-loop bubble diagram of $\Sigma$ contains a 
linear infrared divergence. Even if regulated with a non-zero photon mass
$M$ of order of the Debye mass, this infrared sensitivity implies that
the two-loop self-energy contributes terms to the fermion dispersion relation 
that are comparable to or even larger than the next-to-leading-order (NLO) 
contributions of the one-loop $\Sigma$. Additional evidence for the necessity 
of a systematic restructuring of the loop expansion comes from the explicit 
gauge parameter dependence of the fermion damping rate at both one and 
two-loops. The leading terms in the high temperature expansion of the 
two-loop self-energy for all topologies arise from an explicit hard-soft 
factorization pattern, in which one of the loop integrals is hard ($p \simeq T$), 
nested inside a second loop integral which is soft ($0 \le p \lesssim T$ 
for real parts; $p \simeq e T$ for imaginary parts). There are {\it no} 
hard-hard contributions to the two-loop $\Sigma$ at leading order at high 
$T$. Provided the same factorization pattern holds for arbitrary $\ell$ loops, 
the NLO high temperature contributions to the electron self-energy come 
from $\ell - 1$ hard loops factorized with one soft loop integral.
This hard-soft pattern is a necessary condition for the resummation over 
$\ell$ to coincide with the one-loop self-energy calculated with HTL dressed 
propagators and vertices, and to yield the complete NLO correction to $\Sigma$ 
at scales $\sim eT$, which is both infrared finite and gauge invariant. We 
employ spectral representations and the Gaudin method for evaluating finite 
temperature Matsubara sums, which facilitates the analysis of multi-loop 
diagrams at high $T$. 

\end{abstract}

\pacs{11.10.Wx,\ 11.15.Bt,\ 12.20.Ds}
\date{\today}

\maketitle

\section{Introduction}

The systematics of the usual loop expansion in powers of the weak coupling
$e^2 =1/137 \ll 1$ is well understood in quantum electrodynamics (QED), in 
perturbation theory about the vacuum state. However, when one considers 
extreme states of matter at high temperatures or densities, straightforward 
perturbation theory is of little use, even in a weakly coupled gauge theory
like QED. At very high temperatures ($T \gg m_e$) one signal of the failure 
of the usual loop expansion is the appearance of infrared 
singularities of the form $(e T/\omega)^{2 \ell -2}$, at $\ell$ loop order, 
where $\omega$ is a characteristic frequency, such as that of an 
external line in a self-energy diagram. The appearance of such infrared
singularities implies that higher order terms in the perturbative series 
in $e^2$ are unsuppressed for low energies or momentum scales, 
and hence the usual perturbative loop expansion fails for $\omega \lesssim e T$.

The singular infrared behavior of perturbation theory at finite temperature
can be seen already at the one-loop level.  The leading order one-loop
singularities arise from the ultraviolet or hard region of the momentum
integral, whose power law growth is cut off only by the 
Bose-Einstein or Fermi-Dirac thermal distribution, so that the loop momentum
becomes of order of the temperature $T$.  These leading contributions in the
high temperature expansion of one-loop self-energies and vertices are the
hard thermal loops (HTL's) \cite{pisarski89}, studied extensively in the 
1990's in both QED and QCD (see \cite{thoma95,lebellac96,blaizot02,kapusta06} 
for reviews). It was found that the leading order (LO) HTL 
self-energies and proper vertices enjoy some remarkable properties
\cite{braaten90a,frenkel90,braaten90b}, among which is that they obey the 
same Ward identities as those of the exact theory. This enabled Braaten and 
Pisarski to propose an HTL resummation program \cite{braaten90a,braaten90b},
in which the LO HTL propagators and vertices are used instead of 
the bare ones in any diagram in which the loop integration momentum becomes 
soft ($\sim e T \ll T$).  This resummation algorithm was successfully applied 
to a number of problems, chief among them the first determination of the 
damping rate of gluons at rest in the high temperature QCD plasma, which
is finite and free of any gauge ambiguities \cite{braaten90c}.

The Ward identities obeyed by the lowest order HTL self-energies and vertices, 
and the success of HTL resummation in producing gauge invariant physical 
quantities suggests it should be possible to restructure the perturbative 
loop expansion of gauge theories at high temperatures or densities in a 
systematic way. To do so, however, requires a more complete understanding 
of the infrared singularities of the subleading terms in the high temperature 
expansion at higher loops. The subleading or next-to-leading order (NLO) terms
in the high temperature expansion of the one-loop level fermion self-energy
have been computed \cite{mitra00,wang04}, but there have been only a few
partial calculations of various quantities in different kinematic limits
at two-loop order \cite{qader92,altherr93,schulz94,majumder01,CarMot}.
As a consequence, both the nature of the subleading infrared singularities 
of perturbation theory which HTL resummation includes and those which it 
does not include, and the general systematics of the reordering of the loop 
expansion at high temperatures beyond HTL's, if it exists, has remained 
relatively unexplored. 

The promising beginning of HTL resummation of infrared singularities, 
and the possibility of extending its success to higher orders in a
systematic way is especially interesting in gauge theories. More
than just a formal question about the nature of perturbation theory 
at high temperatures, the lack of systematic expansion methods that 
can capture both collective behavior and infrared physics in hot or 
dense matter, and also preserve gauge invariance and the UV renormalization 
structure of the underlying microscopic theory is a serious obstacle 
to physical applications in processes from the early universe to 
heavy-ion collisions, both in and out of equilibrium.

Motivated by these considerations, we present in this paper a calculation 
of the two-loop electron self-energy $\Sigma$ in high temperature QED. 
Our approach is simply to calculate the two-loop self-energy in the
high temperature limit, systematically extracting the leading terms at
high $T$, without making any assumptions about the relative 
importance or unimportance of different contributions or momentum ranges,
and without assuming {\it a priori} how they should be resummed.
Once these leading terms in the high temperature expansion of the two-loop 
self-energy are known, they may be compared then with the known NLO terms of the 
one-loop self-energy \cite{mitra00,wang04}. By explicit calculation
we shall find that the {\it leading} terms in the high temperature expansion 
of the two-loop self-energy are of the same size, {\it or even larger} than the 
NLO terms of the one-loop $\Sigma_1$ when evaluated on the fermion mass shell. 
This is a direct demonstration of the need to take two and higher loops 
into account in order to obtain the full NLO result.

Dissecting the explicit two-loop self-energy calculation also allows for
a more complete understanding of the relative importance of different
momentum ranges in the loop integrations. It is already well known that
the LO HTL result at one-loop order results from the strictly hard region 
of the loop integration where $p\sim T$. The NLO terms of the one-loop 
$\Sigma$ arise from the region of the loop momentum integral 
which is parametrically less than $T$. For the case of the imaginary part 
of $\Sigma$, the relevant momentum scale is $p \sim e T \ll T$, while the 
real part of the self-energy receives contributions form the entire range 
of loop momentum $0\le p \lesssim T$, and give rise to $\ln (T/\omega)$ 
weak dependence upon the temperature. Although momenta up to $T$ do play 
a role in the real parts, unlike the purely hard contributions which 
grow as a power of $T$, logarithms are a signal of sensitivity to 
the low momentum, infrared region of the loop integration as well. Following
the convention in the field \cite{pisarski89,braaten90a}, we term all 
such contributions where internal momenta of order $eT$ or less
play an important role, {\it soft} loop integrals. We shall see that
the leading order terms at two loops which are comparable to the NLO
terms at one loop all come from momentum integrations where one of 
the loops is hard, while the other is soft, in the sense just described.
Thus these comparable terms coming from one- and two-loop diagrams have in
common that they arise from {\it exactly one soft internal momentum loop}.

An additional signal of the need for a reordering of perturbation theory
comes from the gauge dependence of the one-loop self-energy, even when
evaluated on the fermion mass shell. In the LO HTL limit, {\it i.e.}
when the loop momentum is hard, the one-loop electron dispersion relation 
is independent of the gauge fixing parameter $\xi$. However at NLO in the 
high temperature expansion coming from the soft region of the loop
momentum integral, the one-loop dispersion relation becomes dependent 
upon $\xi$ \cite{mitra00,wang04}. This is sufficient proof that the NLO 
one loop calculation is incomplete and physically meaningless by itself, 
and a systematic resummation method to obtain gauge invariant results 
for physical quantities such as the fermion damping rate at order $e^2T$ 
is necessary. Our calculation will show that this gauge dependence is
not cured at two-loops, and a full resummation of higher loops is necessary 
in order to arrive at the full NLO gauge invariant decay rate of an on-shell 
fermion in the plasma of order $e^2T$.

When higher numbers of loops are considered, we encounter a new 
phenomenon not present at one loop. At two-loop order in addition to 
gauge parameter dependence of the on-shell dispersion relation, we 
encounter uncanceled true infrared and/or collinear divergences, not 
regulated even at finite values of the soft external energy $\omega$, or
momentum $\q$. In particular, a linear infrared divergence appears 
in the bubble diagram Fig.~\ref{Fig:diagrams} of the two-loop fermion 
self-energy, arising from the double photon massless propagator pole
in this diagram. Quite apart from dependence upon the gauge parameter
$\xi$, the appearance of such unregulated divergences from massless,
unscreened photons is itself also sufficient evidence for the breakdown 
of the loop expansion at high temperatures, and the need for a resummation 
of diagrams with still larger numbers of loops. 

In order to regulate the divergences in the two-loop fermion self-energy, 
we introduce a photon mass $M$ in the evaluation of the bubble diagram 
of Fig.~\ref{Fig:diagrams}. In the usual perturbative expansion, the
formal limit $M \rightarrow 0$ should be taken. Since this limit
does not exist at two-loop order, we may anticipate the result that 
after HTL resummation of the photon self-energy the photon acquires a 
non-zero Debye screening mass of order $eT$. Taking $M \sim eT$ allows 
us to characterize the order of all the two-loop contributions to $\Sigma$, 
and compare finite contributions from different loop orders in a meaningful way. 
In other words, when the formal photon regulator mass is replaced by the 
physical Debye screening mass, by a partial resummation of the photon propagator 
for soft momenta, there are neither infrared nor collinear divergences in 
the two-loop dressed self-energy, and all NLO contributions to the fermion 
self-energy are estimated correctly in powers of $e$. The photon mass $M$ 
thus plays a dual role in our considerations, first as a regulator of infrared 
divergences in bare loop diagrams, and second, as a parameter 
whose finite value of order $eT$ after resummation of higher loop diagrams 
enables us to estimate the finite magnitude of resummed amplitudes.

Both the gauge parameter dependence and the explicit infrared
divergences as $M \rightarrow 0$ in the bare loop expansion can be
cured only by systematically including the contributions from higher
numbers of loops $\ell >2$, despite the fact that these contain
additional explicit powers of the coupling. This HTL improved
resummation leads finally to the complete order $e^2T$ corrections to
the fermion dispersion relation in the plasma for both its real and
imaginary parts, compared to the LO result for the fermion thermal
mass, $eT/\sqrt{8}$ which is purely real.  Thus the reordering of
perturbation theory in the plasma is decidedly non-analytic in $\alpha
= e^2/4\pi$.
 
The structure of the explicit two-loop calculation of $\Sigma$ 
shows that the leading contributions at high temperatures arise entirely from 
momentum integrals which {\it factorize} into one momentum integral which is 
hard ($p \sim T$), and a second one which is soft (in the sense defined 
above). Overlapping two-loop momentum integrals which violate this hard-soft 
pattern can be shown explicitly to be suppressed by additional powers of $e$, 
thus justifying their neglect, in ${\cal O}(e^2T)$ calculation of the self-energy 
and damping rate. In particular, the detailed analysis of the two-loop self-energy 
shows that there are {\it no} contributions at this order coming from both loops 
hard, which naive power counting would apparently allow \cite{braaten90b,lebellac96}. 
That such hard-hard contributions from two-loop diagrams at this order
in $e^2T$ are completely absent is in accordance with the an earlier result 
by Schulz for the two-loop gluon polarization in QCD at next to leading order 
\cite{schulz94}. A fact apparently not widely recognized is that the contribution 
at two-loop order of only hard-soft integrals to terms of order $e^2T$ in the
fermion self-energy and the complete absence of hard-hard contributions
is exactly what is {\it required} by the HTL resummation program, when HTL 
propagators and vertices are re-expanded in primitive bare loops of 
perturbation theory, a point which we elaborate in some detail in Sec. \ref{sec:HTL}. 
Thus one result of our detailed evaluation of the two-loop self-energy is a 
refinement of naive power counting rules, and a clarification of the hard-soft 
factorization structure of the high temperature expansion of perturbation theory, 
which underpins resummation of one-loop diagrams with HTL dressed propagators 
and vertices.

The factorization pattern should be expected to continue at $\ell > 2$ 
loops as well, namely all the order $e^2T$ corrections to the fermion
dispersion relation should arise from $\ell - 1$ hard loops inside
only one soft loop. This HTL reordering of the ordinary loop expansion 
is necessary to collect all the infrared sensitive terms of 
relative order $(e T/\omega)^{2\ell -2}$ from $\ell$ loops, which 
make equally important contributions to the fermion dispersion 
relation and damping rate, when $\omega \sim e T$. When all these 
subleading contributions are summed over all $\ell$ loops by the HTL 
resummation of propagators and vertices in an HTL dressed one-loop diagram, 
the gauge parameter dependence at subleading $e^2T$ order is eliminated. 
The curing of both the infrared divergences, and the gauge parameter 
dependence of the ordinary bare loop expansion by the same resummation 
in HTL propagators and vertices, when calculating damping rates and 
plasma frequencies, represents an important check on its physical 
consistency \cite{braaten90c,kobes92,braaten92,schulz94,schulz95,meg07,meg08}.

The technical evaluation of the two-loop diagrams in the imaginary
time formalism \cite{lebellac96,kapusta06} is facilitated considerably
by using Gaudin's method \cite{gaudin65,reinosa06}, which we review in
Appendix \ref{App:gaudins_method}. This very efficient method of performing 
the frequency sums makes calculations of diagrams involving many
finite temperature propagators feasible. The Gaudin method also makes the
hard-soft factorization pattern at two-loops more evident, and makes
feasible the analysis of dressed and/or even higher loop diagrams, at 
least for the purpose of extracting their leading order behavior in the high 
temperature limit.

The outline of the paper is as follows. In Section \ref{sec:one-loop}
we review the one-loop finite temperature electron
self-energy, including both the LO HTL and NLO terms, and the latter's
dependence upon the gauge fixing parameter $\xi$. In Section
\ref{sec:2loop} we present the basic calculation of the three diagrams
of two different topologies in the two-loop electron self-energy by
the Gaudin method in the Feynman gauge, introducing the photon mass
regulator for the most infrared sensitive bubble diagram. In Section
\ref{sec:2loop_anal} we present the detailed analysis of the two-loop
integrals obtained in Section~\ref{sec:2loop}, for a fermion at rest
${\bf q} =0$. We show that the leading contributions in the high
temperature expansion of the two-loop diagrams of both topologies come
from one loop hard, the other soft, thus demonstrating explicitly that
there are no hard-hard contributions, and the soft-soft contributions
are subleading at this order. In Section \ref{sec:HTL} we compare our
results with the insertions of effective HTL self-energies and
vertices in the one-loop $\Sigma$, explicitly verifying the
equivalence of these insertions for both the real and imaginary parts
to order $e^2T$, which is necessary for the HTL resummation program to
work. In Section~\ref{sec:xi} we study the gauge parameter dependence
of the two-loopself-energy, and show that it does not cancel even on
shell, in neither its real or imaginary parts. Section \ref{sec:sum}
contains a summary of the main results for the two-loop self-energy to
order $e^2T$, concluding with a discussion of the lessons for a
systematic reorganization of the loop expansion in high temperature
gauge theories to higher orders.

%------@#$---------- ONE LOOP SIGMA  -----------------

\section{The One-Loop Electron Self-Energy
\label{sec:one-loop}}

\begin{figure}[!t]
\includegraphics[keepaspectratio,width=0.2\textwidth,angle=0]
{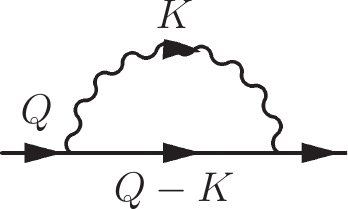}
\caption{One-loop contribution to the fermion self-energy.}
\label{Fig:Sigma1}
\end{figure}

We begin by setting our conventions and reviewing the existing leading order (LO)
\cite{lebellac96,thoma98} and NLO \cite{mitra00,wang04} calculations 
of the one-loop QED electron finite temperature self-energy, 
\be
\Sigma_1(Q)=-i e^2 \int_K \gamma_\mu G(Q-K) \gamma_\nu D^{\mu\nu}(K)
\label{Sig1loop}
\ee
represented in Fig.~\ref{Fig:Sigma1}. For the purposes of this paper, 
we use the Feynman rule conventions of \cite{Peskin-book} wherein
the tree-level electron-photon vertex is $-i e\gamma_\mu.$
In (\ref{Sig1loop}) the tree-level fermion propagator is given by
\be
G(K)=\frac{i}{\slK}=\slK\tilde D(K)\,, \qquad \tilde D(K)=i/(K^2+i\epsilon)\,,
\ee
with $\slK=\gamma^\mu K_\mu=k_0\gamma_0-\k \cdot {\bm \gamma}$.
The tree-level photon propagator in the one-parameter family of covariant
$\xi$-gauges is
\be
D^{\mu\nu}(K)=-\left(g^{\mu\nu}-(1-\xi)\frac{K^\mu K^\nu}{K^2}\right) D(K)\,,
\qquad D(K)=i/(K^2+i\epsilon)\,.
\ee
We employ the somewhat redundant notation of $\tilde D(K)$ and $D(K)$ 
for the same function of $K^2$ in order to distinguish propagators 
with fermionic Matsubara frequencies from those with bosonic Matsubara
frequencies in the imaginary time formalism of finite temperature
field theory \cite{lebellac96,kapusta06}. Thus, in imaginary time 
where the generic four-momentum becomes $K^\mu=(i\omega_n,\k)$,
\begin{subequations}
\bea
&&D(i\omega_n,\k)=-i\Delta(i\omega_n,\k)= \frac{-i}{-(i\omega_n)^2+\k^2}\,,
\qquad \omega_n = 2\pi n T\,;  \\
&&\tilde D(i\omega_n,\k)=-i\tilde\Delta(i\omega_n,\k) =\frac{-i}{-(i\omega_n)^2+\k^2}\,,
\qquad \omega_n = (2n + 1)\pi T\,,
\eea
\label{Eq:Matsubara_prop} 
\end{subequations}
\vspace{-.6cm}

\noindent
The indices of the four-momentum are raised and lowered with the metric 
tensor $g^{\mu\nu}=\textnormal{diag}(1,-1,-1,-1)$. The Dirac
$\gamma$-matrices satisfy the anticommutation relation
$\{\gamma_\mu,\gamma_\nu\}=2g_{\mu\nu}$, so that $(\gamma^0)^{\dagger} =
\gamma^0$ is Hermitian. 

For evaluation of loop integrals such as that in (\ref{Sig1loop})
we use the convention
\be 
\displaystyle \int_K\longrightarrow i T\sum_n\int_\k \equiv 
i T\sum_n\int\frac{d^3\k}{(2\pi)^3}\,.
\ee
Moreover, our conventions are such that the electron and photon proper 
self-energies, $\Sigma$ and $\Pi_{\mu\nu}$, contribute to the full 
inverse propagators in the form
\be
i{\cal G}^{-1}(Q)=i G^{-1}(Q)-\Sigma(Q),\qquad
i{\cal D}_{\mu\nu}^{-1}(Q)=i D_{\mu\nu}^{-1}(Q)-\Pi_{\mu\nu}(Q)\,,
\label{invprop}
\ee
for the electron and photon respectively. The sign convention for $\Sigma$ 
here is the opposite of that of \cite{wang04}.

It will also be convenient to work with the spectral representation of the
propagators, which is given by
\begin{subequations}
\bea
D^{\mu\nu}(i\omega_n,\k)&=&-i\int_{-\infty}^{\infty}\frac{d k_0}{2\pi} 
\frac{\rho^{\mu\nu}(k_0,k)}{k_0-i\omega_n}=-i \Delta^{\mu\nu}(i\omega_n,
\k),
\label{Eq:rho_photon}
\\
G(i\omega_n,\k)&=& -i\int_{-\infty}^{\infty}\frac{d k_0}{2\pi} 
\frac{\rho_F(k_0,k)}{k_0-i\omega_n} = -i\Delta_F(i\omega_n, \k),  \ \ 
\label{Eq:rho_fermion}
\eea
\end{subequations}
where 
\begin{subequations}
\bea
\rho^{\mu\nu}(k_0,\k)&\equiv&-\left(g^{\mu\nu}+(1-\xi)K^\mu K^\nu
\frac{\partial }{\partial k_0^2}\right)  \rho(k_0,\k),  \\
\rho_F(k_0,\k)&\equiv&\slK \rho(k_0,\k).
\eea
\label{rhodefs}   
\end{subequations}
Here $\rho(k_0,\k)$ is the spectral function of a free massless boson,
\be
\rho(K) \equiv \rho(k_0,\k)=2\pi\varepsilon(k_0)\delta(k_0^2-\k^2)\,,
\label{rhoboson}
\ee
and
$\varepsilon(k_0)=\theta(k_0)-\theta(-k_0)$ is the sign function.
Accordingly, the spectral representation of $\Delta(i\omega_n,\k)$ is
\be
\Delta(i\omega_n,\k)= -\frac{1}{K^2} = \int_{-\infty}^{\infty}\frac{d k_0}{2\pi}  
\frac{\rho(k_0,\k)}{k_0-i\omega_n}\,,  
\label{Eq:scalar_prop_spectral} 
\ee
and similarly for $\tilde \Delta(i\omega_n,\k)$, with the only difference
that in this latter case $\omega_n$ is fermionic.

\subsection{Feynman Gauge Evaluation}

We consider first the Feynman gauge condition ($\xi =1$). In this case,
using the spectral representation of the propagators, the self-energy 
(\ref{Sig1loop}) for a fermion with external Matsubara frequency $\nu_q$ becomes  
\be
\Sigma_1(i\nu_q,\q)=-2e^2 \int_{\k}
\int_{-\infty}^\infty \frac{d k_0}{2\pi} 
\int_{-\infty}^\infty \frac{d p_0}{2\pi} \rho(k_0,\k) \rho_F(p_0,\q-\k)\, 
T\sum_n \left[\frac{1}{k_0-i\omega_n}\right]
\left[\frac{1}{p_0-i(\nu_q-\omega_n)}\right]. 
\label{Sig1loopFeyn} 
\ee
The frequency sum can be performed using Gaudin's method \cite{gaudin65},
reviewed recently in \cite{reinosa06} and in Appendix~\ref{App:gaudins_method}. 

Gaudin's method relies first on the spectral representation of finite temperature
propagators in the imaginary time formalism (\ref{Eq:scalar_prop_spectral}),
and second, on a graphical representation of a particular partial fractioning 
decomposition of the products of denominators that arise in finite temperature
perturbation theory into a sum of products, in each of which the only Matsubara 
frequencies which appear are the {\it independent} ones to be summed.  
Since an $\ell$-loop finite temperature diagram contains $\ell$ 
independent Matsubara sums, this means that it can be written as a sum 
of terms each one of which contains exactly $\ell$ independent frequency 
sums, which can be performed simply, no matter how many propagators or
vertices the original Feynman diagram contains. This disentangling
of the independent and dependent Matsubara frequencies greatly 
facilitates the computation of multi-loop finite temperature diagrams,
or even one-loop finite temperature  diagrams with multiple vertex insertions.

For the one-loop diagram of Fig.~\ref{Fig:Sigma1} the Gaudin decomposition
into terms with the one independent Matsubara frequency to be summed 
amounts to the simple partial fraction identity,
\begin{eqnarray}
&&T\sum_n \left[\frac{e^{i\omega_n\tau_1}}{k_0 - i\omega_n}\right]\ 
\left[\frac{e^{i(\nu_q-\omega_n)\tau_2}} {p_0-i(\nu_q-\omega_n)}\right]\nonumber\\
&& \quad = \frac{e^{i\nu_q\tau_2}}{k_0 + p_0 -i\nu_q} 
\left[ T \sum_n \frac{e^{i\omega_n(\tau_1-\tau_2)}}{k_0 - i \omega_n}\right]
+ \frac{e^{i\nu_q\tau_1}}{k_0 + p_0 -i\nu_q} 
\left[ T \sum_n \frac{e^{i(\nu_q -\omega_n)(\tau_2-\tau_1)}}
{p_0 - i (\nu_q -\omega_n)}\right]\,,
\label{Gaudexam}
\end{eqnarray}
so that only one denominator in each term now contains the Matsubara
frequency to be summed. Each of the $\tau_k$ are infinitesimals, 
assumed positive, assigned to each line; for example $\tau_k = k \epsilon$ with 
$\epsilon \rightarrow 0^+$, so that  $\tau_1 - \tau_2$ has a definite 
sign. Shifting then the summation over the bosonic frequency
$\omega_n = 2\pi n T$ in the second term of (\ref{Gaudexam}) to the fermionic
frequency $\tilde \omega_n \equiv \nu_q - \omega_n = (2n+1) \pi T$,
and applying the fundamental summation formulas (\ref{ap_Eq:sumbose})
and (\ref{ap_Eq:sumfermi}) for bosonic and fermionic frequencies
respectively, we obtain for the sum in (\ref{Sig1loopFeyn}) or 
(\ref{Gaudexam}) the result 
\be
T\sum_n \frac{1}{(k_0-i\omega_n)[p_0-i(\nu_q-\omega_n)]}=-
\frac{n_F(p_0)+n_B(-k_0)}{k_0+p_0-i\nu_q},
\label{Eq:Matsubara-1loopFSE}
\ee
in the limit $\tau_i \rightarrow 0^+$ with $\tau_1 - \tau_2 \rightarrow 0^-$.
Here $n_B(E),n_F(E)$ are the Bose-Einstein and Fermi-Dirac
distribution functions given by  
\be
n_B(E) \equiv \frac{1}{\exp(E/T)-1},\qquad  
n_F(E) \equiv \frac{1}{\exp(E/T)+1}\,.  
\label{Eq:BE_FD}  
\ee
The power of the Gaudin method lies in the fact that once the 
propagator denominators with independent Matsubara frequencies 
are isolated, as in (\ref{Gaudexam}), one never needs any
other Matsubara frequency sums other than the fundamental 
ones (\ref{ap_Eq:sumbose}) and (\ref{ap_Eq:sumfermi}), no
matter how many propagators or vertices the original Feynman
diagram contains. This advantage becomes more and more
evident at higher loop orders, or in diagrams with larger
numbers of finite temperature propagators.

We remark also that had we assigned the $\tau_i$ differently, so that, 
{\it e.g.}  $\tau_1 - \tau_2 \rightarrow 0^+$, we would have obtained 
$+[n_F(-p_0)+n_B(k_0)]$ instead of $-[n_F(p_0)+n_B(-k_0)]$
in the numerator of (\ref{Eq:Matsubara-1loopFSE}). However
these two expressions are equal, by use of the identities
\be  
n_B(-E)= -1 - n_B(E),\qquad n_F(-E)= 1 - n_F(E),  
\label{BE_FD_identities}  
\ee
for the Bose-Einstein and Fermi-Dirac statistical distributions.
Hence although one needs {\it some} regulator to define
the frequency sums (\ref{ap_Eq:sumbose}) and (\ref{ap_Eq:sumfermi}),
appearing in intermediate steps in the Gaudin method,
one can check that the results for any diagram containing at least 
two Matsubara propagators (whose frequency sums converge
absolutely) are in fact independent of the regulator used.

Performing next the frequency integrals over $k_0$ and $p_0$ in 
(\ref{Sig1loopFeyn}), by making use of the $\delta$-functions in the 
spectral representations (\ref{rhodefs}) and (\ref{rhoboson}), and 
using the first identity in (\ref{BE_FD_identities}), we obtain
\be
\Sigma_1(i\nu_q,\q)=2e^2\int\frac{d^3 \k}{(2\pi)^3}\frac{1}{4 E_1 E_2}
\sum_{s_1,s_2=\pm 1}\left[
\left(\gamma^0 E_2 s_1 -(\q-\k)\cdot\bm{\gamma}\, s_1 s_2\right)  
\frac{1+n_B(s_1 E_1)-n_F(s_2 E_2)}{i\nu_q-s_1 E_1-s_2 E_2}
\right],  
\label{Eq:freq_sum}  
\ee   
where $E_1=k,$ $E_2=|\q-\k|.$ For each particular value of $s_1$ and
$s_2$ the terms in the sum can be associated with physical processes
like pair creation and annihilation or scattering off of particles in
the heat bath \cite{weldon83}.

Examining (\ref{Eq:freq_sum}) one sees that for soft external
four-momentum there are two types of terms: one with $s_2=-s_1$ and the
other with $s_2=s_1$. In the first case the energy denominator is
small for $|\q| \ll |\k|$, since $E_1-E_2\simeq\k\cdot\q/k$,
and the $\k$ integration is dominated by hard $k = |\k| \sim T$.
These hard momentum integrals are proportional to $T^2$ and give the 
LO result in the HTL approximation. Physically these terms correspond to 
Landau damping in the plasma, which includes processes possible only 
at nonzero temperature: particle absorption from the heat bath 
and emission into the heat bath.

In the second case, that is with $s_2=s_1=\pm 1,$ the two energies
$E_1$ and $E_2$ add rather than cancel for large $k$ in the
denominator of (\ref{Eq:freq_sum}), and hence there are additional
powers of $k$ in the denominator which makes the integral over $\k$
more sensitive to its small $k$ range.  Since $1+n_B(s_1 E_1)-n_F(s_1
E_2)=\varepsilon(s_1)(1+n_B(E_1)-n_F(E_2))$, the appearance of the
``$1$'' shows that here one encounters processes which are possible
also at zero temperature, {\it e.g.} pair creation. The NLO terms in
the high temperature expansion of the one-loop self-energy come from
these types of processes.

Focusing on a fermion with vanishing spatial three-momentum ($\q=0$) one has
$E_1=E_2=k$ and the above two possibilities give  
\begin{alignat}{4}  
\label{Eq:LO_HTL_region}  
\textrm{  LO HTL:}\qquad   & \textrm{from} \sum_{s_2=-s_1=\pm 1} 
&&\longrightarrow  &-&\frac{1}{2k}\big[n_B(k)+n_F(k)\big] \frac{1}{i\nu_q},\\  
\label{Eq:NLO_HTL_region}  
\textrm{NLO HTL:}\qquad  & \textrm{from} \sum_{s_2=s_1=\pm 1} 
&&\longrightarrow  &-&\frac{1}{4k}\big[1+n_B(k)-n_F(k)\big] 
\left(\frac{1}{i\nu_q-2k}+\frac{1}{i\nu_q+2k}\right).  
\end{alignat}   
Using (\ref{Eq:LO_HTL_region}) and (\ref{Eq:NLO_HTL_region}) one obtains then
\be
\Sigma_1(i\nu_q,\q ={\bf 0})=\frac{e^2}{2\pi^2i\nu_q}\gamma_0  
\int_0^\infty d k\, k \big[n_B(k)+n_F(k)\big]- \frac{e^2 i\nu_q}{8\pi^2} \gamma_0 
\int_0^\infty d k  \frac{k}{k^2-(i\nu_q)^2/4}\, \big[1+n_B(k)-n_F(k)\big].
\label{Eq:sigma-1loop_final}
\ee
In the last expression the ``$1$'' in square brackets is the logarithmically
UV divergent vacuum contribution, which must be renormalized in the standard 
manner. Since we are interested only in the high $T$ dependence of the
self-energy in the hot plasma, we may discard such vacuum contributions.
It is important to point out that because of the identities 
(\ref{BE_FD_identities}), identifying the vacuum contribution unambiguously
requires first expressing all statistical distributions as functions of
energy arguments which are {\it positive}. This can be done most conveniently
before doing the frequency integrals by using the following expressions:
\be
\begin{split}
&n_B(E)=-\theta(-E)+\varepsilon(E) n_B(|E|),\,\quad  
n_B(-E)=-\theta(E)-\varepsilon(E) n_B(|E|),\\
&n_F(E)=\theta(-E)+\varepsilon(E) n_F(|E|),\qquad
n_F(-E)=\theta(E)-\varepsilon(E) n_F(|E|).
\end{split}
\label{Eq:BE_FD_decomp}   
\ee

Clearly the $k$ integral in the first term of (\ref{Eq:sigma-1loop_final})
is hard in the sense that a positive power of $k$ in the integrand is cut off 
only by a statistical distribution function $n_B$ or $n_F$ at $k \sim T$,
and we obtain
\be
\int_0^\infty d k\, k \big[n_B(k)+n_F(k)\big]=\frac{\pi^2 T^2}{6} 
+\frac{\pi^2 T^2}{12} = \frac{\pi^2 T^2}{4}\,. 
\label{Eq:hard_integral}  
\ee    
In contrast, in the second integral of (\ref{Eq:sigma-1loop_final})
the integrand is a decreasing function of $k$ for large $k$, even 
without the statistical distribution functions. Integrals of this 
kind are soft, in the sense that the small $k$ range is important, 
and the $n_B$ or $n_F$ thermal functions serve only to cut off at
most an otherwise logarithmically diverging integral at large $k$,
in which the argument of the logarithm is sensitive to the soft
momentum region. The high temperature limit of the second soft term 
in (\ref{Eq:sigma-1loop_final}) is evaluated in Appendix
\ref{App:mitras_integral}, where it is shown that
\begin{subequations}
\bea
&& \int_0^\infty d k  \frac{k}{k^2-(i\nu_q)^2/4}\, n_B(k)\bigg\vert_{i\nu_q = \omega + i 0^+}
 = \frac{i\pi T}{\omega} -\frac{1}{2} \ln \left(\frac{T}{\omega}\right) + \dots\,,\\
&&\int_0^\infty d k  \frac{k}{k^2-(i\nu_q)^2/4}\, n_F(k)\bigg\vert_{i\nu_q = \omega + i 0^+}
 = 0 \ + \ \frac{1}{2} \ln \left(\frac{T}{\omega}\right) + \dots \,,
\eea
\label{softNLO1}
\end{subequations}
\vspace{-.6cm}

\noindent
where only the high temperature asymptotic values of the integrals are displayed,
neglecting terms that do not grow with $T$. Thus, in Feynman gauge,
\be
\Sigma_1 (\omega,{\bf 0})\simeq  
\frac{e^2 T^2}{8\omega}\gamma_0
+\frac{e^2}{8\pi^2}\gamma_0 \left(
\omega\ln\frac{T}{\omega} -i\pi T\right)\,,\qquad (\xi =1)\,. 
\label{Eq:1-loop-RI}
\ee
The first term is the leading order (LO) HTL result, which receives
contributions from both the electron and photon thermal
distributions in (\ref{Eq:hard_integral}). It is purely real, and
in fact, independent of the gauge parameter $\xi$.
The second two terms in (\ref{Eq:1-loop-RI}) are the next-to-leading
order (NLO) or subleading terms in the high temperature expansion, which
come from the soft momentum region in the integrals (\ref{softNLO1}), 
and yield both a real and imaginary part. The contribution of the 
Fermi-Dirac statistical factor to the imaginary part and of the zero 
temperature vacuum part are subleading to this (NNLO) in the high 
temperature expansion.

The real part of the off-shell one-loop self-energy $\Sigma_1$ in
(\ref{Eq:1-loop-RI}) illustrates the infrared divergence at small 
external frequencies $\omega \rightarrow 0$. The on-shell condition 
at LO is obtained by finding the zeroes of the one-loop corrected 
inverse propagator $i G^{-1} (Q) - \Sigma_1(Q)$ in (\ref{invprop}). 
For vanishing spatial momentum $\q =0$, this on-shell condition is
\be
\omega\gamma_0 = \Sigma(\omega,\q =0 )
\label{Eq:uj_on-shell}
\ee
or to leading order,
\be
\omega = \frac{e^2T^2}{8\omega} + \dots
\label{Eq:LO_dispersion}
\ee
where the ellipsis denotes the NLO terms. Thus on the fermion mass
shell the tree level inverse Dirac propagator is set equal to the one-loop
result and $\omega \simeq e T/\sqrt{8}$ \cite{klimov,weldon,lebellac96}. 
It is the fact that $\omega \sim eT$ on-shell owing to the infrared enhancement 
of the one-loop self-energy at small $\omega$ that leads to inverse powers 
of the coupling at higher loop orders, re-ordering of the perturbative 
expansion, and eventually the need for further resummation.

\subsection{Gauge Parameter Dependence of $\Sigma_1$}

The general expression for the one-loop self-energy at NLO in the HTL
approximation and for $\q\ne 0$, in a general $\xi\ne 1$ gauge can be 
found in \cite{mitra00} (real part) and \cite{wang04} (both real 
and imaginary parts). Explicitly, the $\xi$ dependent term is
\be
^{\xi}\Sigma_1(Q) = (1-\xi)i e^2\int_K \gamma_\mu\frac{1}{\slQ-\slK} 
\gamma_\nu \frac{K^\mu K^\nu}{(K^2)^2}.
\ee
Making the substitution $\slK = -(\slQ-\slK) + \slQ$ we can rewrite this as
\be
^{\xi}\Sigma_1(Q) = (1-\xi)i e^2 \int_K \frac{1}{(K^2)^2}
\left[\slQ \frac{1}{\slQ-\slK} \slQ - (\slQ + \slK)\right]\,.
\label{oneloopxi}
\ee
Since the very last term here involving only $\slK$ vanishes by symmetry,
in this form one can see that if the lowest order tree level on-shell condition 
for a massless fermion $\slQ = 0$ is used, the gauge parameter dependence of
the one-loop self-energy vanishes, provided that there are no infrared
divergences in this limit. This is indeed the result that one obtains in 
the well-known QED perturbation theory at zero temperature with a finite 
electron mass.

At high temperatures, the existence of the $\omega^{-1}$ infrared 
behavior of the LO HTL self-energy in (\ref{Eq:1-loop-RI}), and imposition of 
the modified HTL on-shell condition (\ref{Eq:LO_dispersion}) implies that 
$\slQ = \omega\gamma_0 \neq 0$ for a fermion at rest in the plasma.
Hence (\ref{oneloopxi}) no longer vanishes for HTL dressed fermions
and the gauge parameter independence of the modified fermion dispersion 
relation at any given loop order is no longer guaranteed, even on-shell.
This is clearly the result of using an on-shell condition 
such as (\ref{Eq:LO_dispersion}) which mixes loop orders.

To evaluate $^{\xi}\Sigma_1(Q)$ explicitly in the high temperature limit,
we group the terms in a somewhat different way. Using the fact that 
$\slK (\slQ-\slK)\slK= Q^2\slK - K^2\slQ -(Q-K)^2\slK$, and that
$\frac{\partial}{\partial M^2}\int_K \slK D_{M^2}(K)$ vanishes, we 
may also write $^{\xi}\Sigma_1(Q)$ in the form
\be
^{\xi}\Sigma_1(Q) = (1-\xi)i e^2\left\{
-i Q^2\frac{\partial}{\partial M^2}
\int_K \slK D_{M^2}(K)\tilde D(Q-K)\bigg|_{M^2=0}
+ i\slQ\int_K D(K)\tilde D(Q-K) \right\}\,,
\label{Eq:1loop_xi_prop}
\ee
where
\be
D_{M^2}(K) \equiv \frac{i}{K^2 - M^2}
\label{DM2}
\ee
is the bosonic propagator with finite $M^2$. The corresponding
spectral representation is
\be
D_{M^2}(K) = -i \int_{-\infty}^{\infty}\frac{d k_0}{2\pi} 
\frac{\rho_{M^2}(k_0,\k)}{k_0-i\omega_n}\,,\\ 
\label{Eq:DM2_spectral}
\ee
with
\be
\rho_{M^2}(k_0,\k)=2\pi\varepsilon(k_0)\delta(k_0^2 - \k^2 - M^2)\,.
\label{rhoMdef}
\ee
The technique of introducing a photon mass $M$, differentiating with 
respect to it, and taking the $M\rightarrow 0$ limit at the end of 
the calculation is a useful method for defining the double pole
term, $(K^2)^{-2}$ appearing in both $^{\xi}\Sigma_1$ and the
two-loop self-energy bubble diagram of the next section. At this
point the photon mass $M$ is introduced as a mathematical convenience
and infrared regulator only, and the limit $M \rightarrow 0$ should be taken at
the end of the calculation. A finite photon self-energy of order $M \sim eT$
is obtained only after HTL resummation. 

The reason for the rearrangement of $^{\xi}\Sigma_1$ into the form
(\ref{Eq:1loop_xi_prop}) is that each of the terms in (\ref{Eq:1loop_xi_prop}) 
are finite in the limit $M\rightarrow 0$, whereas as we shall show in Sec.
\ref{sec:xi}, each of the terms in (\ref{oneloopxi}) is $M$ dependent and
divergent in this limit, with the sum of terms in $^{\xi}\Sigma_1$ finite.
In imaginary time, using the spectral representation of the propagators 
(\ref{Eq:scalar_prop_spectral}) and (\ref{Eq:DM2_spectral}) we have
\bea
\nonumber
&&^{\xi}\Sigma_1(i\nu_q,\q) = (1-\xi)e^2\gamma_0 \times\nonumber\\
&& \Bigg\{
-Q^2\frac{\partial}{\partial M^2}\int_\k 
\int_{-\infty}^\infty \frac{d k_0}{2\pi}\,k_0\,
\rho_{M^2} (k_0,\k) \int_{-\infty}^\infty \frac{d p_0}{2\pi}
\rho(p_0,\q-\k)
T\sum_n\frac{1}{(k_0-i\omega_n)[p_0-i(\nu_q-\omega_n)]}
\Bigg|_{M^2=0}\nonumber\\
&&+\slQ \int_\k
\int_{-\infty}^\infty \frac{d k_0}{2\pi}
\int_{-\infty}^\infty \frac{d p_0}{2\pi}
\rho(k_0,\k)\rho(p_0,\q-\k)
T\sum_n\frac{1}{(k_0-i\omega_n)[p_0-i(\nu_q-\omega_n)]}
\Bigg\}\,.\ \ 
\label{Eq:1loop_xi}
\eea
Since $\omega_n$ is bosonic and $\nu_q-\omega_n$ is fermionic, the
Matsubara sum here is identical to that given in
(\ref{Eq:Matsubara-1loopFSE}). One next performs the frequency
integrals over $k_0$ and $p_0$ in (\ref{Eq:1loop_xi}). For the
integrals over the frequencies which are {\it not} arguments of the
statistical factors, the spectral decomposition
(\ref{Eq:scalar_prop_spectral}) may be undone, re-obtaining Matsubara
propagators. The remaining frequency integrals are performed using the
Dirac $\delta$ functions in the spectral functions after using 
the relations (\ref{Eq:BE_FD_decomp}), which allows the separation 
of the vacuum and thermal pieces by making the arguments of 
the statistical distributions explicitly positive. 
Keeping only the thermal part and taking $\q=0,$ in this way one obtains
\bea
\nonumber
^{\xi}\Sigma_1(i\nu_q,\q ={\bf 0})=\frac{(1-\xi)e^2}{8\pi^2}\gamma_0
&\!\!\Bigg\{& \!\! (i\nu_q)^2
\frac{\partial}{\partial M^2}
\Bigg[\frac{2}{i\nu_q}\mu_-\mu_+
\int_0^\infty d k \frac{k\, n_F(k)}{k^2-\mu_-^2}
-\frac{2}{i\nu_q} \int_0^\infty d k\frac{k^2 E_k\, n_B(E_k)}{E_k^2-\mu_+^2}
\Bigg]\Bigg|_{M^2=0}
\\
&&
+i\nu_q \int_0^\infty d k\, \frac{k}{k^2-(i\nu_q)^2/4} 
\big[n_B(k)-n_F(k)\big] \Bigg\},
\label{Eq:1loop_xi_final}
\eea
where $E_k^2 \equiv k^2 + M^2$ and $\mu_\pm=((i\nu_q)^2\pm M^2)/(2 i\nu_q)$.
We have omitted from the square bracket terms independent of $M^2$ which give
a vanishing contribution upon differentiation. 

The last integral in (\ref{Eq:1loop_xi_final}) is the same one
obtained in the Feynman gauge calculation, {\it c.f.} (\ref{Eq:sigma-1loop_final}).
After analytical continuation ($i\nu_q\to \omega+i 0^+$), the $\xi$-dependent 
real part of the self-energy comes only from this term. From (\ref{softNLO1})
it also contributes to the imaginary part. The first integral of 
(\ref{Eq:1loop_xi_final}) is of the form given in (\ref{Ipm}) with the 
replacement $i a\to \mu_-/T$. Using (\ref{Eq:MitraF}) one can readily 
check that the term containing this integral will give a contribution 
which is subleading to (\ref{Eq:1loop_xi_RI}). The second integral in 
(\ref{Eq:1loop_xi_final}) contributes to the imaginary part of 
$^{\xi}\Sigma_1$ as shown by (\ref{Eq:int_in_1-loop_xi}). 
All these $\xi$ dependent terms arise from the soft momentum region. 
The final result for the high temperature expansion of the $\xi$ dependent 
part of the one-loop fermion self-energy is 
\be
^{\xi}\Sigma_1(\omega,{\bf 0})= (1-\xi)\frac{e^2}{8\pi^2}\gamma_0
\left(- \omega \ln\frac{T}{\omega} + \frac{3\pi i T}{2} \right).
\label{Eq:1loop_xi_RI}
\ee
Unlike the LO HTL result, the NLO terms and therefore the imaginary part 
of the one-loop self-energy for a fermion at rest in the plasma
is gauge-parameter dependent, even after the LO HTL on-shell condition
(\ref{Eq:LO_dispersion}) has been utilized. This clearly indicates
that the imaginary part obtained from the NLO one-loop self-energy 
calculation alone cannot be the full answer for the dressed quasi-particle
damping rate, and both the real and imaginary parts of $\Sigma$ should 
receive additional contributions of the same magnitude in the weak coupling
$e$ at higher loops, in order to obtain a physical result independent
of $\xi$. We show next that terms of the same or larger magnitude to
(\ref{Eq:1loop_xi_RI}) are indeed generated at two-loop order, for 
$\omega \sim eT$ due to increased sensitivity to the infrared of the 
two-loop self-energy.

%------@#$---------- TWO LOOP SIGMA  -----------------

\section{The Two-Loop Electron Self-Energy
\label{sec:2loop}}

In this section we compute the two-loop electron self-energy in the Feynman
gauge $\xi =1$, perform the Matsubara sums in the imaginary time formalism
by Gaudin's method, and reduce the expressions to integrals over two spatial 
loop momenta $\k$ and $\p$, for general external electron energy and momentum
$(i\nu_q, \q)$. The leading terms in the high temperature expansion arise from 
only those terms with two statistical factors, $n_B$ or $n_F$. The analysis 
of the integrals, the explicit extraction of their leading high temperature, 
most infrared sensitive terms, and the analytic continuation to real electron 
energies $i\nu_q \rightarrow \omega + i0^+$ for  $\q = 0$ electrons at rest 
in the plasma are performed in the next section.

\begin{figure}[t]
\includegraphics[keepaspectratio, width=0.8\textwidth,angle=0]{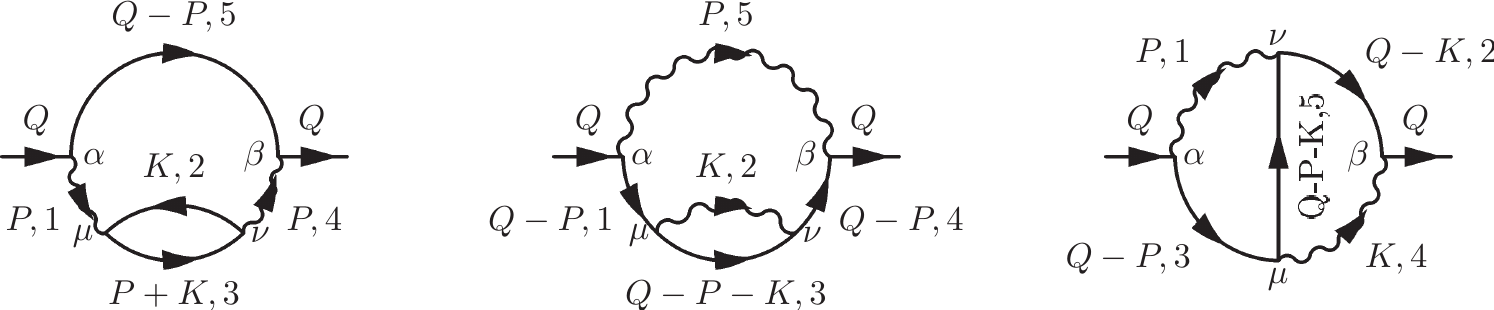}
\caption{Two-loop contributions to the fermion self-energy: from left to right, 
the bubble (B) diagram, rainbow (R) diagram, and the crossed photon (C) 
diagram.}
\label{Fig:diagrams}
\end{figure}

At two-loops the electron self-energy is composed of the three diagrams pictured 
in Fig.~\ref{Fig:diagrams}, which we call the bubble (B) diagram, the rainbow (R) 
diagram, and the crossed photon (C) diagram. The corresponding contributions 
to $\Sigma$ will be denoted by  $\Sigma_\textrm{\tiny B}$, $\Sigma_\textrm{\tiny R}$ 
and $\Sigma_\textrm{\tiny C}$ respectively.

An earlier partial calculation of the two-loop electron self-energy at finite temperatures 
has been given the literature in Ref. \cite{qader92}. The authors of this work were 
interested in the ultraviolet renormalizability of QED and electron mass renormalization 
at finite temperatures. For that reason they used real time Feynman propagators
(rather than the appropriate causal real time Schwinger-Keldysh formalism), and did not
calculate the imaginary parts of any of the two-loop diagrams. They also did
not calculate the contribution of the bubble diagram (B), which turns out to give 
the dominant contribution to the real part at leading order in the high 
temperature expansion, in which we are interested. 

The calculation of such multi-loop diagrams in the imaginary time
formalism is greatly facilitated by Gaudin's method for performing 
the Matsubara sums \cite{gaudin65}.  Since each of the two-loop
self-energy diagrams contains five finite temperature propagators,
finding the appropriate partial fraction identity analogous to
(\ref{Gaudexam}) for disentangling the two independent Matsubara
frequencies is already quite non-trivial. Gaudin's method for doing
this involves a graphical decomposition of each original two-loop 
Feynman diagram in $\Sigma$ into a sum of tree diagrams, each 
containing three propagators, and the complement of the tree 
containing the other two propagators which carry the two 
independent Matsubara frequencies. This graphical decomposition 
is illustrated in Fig.~\ref{Fig:b_tree}, for the case of the bubble (B) 
self-energy diagram.  Calculations in scalar field theory using Gaudin's 
method were presented in \cite{reinosa06} on a two-loop diagram 
having the topology of diagram (C). Diagrams of type (R) or (B) were 
outside the interest of the authors of \cite{reinosa06} because, having 
a self-energy insertion on an internal line, they are not in the class
of two-particle irreducible (2PI) diagrams. The method works equally
well on diagrams of all topologies.  However, since two fermion
propagators (diagram R) or two photon propagators (diagram B) 
have the same four-momentum, the two diagrams suffer from a double 
pole problem (in any method). In Gaudin's method this problem manifest 
itself in two different ways, as one sees from the Gaudin tree graphs 
of Fig.~\ref{Fig:b_tree}.  The first way is illustrated by the first four
graphs, where one encounters the product of a spectral function
corresponding to a line which acquires a statistical factor and a
Matsubara propagator with the same four-momentum, that is $\delta
(p_0^2-\p^2)/(\p^2-p_0^2)$. The second way in which the double pole
problem arises is found in the last three Gaudin tree graphs, where
one encounters the product of two Matsubara propagators sharing the
same 4-momentum.  In both cases the expressions can be regularized 
by a massive regulator, which effectively defines ill-defined double pole
quantities by the replacements
\begin{subequations}
\bea
&&\rho(p^0,\p) \Delta(p^0,\p) \rightarrow \frac{1}{2} 
\frac{\partial}{\partial M^2} \rho_{M^2} (p^0,\p) \equiv  
\frac{1}{2} \frac{\partial}{\partial M^2}
\bigg\{2\pi \varepsilon(p^0) \delta \Big((p^0)^2 -\p^2 - M^2\Big)\bigg\}\,,
\label{rhoDel}\\
&&\big[\Delta (p^0, \p)\big]^2 \rightarrow  -\frac{\partial}{\partial M^2} 
\Delta_{M^2} (p^0,\p)
 \equiv   -\frac{\partial}{\partial M^2} 
\left\{\frac{1}{-(p^0)^2 +\p^2 + M^2}\right\}\,.
\label{Del2}
\eea
\label{Mreg}
\end{subequations}

\vspace{-.5cm}
\noindent
The $M^2 \rightarrow 0$ limit can be examined then at the end of the calculation.
As a technical remark, $M^2$ may be considered to have an infinitesimal
negative imaginary part $M^2 \rightarrow M^2 - i 0^+$, if necessary in
any intermediate steps of the calculation.

\subsection{The Bubble Diagram 
\label{ss:bubble}}

The contribution of the bubble graph of Fig.~\ref{Fig:diagrams} reads
\be
\Sigma_\textrm{\tiny B}(Q)=-i e^4\int_K\int_P \gamma_\alpha G(Q-P)
\gamma_\beta D^{\alpha\mu}(P)
\textnormal{tr}\left[\gamma_\mu G(P+K)\gamma_\nu G(K) \right] D^{\nu\beta}(P).
\label{Eq:B_integral}
\ee
The double pole of this diagram leads both to possible infrared and collinear 
divergences (see e.g. \cite{blaizot97} and references therein), which requires
regularization. For this reason we introduce a mass $M$ for the photon
which will be kept to the end of the calculation, in order both to regulate the
divergences of the perturbative two-loopself-energy and also to estimate the 
coupling and temperature dependence of the resummed electron self-energy when 
$M$ is replaced by the Debye screening mass $\sim eT$.

\begin{figure}[!t]
\includegraphics[keepaspectratio, width=0.7\textwidth, angle=0]{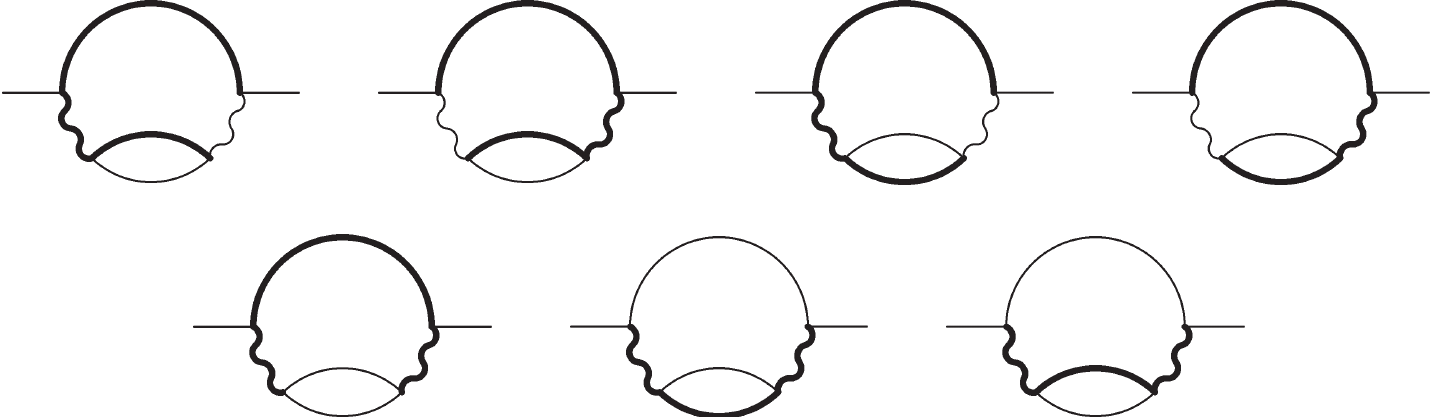}
\caption{The Gaudin tree graphs corresponding to the bubble diagram of 
Fig.~\ref{Fig:diagrams}. The thin lines carry the independent Matsubara frequencies 
while the thick lines belong to the tree (see Appendix~\ref{App:gaudins_method}). 
The assignment of the momenta of the lines are the same as in the 
corresponding bubble Feynman diagram of Fig.~\ref{Fig:diagrams}.}
\label{Fig:b_tree}
\end{figure}

In imaginary time, using the spectral representation of the
propagators, we have
\bea
\Sigma_\textrm{\tiny B}(i\nu_q,\q)&=&e^4 \int_\k\int_\p
\left(\prod_{i=1}^5\int \frac{d p_i^0}{2\pi}\right)
\gamma_{\alpha}\rho_F(P_5)\gamma_{\beta}\rho^{\alpha\mu}(P_1) \rho^{\nu\beta}(P_4)
\textnormal{tr}\left\{\gamma_\mu \rho_F(P_3)\gamma_\nu \rho_F(P_2) \right\}
\nonumber\\
&& \qquad \qquad\times\left( T^2\sum_{n,m} \prod_{i=1}^5\frac{1}{p_i^0-i\omega_i}\right),
\label{Eq:Sigma_b}
\eea
where $P_1=P_4=P,$ $P_2=K,$ $P_3=P+K,$ and $P_5=Q-P$ with
$Q=(i\nu_q,\q),$ $P=(i\omega_n,\p),$ and $K=(i\omega_m,\k)$.
The Matsubara sums performed with Gaudin's method gives for the last factor of
(\ref{Eq:Sigma_b}),
\bea
\nonumber
&&\hspace{-.5cm}T^2\sum_{n,m} \prod_{i=1}^5\frac{1}{p_i^0-i\omega_i}=
T^2\sum_{n,m}\left(\frac{1}{p_1^0 - i \omega_n}\right)\left(\frac{1}{p_2^0 -i\omega_m}\right)
\left(\frac{1}{p_3^0 - i \omega_m -i\omega_n}\right) \left(\frac{1}{p_4^0 -i\omega_n}\right)
\left(\frac{1}{p_5^0-i\nu_q + i\omega_n}\right)\\
\nonumber
&& \qquad = \frac{1}{(p_1^0-p_4^0)(p_5^0+p_4^0-i\nu_q)}
\left[\frac{n_F(p_3^0) n_B(-p_4^0)}{p_2^0+p_4^0-p_3^0}\right]
+\frac{1}{(p_4^0-p_1^0)(p_5^0+p_1^0-i\nu_q)}
\left[\frac{n_B(-p_1^0)n_F(p_3^0)}{p_2^0+p_1^0-p_3^0}\right]\\
\nonumber
&& \qquad \quad-\ \frac{1}{(p_1^0-p_4^0)(p_5^0+p_4^0-i\nu_q)}
\left[\frac{n_F(p_2^0) n_B(p_4^0)}{p_3^0-p_2^0-p_4^0}\right]
-\frac{1}{(p_4^0-p_1^0)(p_5^0+p_1^0-i\nu_q)}
\left[\frac{n_B(p_1^0)n_F(p_2^0)}{p_3^0-p_1^0-p_2^0}\right]\\
\nonumber
&&\qquad\qquad+\ \frac{n_F(p_2^0)n_F(p_3^0)}
{(p_1^0+p_2^0-p_3^0)(p_4^0+p_2^0-p_3^0)(p_5^0+p_3^0-p_2^0-i\nu_q)}\\
&&\qquad\quad+\ \frac{1}{(p_1^0+p_5^0-i\nu_q)(p_4^0+p_5^0-i\nu_q)(p_2^0-p_3^0-p_5^0+i\nu_q)}
\Big[n_F(p_2^0) n_F(-p_5^0) + n_F(p_3^0)n_F(p_5^0)\Big]\,,
\label{Eq:uj_Sigma_b_sum}
\eea
where each of the seven terms corresponds respectively to each 
of the seven Gaudin tree graphs shown in Fig.~\ref{Fig:b_tree}.

For each of these terms two of the lines (the thin lines in
Fig.~\ref{Fig:b_tree}) have associated statistical factors $n_B$ or
$n_F$, while the remaining three lines (the thick lines in
Fig.~\ref{Fig:b_tree}) have no statistical factors associated with
them. The integrations over the $p_i^0$ for these latter three
propagators can be performed simply by undoing the substitution of the
spectral representation (\ref{Eq:rho_photon}) and (\ref{Eq:rho_fermion}) 
in favor of the original imaginary time propagators, $\Delta^{\mu\nu}$ 
or $\Delta_F$. Then the first and second of the seven terms give 
identical contributions with $p_1^0$ replaced by $p_4^0$, as do the 
third and fourth terms with the same replacement. Dropping the $^0$ 
superscript on the remaining integration variables $p_i^0$ to 
simplify the notation, we obtain then from (\ref{Eq:Sigma_b}) and
(\ref{Eq:uj_Sigma_b_sum})
\bea
&&\Sigma_\textrm{\tiny B}(i\nu_q,\q)=e^4 \int_\k\int_\p 
\left\{2\int\frac{d p_3}{2\pi} \int\frac{d p_4}{2\pi}
n_F(p_3) n_B(-p_4)\times\right.\nonumber\\
&&\qquad  \gamma_{\alpha} \Delta_F(i\nu_q - p_4, \q - \p)
\gamma_{\beta} \Delta^{\alpha\mu}(p_4, \p) \rho^{\nu\beta}(p_4, \p)
\textnormal{tr}\left\{\gamma_{\mu}\rho_F(p_3, \p + \k) \gamma_{\nu}
\Delta_F(p_3 - p_4, \k)\right\}\nonumber\\
&&- 2\int\frac{d p_2}{2\pi} \int\frac{d p_4}{2\pi}
n_F(p_2) n_B(p_4) \times\nonumber\\
&& \qquad \gamma_{\alpha} \Delta_F(i\nu_q - p_4, \q - \p)
\gamma_{\beta} \Delta^{\alpha\mu}(p_4, \p) \rho^{\nu\beta}(p_4, \p)
\textnormal{tr}\left\{\gamma_{\mu}\Delta_F(p_2 + p_4, \p + \k) \gamma_{\nu}
\rho_F(p_2, \k)\right\}\nonumber\\
&&+ \int\frac{d p_2}{2\pi} \int\frac{d p_3}{2\pi}
n_F(p_2)n_F(p_3)\times\nonumber\\
&&\qquad \gamma_{\alpha} \Delta_F(i\nu_q + p_2 - p_3, \q - \p)
\gamma_{\beta} \Delta^{\alpha\mu}(p_3 - p_2, \p) \Delta^{\nu\beta}(p_3 - p_2, \p)
\textnormal{tr}\left\{\gamma_{\mu}\rho_F(p_3, \p + \k) \gamma_{\nu}
\rho_F(p_2, \k)\right\}\nonumber\\
&& -\int\frac{d p_2}{2\pi} \int\frac{d p_5}{2\pi}
n_F(p_2)n_F(-p_5)\times\nonumber\\
&& \qquad \gamma_{\alpha} \rho_F(p_5, \q - \p)
\gamma_{\beta} \Delta^{\alpha\mu}(i\nu_q - p_5, \p) \Delta^{\nu\beta}(i\nu_q-p_5, \p)
\textnormal{tr}\left\{\gamma_{\mu}\Delta_F(i\nu_q +p_2 -p_5, \p + \k) \gamma_{\nu}
\rho_F(p_2, \bf k)\right\}\nonumber\\
&&+ \int\frac{d p_3}{2\pi} \int\frac{d p_5}{2\pi}
n_F(p_3)n_F(p_5) \times\nonumber\\
&& \left. \gamma_{\alpha} \rho_F(p_5, \q - \p)
\gamma_{\beta} \Delta^{\alpha\mu}(i\nu_q - p_5, \p) \Delta^{\nu\beta}(i\nu_q-p_5, \p)
\textnormal{tr}\left\{\gamma_{\mu}\rho_F(p_3, \p + \k) \gamma_{\nu}
\Delta_F(p_3+p_5-i\nu_q, \k)\right\}\right\}.
 \label{SigB1}
\eea
Expressing the fermionic and gauge Euclidean propagators 
$\Delta_F$ and $\Delta^{\mu\nu},$ and the spectral densities 
$\rho_F$ and $\rho^{\mu\nu}$ in terms of the bosonic forms 
$\Delta$ and $\rho$ defined in (\ref{Eq:scalar_prop_spectral}) and 
(\ref{rhoboson}), and evaluating the expression (\ref{SigB1}) 
in the $\xi = 1$ gauge, we obtain the covariant form for the bubble diagram,
\bea
&&\Sigma_\textrm{\tiny B}(i\nu_q,\q)\big|_{\xi=1}=e^4 \int_\k\int_\p 
\int_{-\infty}^{\infty}\frac{d k^0}{2\pi} 
\int_{-\infty}^{\infty}\frac{d p^0}{2\pi}\,\rho(K)\rho(P)\times\nonumber\\
&&\Big\{2 \Big[n_F(-k^0) n_B(-p^0) + n_F(k^0)n_B(p^0)\Big]  
B(K; K + P; Q-P)\Delta(P) \Delta(K+P) \Delta(Q-P)\nonumber\\
&&\quad +\ n_F(k^0) n_F(p^0) B(K; P; K+Q-P) 
\Delta^2(K-P)  \Delta(K+Q-P) \nonumber\\
&&- \Big[n_F(k^0) n_F(-p^0) + n_F(-k^0)n_F(p^0)\Big]
B(K; K+Q-P; P) \Delta^2(Q-P) \Delta(K+Q-P) \Big\}\,,
\label{SigB2}
\eea
where we have defined the factor arising from the Dirac algebra,
\be
B(P_2;P_3;P_5) \equiv B(p_2^0,\p_2;p_3^0, \p_3;p_5^0, \p_5) 
\equiv \gamma^\mu\slP_5\gamma^\nu 
\textnormal{tr}\left\{\gamma_\mu\slP_3\gamma_\nu\slP_2\right\}=
8[\slP_2 (P_3\cdot P_5) +\slP_3 (P_2\cdot P_5)] \,,
\label{Eq:Dirac_bubble}
\ee
and relabeled in the first term of (\ref{SigB1}) 
$p_3 \rightarrow - k^0, p_4 \rightarrow p^0,  \k \rightarrow - \k - \p$ ; 
in the second term of (\ref{SigB1}) $p_2 \rightarrow k^0, p_4 \rightarrow p^0$;  
in the third term of (\ref{SigB1}) $p_2 \rightarrow k^0, p_3 \rightarrow p^0, \p \rightarrow \p -\k$;
in the fourth term of (\ref{SigB1}) $p_2 \rightarrow k^0, p_5 \rightarrow p^0, \p \rightarrow -\p +\q$; and finally in the fifth term of (\ref{SigB1}), $p_3 \rightarrow -k^0, p_5 \rightarrow p^0,
 \p \rightarrow -\p +\q, \k \rightarrow -\k + \p -\q$,  in order to arrive at (\ref{SigB2}).
 
From the form of (\ref{SigB2}) we observe the double pole structure of
the bubble diagram as discussed at the beginning of this section.  The
first term of (\ref{SigB2}) within curly brackets contains the
singular product $\rho(P) \Delta(P)$, which is regulated by the
replacement (\ref{rhoDel}).  This term will turn out to give the most
singular contribution to the two-loop $\Sigma$ in this limit. The
second and third terms of (\ref{SigB2}) within the curly brackets 
involve the square of a zero temperature bosonic propagator
$\Delta^2(K-P)$ or $\Delta^2(Q-P)$, both of which are regulated by the
replacement (\ref{Del2}).

In order to perform the detailed analysis of the integrals appearing (\ref{SigB2})
and exhibit explicitly their hard-soft pattern, it turns out to be more convenient to express
(\ref{SigB2}) in a slightly different form. Leaving the first term as it is, in the second term 
of (\ref{SigB2}) we shift $P \rightarrow P + K$ ({\it i.e.} $p^0\to p^0+k^0$ and $\p\to \p+\k$), 
while in the third term of (\ref{SigB2}) we undo the shifts defined in the passage
from (\ref{SigB1}) to (\ref{SigB2}) which were necessary in order to express
(\ref{SigB1}) in the compact covariant form (\ref{SigB2}), that is, simply relabel
$p_2 = k^0, p_5 = p^0$ and $p_3 = k^0, p_5 = p^0$ in the last two terms
of (\ref{SigB1}), without changing the integration variables $\p$ and $\k$.
After introducing the $M^2$ regulator (\ref{rhoDel}) of the double pole terms,
and evaluating the $k^0$ integrals, one obtains from either (\ref{SigB1}) or (\ref{SigB2})
the three terms
\be
\Sigma_\textrm{\tiny B}(i\nu_q,\q)\big|_{\xi=1}=
\Sigma_\textrm{\tiny B1}(i\nu_q,\q)+
\Sigma_\textrm{\tiny B2}(i\nu_q,\q)+
\Sigma_\textrm{\tiny B3}(i\nu_q,\q),
\label{SigBubble}
\ee
with
\begin{subequations}
\bea
\nonumber
\Sigma_\textrm{\tiny B1}(i\nu_q,\q)&=&e^4 \lim_{M^2\to 0}
\frac{\partial}{\partial M^2}\int_\k\int_\p\int_{-\infty}^\infty
\frac{d p^0}{2\pi}\frac{1}{2k} \Delta(i\nu_q-p^0,\q - \p)\rho_{M^2}(p^0,\p)
\sum_{r=\pm 1} r \Bigg\{ \Delta(r k+p^0,\k+\p)
\\
&&
\qquad
\Big[n_F(-r k) n_B(-p^0) + n_F(r k)n_B(p^0)\Big]  
B(r k,\k; r k+p^0, \k + \p ;i\nu_q-p^0,\q-\p)\Bigg\},
\label{SigBubble1}
\\
\Sigma_\textrm{\tiny B2}(i\nu_q,\q)&=&-e^4 \lim_{M^2\to 0}
\frac{\partial}{\partial M^2}\int_\k\int_\p\int_{-\infty}^\infty
\frac{d p^0}{2\pi}\frac{1}{2k} \Delta(i\nu_q-p^0,\q-\p)\Delta_{M^2}(p^0,\p)
\sum_{r=\pm 1} r \Bigg\{ \rho(p^0 + r k,\p +\k)
\nonumber
\\
&&
\qquad
n_F(r k) n_F(p^0+rk) B(r k,\k; p^0 + r k, \p + \k;i\nu_q-p^0,\q-\p)\Bigg\},
\label{SigBubble2}
\\
\nonumber
\Sigma_\textrm{\tiny B3}(i\nu_q,\q)&=&e^4 \lim_{M^2\to 0}
\frac{\partial}{\partial M^2}\int_\k\int_\p\int_{-\infty}^\infty
\frac{d p^0}{2\pi}
\rho(p^0,\q-\p) \Delta_{M^2}(i\nu_q-p^0,\p)\sum_{r=\pm 1} r \Bigg\{
\\
&&
\frac{n_F(r k)}{2 k}\, n_F(-p^0) 
B(r k,\k;i\nu_q+r k-p^0,\p + \k ;p^0,\q-\p) \Delta(i\nu_q+r k-p^0,\p + \k)
\nonumber\\
&&
-\frac{n_F(r |\p + \k|)}{2 |\p + \k|}\, n_F(p_0) 
B(r k + p^0-i\nu_q,\k;r k,\p + \k;p^0,\q-\p) \Delta(r k+p^0-i\nu_q,\k)
\Bigg\},
\label{SigBubble3}
\eea
\label{SigBubble123}
\end{subequations}

\vspace{-.5cm}
\noindent
where $k \equiv  |\k|, p \equiv |\p|$. We will evaluate the leading order terms 
in the high temperature expansion of (\ref{SigBubble123}) for $\bf q =0$ in the next 
section.

\subsection{The Rainbow Diagram
\label{ss:double-rainbow}}

The contribution of the rainbow (R) graph of Fig.~\ref{Fig:diagrams}
is given by
\be
\Sigma_\textrm{\tiny R}(Q)=i e^4\int_K\int_P
\gamma_\alpha G(P_1)\gamma_\mu G(P_3)\gamma_\nu
D^{\nu\mu}(P_2) G(P_4)\gamma_\beta D^{\beta\alpha}(P_5),
\label{Eq:rainbow}
\ee
where we have introduced the labels, $P_i$ for the five different four-momenta,
$P_1=P_4 =Q-P,$ $P_2=K,$ $P_3=Q-P-K$, and $P_5=P$, with 
$Q=(i\nu_q,\q),$ $P=(i\omega_n,\p),$ and $K=(i\omega_m,\k)$, as in the
previous B diagram.

It is possible to evaluate this rainbow diagram in a completely analogous manner
as the bubble diagram. Since the topology of the two graphs is the same, one
will obtain seven tree diagrams by the Gaudin method for the rainbow self-energy,
analogous to the seven tree diagrams of Fig.~\ref{Fig:b_tree}. In the case of the
rainbow diagram the differing Dirac structure allows for some simplification. In
the Feynman gauge $\xi = 1$, we have
\be
\gamma_\alpha\slP_1\gamma_\mu\slP_3\gamma^\mu\slP_4\gamma^\alpha=
4\slP_4\slP_3\slP_1=
4\left\{
\left[P_1^2+P_3^2-(P_1-P_3)^2\right]\slP_4-P_1^2\slP_3\right\}\,,
\ee
since $P_1=P_4$. Using also that $P_1-P_3=P_2$ we have
\bea
\frac{\gamma_\alpha\slP_1\gamma_\mu\slP_3\gamma^\mu\slP_4\gamma^\alpha}
{P_1^2 P_2^2 P_3^2 P_4^2 P_5^2}=
4\left[\frac{\slP_4}{P_2^2P_3^2P_4^2P_5^2}-
\frac{\slP_3}{P_2^2P_3^2P_4^2P_5^2}\right]+
4\frac{\slP_4}{(P_4^2)^2}\frac{1}{P_5^2}\left(\frac{1}{P_2^2}-
\frac{1}{P_3^2}\right).
\label{Eq:Rainbow-decomposition}
\eea
As in the bubble diagram there is a potential double pole problem which shows up
only in the second set of terms in (\ref{Eq:Rainbow-decomposition}), and can most 
conveniently handled by introducing a fermion mass $m$, and making
the replacement
\be
\frac{i\slP_4}{(P_4^2)^2} \rightarrow \frac{\partial}{\partial  m^2}
\left(\frac{i\slP_4}{P_4^2- m^2}\right) \equiv
\frac{\partial G_{m^2}(P_4)}{\partial m^2}\,,
\label{Eq:handle_dp_in_rainbow}
\ee
and taking the limit $m^2 \rightarrow 0$ at the end of the calculation.
Because this is a fermionic double pole rather than the bosonic one in
the bubble diagram, the limit will be less singular, and in fact not
give rise to divergences in the leading order of the high temperature
expansion.

The two sets of terms in (\ref{Eq:Rainbow-decomposition}), denoted by R1 and R2
give for the full rainbow two-loop self-energy the expression
\be
\Sigma_\textrm{\tiny R}(Q)\big|_{\xi=1}=\Sigma_\textrm{\tiny R1}(Q)+
\Sigma_\textrm{\tiny R2}(Q)\,,
\ee
with
\begin{subequations}
\bea
\Sigma_\textrm{\tiny R1}(Q)&=&-4e^4\int_K\int_P\left[
D(P_2) \tilde D(P_3) G(P_4) D(P_5)-D(P_2) G(P_3) \tilde D(P_4) D(P_5)
\right],
\label{Eq:rainbow_term_A_rt}
\\
\Sigma_\textrm{\tiny R2}(Q)&=&4i e^4
\left(\int_K \tilde D(K)-\int_K D(K)\right)\ 
\frac{\partial }{\partial m^2}\int_P G_{m^2}(P) D(Q-P)\bigg|_{m^2=0}.
\label{Eq:rainbow_term_B_rt}
\eea
\label{rain}
\end{subequations}
\vspace{-.6cm}

\noindent
To obtain the second R2 term we have used for the last term inside the 
parentheses of (\ref{Eq:Rainbow-decomposition}) the shift $K\to Q-P-K$ and
then for both terms within the parentheses the shift $P\to Q-P$. 
The advantage of
the form (\ref{rain}) is that whereas the original expression
(\ref{Eq:rainbow}) has five propagators, the R1 terms have only 
four propagators, and the R2 terms factorize into the product of a 
difference of two tadpoles and a derivative of a one-loop self-energy with
only two propagators. In these expressions it is important to keep track 
of which momenta are bosonic and which are fermionic. The latter propagators 
are denoted with a tilde. 

\begin{figure}[!t]
\includegraphics[keepaspectratio,width=0.98\textwidth,angle=0]
{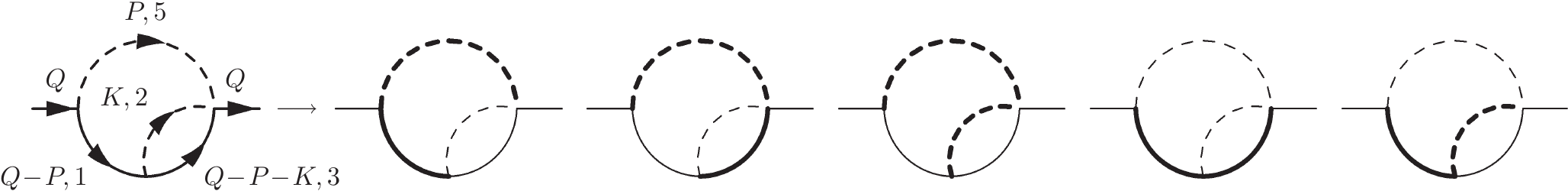}
\vspace*{-0.5cm}
\caption{The diagrammatic representation of the integrals in
(\ref{Eq:rainbow_term_A_rt}) obtained by reducing the original
rainbow diagram and its decomposition in Gaudin tree graphs. 
The dashed (solid) line means bosonic (fermionic) propagator. 
As in Fig.~\ref{Fig:b_tree} the two thin lines carry the two independent 
Matsubara frequencies to be summed, while the thick lines belong to the 
Gaudin tree and represent propagators that contain the time component 
variable of the appropriate momenta $p_i^0$ in place of the Matsubara 
frequencies.}
\label{Fig:reduced_rainbow}
\end{figure}

In imaginary time, using the spectral representation of the propagators, we have
for the two contributions
\begin{subequations}
\bea
&&
\Sigma_\textrm{\tiny R1}(i\nu_q,\q)=4 e^4\int_\k\int_\p 
\left(\prod_{i=2}^{5}\int\frac{d p_i^0}{2\pi}\right)\!
\big[\rho(p_3^0,\p_3)\rho_F(p_4^0,\p_4)-\rho_F(p_3^0,\p_3) \rho(p_4^0,\p_4)\big]
 \times \nonumber\\
&&
\qquad\qquad\rho(p_2^0,\p_2) \rho(p_5^0,\p_5)
\left(T^2\sum_{n,m}\prod_{i=2}^{5} \frac{1}{p_i^0-i\omega_i}\right),
\label{Eq:rainbow_term_A}\\
&&
\Sigma_\textrm{\tiny R2}(i\nu_q,\q)=
4e^4T\int_\k \left(\sum_m\tilde\Delta(i\omega_m,\k)-
\sum_m\Delta(i\omega_m,\k) \right) \times
\nonumber\\
&&
\quad
\frac{\partial}{\partial m^2} \int_\p
\int_{-\infty}^\infty\frac{d p^0}{2\pi}
\int_{-\infty}^\infty\frac{d k^0}{2\pi}
\rho_F(p^0,E_\p) \rho(k^0,\q-\p) T\sum_n\frac{1}{(p^0-i\omega_n)
(k^0-i(\nu_q-\omega_n))}\bigg|_{ m^2=0}\,. \ \ 
\label{Eq:rainbow_term_B}
\eea
\end{subequations}
\vspace{-.6cm}

\noindent
The Matsubara frequency sums in the R1 terms may now be performed by
Gaudin's method. The Gaudin relevant tree diagrams represented diagrammatically 
in Fig.~\ref{Fig:reduced_rainbow}. Having one propagator less
compared to the original graph, it is somewhat easier to compute the sums,
and we obtain
\bea
\nonumber
T^2\sum_{n,m} \prod_{i=2}^{4}\frac{1}{p_i^0-i\omega_i}&=&
-\frac{n_B(-p_2^0)n_F(-p_3^0)}
{(p_4^0-p_2^0-p_3^0)(p_5^0+p_2^0+p_3^0-i\nu_q)}-
\frac{n_F(-p_4^0)n_B(-p_2^0)}
{(p_3^0-p_4^0+p_2^0)(p_5^0+p_4^0-i\nu_q)}\\
\nonumber
&&-\frac{n_F(-p_4^0)n_F(p_3^0)}
{(p_2^0-p_4^0+p_3^0)(p_5^0+p_4^0-i\nu_q)}-
\frac{n_B(-p_2^0)n_B(p_5^0)}
{(p_4^0+p_5^0-i\nu_q)(p_3^0+p_2^0+p_5^0-i\nu_q)}\\
&&-\frac{n_F(p_3^0)n_B(p_5^0)}
{(p_4^0+p_5^0-i\nu_q)(p_2^0+p_3^0+p_5^0-i\nu_q)}\, ,
\label{R1sums}
\eea
with each term corresponding to a Gaudin tree graph of 
Fig.~\ref{Fig:reduced_rainbow}. In (\ref{Eq:rainbow_term_B}), 
$\omega_n$ is fermionic and $\nu_q-\omega_n$ is bosonic, and we obtain
\be
T\sum_n\frac{1}{(p^0-i\omega_n)(r^0-i(\nu_q-\omega_n))}
=\frac{n_F(-p^0)+n_B(r^0)}{p^0+r^0-i\nu_q}
\label{Eq:Matsubara_term_B}
\ee
for the Matsubara sum.

Substituting (\ref{R1sums}) into (\ref{Eq:rainbow_term_A}), the integrations over
the two $p_i^0$ which do not appear in the statistical distributions can be
performed simply by using the definition (\ref{Eq:scalar_prop_spectral}) of
the spectral representation of the Euclidean propagator function $\Delta$.
By relabeling $p_2^0 \rightarrow k^0, p_3^0 \rightarrow p^0, 
\p \rightarrow -\p +\q -\k$ in the first of the five terms arising from 
(\ref{R1sums}), $p_2^0 \rightarrow k^0, p_4^0 \rightarrow p^0, 
\p \rightarrow -\p +\q$ in the second term, 
$p_3^0 \rightarrow k^0, p_4^0 \rightarrow p^0, \k \rightarrow -\k +\p, 
\p \rightarrow -\p +\q$ in the third term, $p_2^0 \rightarrow k^0, 
p_5^0 \rightarrow p^0$ in the fourth term,
and $p_3^0 \rightarrow k^0, p_5^0 \rightarrow p^0, \k \rightarrow -\k +\q - \p$ in
the fifth term arising from (\ref{R1sums}), we arrive at the covariant form,
\bea
&&\Sigma_\textrm{\tiny R1}(i\nu_q,\q)=-4 e^4\int_\k\int_\p \int_{-\infty}^{\infty} \frac{dk^0}{2\pi}
\int_{-\infty}^{\infty} \frac{dp^0}{2\pi} 
\bigg\{\slK  n_B(-k^0)n_F(-p^0) \big[\Delta(K+P) \Delta (Q-K- P)\nonumber\\
&& + \ \Delta(P-K) \Delta (Q-P)\big] 
+ (\slP - \slK) n_F(k^0) n_F(-p^0) \Delta(P-K) \Delta(Q-P) \nonumber\\
&&  +\ n_B(p^0) \Delta(Q-P) \Delta(Q-P-K) \left[\slP n_B(-k^0) +(\slQ -\slP -\slK)n_F(k^0)\right]
\bigg\} \rho(k^0,\k) \rho(p^0,\p)\,,
\label{SigR1}
\eea
for (\ref{Eq:rainbow_term_A}).

For (\ref{Eq:rainbow_term_B}), we evaluate first
\be
T\sum_n\int_\k \left(\tilde\Delta(i\omega_n,\k)-\Delta(i\omega_n,\k) \right)=
-\frac{1}{2\pi^2}\int_0^\infty d k\,k \big[ n_B(k)+n_F(k)\big]=
-\frac{T^2}{8}\,.
\label{Eq:diff_of_tads}
\ee
Then using (\ref{Eq:Matsubara_term_B}), we obtain
\be
\Sigma_\textrm{\tiny R2}(i\nu_q,\q)=- \frac{e^4T^2}{2}
\frac{\partial}{\partial m^2}
\int_\p \int_{-\infty}^{\infty} \frac{dp^0}{2\pi}
\Big\{\slP  n_F(-p^0) \rho_{m^2}(P) \Delta(Q-P) +
(\slQ - \slP) n_B(p^0) \rho (P) \Delta_{m^2}(Q-P)\Big\}\bigg|_{m^2=0}\,,
\label{SigR2}
\ee
where the substitutions $r^0 \rightarrow p^0$ and $\p \rightarrow \q - \p$ have
been made in the second term to arrive at (\ref{SigR2}).

Finally one may perform the $k^0$ and $p^0$ integrations in (\ref{SigR1}) 
and the $p^0$ integration in (\ref{SigR2}) by using the $\delta$ functions 
in the spectral representations to obtain
\bea
&&\Sigma_\textrm{\tiny R1}(i\nu_q,\q)=- e^4\int_\k\int_\p \sum_{r,s = \pm 1}
\bigg\{ \frac{n_B(k) n_F(p)}{kp}(kr\gamma_0 - \k \cdot {\bm \gamma})\times\nonumber\\
&& \qquad\qquad\big[\Delta(kr +ps, \k + \p) \Delta(i\nu_q -kr-ps, \q - \k -\p)
 + \Delta(ps - kr, \p -\k) \Delta(i\nu_q - ps, \q -\p) \big] \nonumber\\
&&\qquad + \frac{n_F(k) n_F(p)}{kp} \big[ (ps-kr)\gamma_0 - (\p-\k)\cdot{\bm \gamma}\big]
\Delta(ps-kr, \p- \k) \Delta(i\nu_q - ps, \q - \p)
\nonumber\\
&&\qquad + \ \Big[\frac{n_B(k)n_B(p)}{kp} (ps\gamma_0 - \p\cdot{\bm \gamma})
+\ \frac{n_F(k)n_B(p)}{kp} \big[ (i\nu_q -ps-kr)\gamma_0
+ (\q - \p - \k)\cdot{\bm \gamma}\big]\Big] \times\nonumber\\
&& \qquad\qquad\Delta(i\nu_q - ps, \q - \p) 
\Delta (i\nu_q -ps -kr, \q - \p - \k) \bigg\} + \ \dots
\label{SigR1f}
\eea
and
\bea
\Sigma_\textrm{\tiny R2}(i\nu_q,\q)&=&-\frac{e^4 T^2}{4}
\frac{\partial }{\partial m^2}\int_{\p}\sum_{s=\pm 1}
\Bigg\{ \Big[(i\nu_q-s p)\gamma_0-(\q-\p)\cdot{\bm \gamma}\Big]
\Delta_{m^2}(i\nu_q-s p,\q-\p)\frac{n_B(p)}{p}
\nonumber
\\&& 
\qquad\qquad-(s E_p\gamma_0-\p\cdot{\bm\gamma}) \Delta (i\nu_q - s E_p,\q-\p) 
\frac{n_F(E_p)}{E_p}
\Bigg\}\Bigg|_{m^2=0} + \ \dots\,,
\label{Eq:R_B_2stat}
\eea
where the ellipsis in each case denotes the terms with fewer than two
statistical factors, which are subleading in the high temperature limit.
We again postpone to the next section the detailed evaluation 
of (\ref{SigR1f}) and (\ref{Eq:R_B_2stat}) 
in Feynman gauge for a fermion at rest.

\subsection{The Crossed Photon Diagram 
\label{ss:crossed-rainbow}}

\begin{figure}[t]
\includegraphics[keepaspectratio, width=0.7\textwidth, angle=0]
{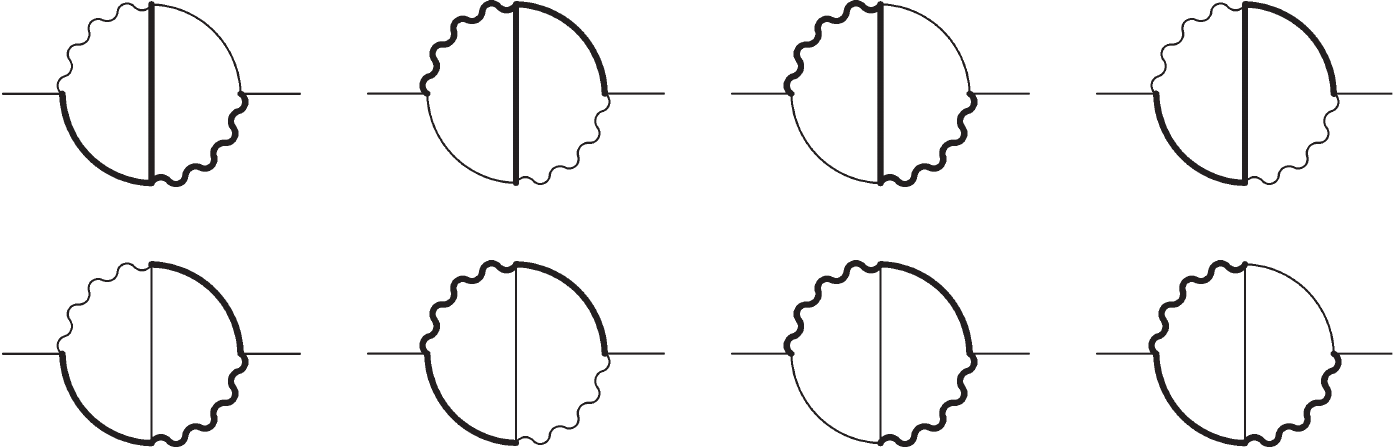}
\caption{
The Gaudin tree graphs corresponding to the crossed photon diagram of 
Fig.~\ref{Fig:diagrams}. A tree is represented by thick lines. 
As in Figs.~\ref{Fig:b_tree} and \ref{Fig:reduced_rainbow} the two thin lines carry the two independent Matsubara frequencies to be summed, while the thick lines belong to the Gaudin tree and contain no Matsubara frequencies.}
\label{Fig:cr_tree}
\end{figure}

The third and final two-loop diagram is the crossed photon diagram of 
Fig.~\ref{Fig:diagrams}, which topologically has the form of a vertex 
correction. The contribution of this crossed diagram to the two-loop
electron self-energy is
\be
\Sigma_\textrm{\tiny C}(Q)=i e^4\int_K\int_P
\gamma_\alpha G(Q-P)\gamma_\mu G(Q-P-K)\gamma_\nu
D^{\nu\alpha}(P) G(Q-K)\gamma_\beta D^{\beta\mu}(K).
\label{Eq:CR_integral}
\ee
In imaginary time, using the spectral representation of the
propagators, this becomes
\be
\Sigma_\textrm{\tiny C}(i\nu_q,\q)=-e^4 \int_\k\int_\p
\left(\prod_{i=1}^5\int \frac{d p_i^0}{2\pi}\right)
\gamma_\alpha\rho_F(P_3) \gamma_\mu
\rho_F(P_5)\gamma_\nu \rho^{\nu\alpha}(P_1)
\rho_F(P_2)\gamma_\beta  \rho^{\beta\mu}(P_4)
\left( 
T^2\sum_{n,m} \prod_{i=1}^5\frac{1}{p_i^0-i\omega_i}\right),
\label{Eq:Sigma_cr}
\ee
where $P_1=P,$ $P_2=Q-K,$ $P_3=Q-P,$ $P_4=K$ and $P_5=Q-P-K$ with
again $Q=(i\nu_q,\q),$ $P=(i\omega_n,\p),$ and $K=(i\omega_m,\k)$. The
Matsubara sums performed with Gaudin's method gives for the last factor of
(\ref{Eq:Sigma_cr}),
\bea
\nonumber
&&T^2\sum_{n,m} \prod_{i=1}^5\frac{1}{p_i^0-i\omega_i}=\\&&
\frac{1}{(p_1^0+p_3^0-i\nu_q)(p_2^0+p_4^0-i\nu_q)}\left[
\frac{n_B(-p_1^0)n_F(p_2^0)}{p_5^0+p_1^0-p_2^0}
+\frac{n_B(-p_4^0) n_F(p_3^0)}{p_5^0+p_4^0-p_3^0}\right]
\nonumber\\
&&+\frac{1}{(p_1^0+p_3^0-i\nu_q)(p_2^0+p_4^0-i\nu_q)}
\left[\frac{n_F(p_2^0)n_F(p_3^0)}{p_5^0-p_2^0-p_3^0+i\nu_q}
+\frac{n_B(-p_1^0)n_B(-p_4^0)}{p_5^0+p_1^0+p_4^0-i\nu_q}\right]\nonumber\\
&&+\frac{1}{p_1^0+p_4^0+p_5^0-i\nu_q}\left[
\frac{n_B(-p_1^0)n_F(p_5^0)}
{(p_2^0-p_1^0-p_5^0)(p_3^0+p_1^0-i\nu_q)}
-\frac{n_B(p_4^0)n_F(p_5^0)}{(p_2^0+p_4^0-i\nu_q)(p_3^0-p_4^0-p_5^0)}
\right]
\nonumber
\\
&&+\frac{1}{p_2^0+p_3^0-p_5^0-i\nu_q}
\left[\frac{n_F(p_3^0)n_F(p_5^0)}
{(p_4^0+p_5^0-p_3^0)(p_1^0+p_3^0-i\nu_q)}
-\frac{n_F(-p_2^0)n_F(p_5^0)}
{(p_1^0+p_5^0-p_2^0)(p_4^0+p_2^0-i\nu_q)}
\right],
\label{Eq:CR_Matsubara}
\eea
where the corresponding eight Gaudin tree graphs are depicted in
Fig.~\ref{Fig:cr_tree}. 

Substituting (\ref{Eq:CR_Matsubara}) into (\ref{Eq:Sigma_cr}),
the integrations over the three $p_i^0$ which do not involve
the statistical factors $n_B$ or $n_F$ in each of the eight terms
can be performed using the definition, Eqs.~(\ref{Eq:rho_photon}) and 
(\ref{Eq:rho_fermion}), simply undoing the spectral decomposition in 
favor of the Matsubara propagator (\ref{Eq:Matsubara_prop}). The first and 
second of the eight terms are identical after a relabeling of momenta, while 
the fifth and sixth terms and the seventh and eight terms can also be grouped 
together. Suppressing the $^0$ superscript on the timelike integration 
variables, we obtain
\bea
&&\Sigma_\textrm{\tiny C}(i\nu_q,\q)=-e^4 \int_\k\int_\p 
\left\{2\int\frac{d p_1}{2\pi} \int\frac{d p_2}{2\pi}
n_B(-p_1) n_F(p_2)\right. \times\nonumber\\
&& \qquad\gamma_{\alpha} \Delta_F(i\nu_q - p_1, \q - \p)
\gamma_{\mu} \Delta_F(p_2 - p_1, \q - \p - \k)\gamma_{\nu}
\rho^{\nu\alpha}(p_1, \p)\rho_F(p_2, \q - \k) \gamma_{\beta}
\Delta^{\beta\mu}(i\nu_q - p_2, \k) \nonumber\\
&&\qquad + \int\frac{d p_2}{2\pi} \int\frac{d p_3}{2\pi}
n_F(p_2) n_F(p_3) \times\nonumber\\
&&\qquad \gamma_{\alpha} \rho_F(p_3, \q - \p)
\gamma_{\mu}\Delta_F(i\nu_q - p_2 - p_3, \q - \p - \k)\gamma_{\nu}
\Delta^{\nu\alpha}(i\nu_q - p_3, \p) \rho_F(p_2, \q - \k)
\gamma_{\beta} \Delta^{\beta\mu}(i\nu_q - p_2, \k)
\nonumber\\
&&\qquad + \int\frac{d p_1}{2\pi} \int\frac{d p_4}{2\pi}
n_B(-p_1) n_B(-p_4)\times\nonumber\\
&&\qquad  \gamma_{\alpha} \Delta_F(i\nu_q - p_1, \q - \p)
\gamma_{\mu}\Delta_F(i\nu_q - p_1- p_4, \q - \p - \k)\gamma_{\nu}
\rho^{\nu\alpha}(p_1, \p)\Delta_F(i\nu_q - p_4, \k)
\gamma_{\beta} \rho^{\beta\mu}(p_4, \k)
\nonumber\\
&& \qquad +\int\frac{d p_1}{2\pi} \int\frac{d p_5}{2\pi}
n_B(-p_1) n_F(-p_5) \times\nonumber\\
&& \qquad\gamma_{\alpha} \Delta_F(i\nu_q -p_1, \q - \p)
\gamma_{\mu} \rho_F(p_5, \q - \p - \k)\gamma_{\nu}
\rho^{\nu\alpha}(p_1, \p)\Delta_F(p_1 + p_5, \q - \k)
\gamma_{\beta} \Delta^{\beta\mu}(i\nu_q - p_1 - p_5, \k) 
\nonumber\\
&& \qquad- \int\frac{d p_4}{2\pi} \int\frac{d p_5}{2\pi}
n_B(p_4)n_F(p_5) \times\nonumber\\
&&\qquad \gamma_{\alpha} \Delta_F(p_4 + p_5, \q - \p)
\gamma_{\mu}\rho_F(p_5, \q - \p - \k) \gamma_{\nu}
 \Delta^{\nu\alpha}(i\nu_q-p_4 -p_5, \p)\Delta_F(i\nu_q -p_4, \q -\k)
\gamma_{\beta} \rho^{\beta\mu}(p_4, \k)
\nonumber\\
&& \qquad +\int\frac{d p_3}{2\pi} \int\frac{d p_5}{2\pi}
n_F(p_3) n_F(p_5) \times\nonumber\\
&&\qquad\gamma_{\alpha}\rho_F(p_3, \q - \p)
\gamma_{\mu}\rho_F(p_5, \q - \p - \k) \gamma_{\nu}
\Delta^{\nu\alpha}(i\nu_q - p_3, \p) \Delta_F(i\nu_q + p_5 - p_3, \q - \k)
\gamma_{\beta}\Delta^{\beta\mu}(p_3 - p_5, \k)
\nonumber\\
&& \qquad -\int\frac{d p_2}{2\pi} \int\frac{d p_5}{2\pi}
n_F(-p_2) n_F(p_5) \times\nonumber\\
&&\qquad \left.\gamma_{\alpha}\Delta_F(i\nu_q + p_5 - p_2, \q - \p)
\gamma_{\mu}\rho_F(p_5, \q - \p - \k) \gamma_{\nu}
\Delta^{\nu\alpha}(p_2 - p_5, \p)\rho_F(p_2, \q - \k)\gamma_{\beta}
\Delta^{\beta\mu}(i\nu_q - p_2, \k)\right\}\,. \nonumber\\
&&\label{SigC1}
\eea
Expressing the fermionic and gauge Euclidean propagators 
$\Delta_F$ and $\Delta^{\mu\nu},$ and the spectral densities 
$\rho_F$ and $\rho^{\mu\nu}$ in terms of the bosonic forms 
$\Delta$ and $\rho$ 
defined in (\ref{Eq:scalar_prop_spectral}) and (\ref{rhoboson}), and evaluating the expression (\ref{SigC2}) 
in the $\xi = 1$ Feynman gauge, we get the covariant form 
for the crossed photon diagram,
\bea
&&\Sigma_\textrm{\tiny C}(i\nu_q,\q)\big|_{\xi=1}= -e^4 \int_\k\int_\p 
\int_{-\infty}^{\infty}\frac{d k^0}{2\pi} 
\int_{-\infty}^{\infty}\frac{d p^0}{2\pi}\,\rho(K)\rho(P)\times
\nonumber\\
&&\Big\{2 n_F(k^0) n_B(-p^0) C(K; Q- P; K-P)\Delta(Q-P) \Delta(Q-K) \Delta(K-P)\nonumber\\
&&\qquad +\ n_F(k^0) n_F(p^0) C(K; P; K+ P - Q) 
\Delta(Q-P) \Delta(Q-K) \Delta(K+ P - Q) \nonumber\\
&&\qquad +\ n_B(-k^0) n_B(-p^0) C(Q-K;Q- P; Q -P-K) 
\Delta(Q-P) \Delta(Q-K) \Delta(K+ P - Q) \nonumber\\
&&+\Big[n_F(k^0) n_B(-p^0) - n_F(k^0)n_B(p^0)\Big] C(K + P;Q- P; K) 
\Delta(K+ P) \Delta(Q-P) \Delta(Q -K-P)\nonumber\\
&&+\Big[n_F(k^0) n_F(p^0) - n_F(k^0) n_F(-p^0)\Big] C(Q-P+K; P; K) 
\Delta(Q-P) \Delta(Q-P+K) \Delta(K-P)\Big\}\,,\ \ \ \ \ 
\label{SigC2}
\eea
where we have defined the factor arising from the Dirac algebra,
\be
C(P_2;P_3;P_5) \equiv C(p_2^0,\p_2;p_3^0, \p_3;p_5^0, \p_5) 
\equiv \gamma_\alpha\slP_3\gamma_\mu 
\slP_5\gamma^\alpha\slP_2\gamma^\mu=
-8(P_2\cdot P_3)\slP_5\,,
\label{Cdef}
\ee
and relabeled $p_2 \rightarrow k^0, p_1 \rightarrow p^0, \k \rightarrow - \k +\q$
in the first term of (\ref{SigC1}), $p_2 \rightarrow k^0, p_3 \rightarrow p^0,
\k\rightarrow -\k + \q, \p\rightarrow \p +\q$ in the second term, 
$p_4 \rightarrow k^0, p_1 \rightarrow p^0$ in the third term, 
$p_5 \rightarrow k^0, p_1 \rightarrow p^0, \k \rightarrow -\k +\q -\p$
in the fourth term, $p_5 \rightarrow k^0, p_4 \rightarrow p^0, 
\k \rightarrow \p, \p \rightarrow -\k +\q -\p$ in the fifth term,
$p_5 \rightarrow k^0, p_3 \rightarrow p^0, \k \rightarrow -\k +\p, \p \rightarrow -\p +\q$ 
in the sixth term, and $p_5 \rightarrow k^0, p_2 \rightarrow p^0, 
\k \rightarrow -\p + \q, \p \rightarrow \p -\k$ in the seventh term of (\ref{SigC1}), 
in order to arrive at (\ref{SigC2}).

Proceeding as in the previous diagram, for each term in (\ref{SigC2})
the remaining two frequency integrals over $k^0$ and $p^0$ are performed
next, using the Dirac $\delta$ functions in the remaining two spectral functions. 
Then retaining only those terms which have two statistical distribution factors, we obtain
\bea
&&\Sigma_\textrm{\tiny C}(i\nu_q,\q)\big|_{\xi=1}= -e^4 \int_\k\int_\p\,\sum_{r,s = \pm 1} 
\left\{-\frac{n_F(k) n_B(p)}{2kp} C(kr,\k; i\nu_q - ps, \q -\p; kr-ps, \k - \p)
\times\right.\nonumber\\
&&\qquad \qquad\Delta(i\nu_q - ps; \q - \p) \Delta(i\nu_q - kr, \q - \k) \Delta(kr -ps, \k - \p)
\nonumber\\
&&+\ \frac{n_F(k) n_F(p)}{4kp} C(kr, \k; ps, \p; kr+ ps - i\nu_q, \k +\p -\q) \times\nonumber\\
&&\qquad\qquad\Delta(i\nu_q-ps, \q -\p)  \Delta(i\nu_q -kr,\q-\k) 
\Delta(kr+ps-i\nu_q, \k + \p -\q) 
\nonumber\\
&&+\ \frac{n_B(k) n_F(p)}{4kp}
C(i\nu_q -kr, \q - \k; i\nu_q - ps, \q - \p; i\nu_q -ps -kr, \q - \p - \k) 
\times\nonumber\\
&& \qquad \qquad\Delta(i\nu_q - ps, \q - \p)  \Delta(i\nu_q -kr, \q - \k) 
\Delta(kr + ps - i\nu_q, \k + \p - \q) \nonumber\\
&&- \frac{n_F(k) n_B(p)}{2kp} C(kr + ps, \k + \p ; i\nu_q - ps, \q - \p; kr, \k)\times\nonumber\\
&&\qquad\qquad \Delta(kr + ps, \k + \p)  \Delta(i\nu_q - ps, \q - \p) 
\Delta(i\nu_q - kr-ps, \q - \k -\p)\nonumber\\
&&+ \frac{n_F(k) n_F(p)}{2kp}  C(i\nu_q -ps + kr, \q - \p + \k; ps, \p; kr, \k)\times
\nonumber\\
&&\qquad \qquad  \Delta(i\nu_q - ps, \q - \p) \Delta(i\nu_q - ps + kr, \q - \p + \k) 
\Delta(kr-ps, \k - \p)\Bigg\} + \ \dots \,,
\label{SigCf}
\eea
where the ellipsis denotes terms with fewer than two statistical factors, 
which are subleading  in the high temperature expansion. We turn next to the 
explicit extraction of the leading high temperature behavior of all the 
two-loop terms in the self-energy.

%-------@#$--------- TWO LOOP ANALYSIS -----------------

\section{High Temperature Behavior of the Two-Loop Integrals
\label{sec:2loop_anal}}

In this section we analyze the integrals which result from the
evaluation of the three two-loop diagrams of the previous section,
extracting the leading order terms in the high temperature 
expansion $T \gg \nu_q, M, m$, for fermions at rest 
with respect to the plasma ({\it i.e.} $\q=0$), and perform the
analytic continuation to real fermion energies $i\nu_q
\rightarrow \omega + i 0^+$ in order to obtain the real
and imaginary parts of the self-energy $\Sigma$.
The behavior of the high temperature asymptotics for $\q=0$
is sufficient for our purposes of analyzing the systematics
of the two-loop self-energy. When $\q = 0$ only the $\gamma_0$
dependent part of the self-energy is non-zero. In this case
the integrals over spatial momentum also simplify to give
\be
\int_\k\int_\p\longrightarrow \frac{1}{8\pi^4}
\int_0^\infty d k \int_0^\infty d p\, k^2\,p^2 \int_{-1}^1 d x\, ,
\label{Eq:spatial_integral}
\ee 
where $k=|\k|,$ $p=|\p|$ and $x=\cos \theta = \k\cdot\p/(k p)$. 

\subsection{The Bubble Diagram}

We start with $\Sigma_{\textrm{\tiny B1}}(i\nu_q)$ defined by
(\ref{Eq:Dirac_bubble})-(\ref{SigBubble123}) by evaluating 
the function $B$ for $\q=0$ and performing the $p^0$ integral with 
the help of the $\delta$ function from the spectral density $\rho_{M^2}(p^0, \p)$. 
The leading temperature dependence comes from those terms
containing a product of two thermal distributions evaluated at 
{\it  positive} arguments, {\it i.e.}, substituting the identities
(\ref{Eq:BE_FD_decomp}), we may discard the zero temperature
$\theta(\pm E)$ terms and retain only the terms $n_F(|k|)$ and
$n_B(|E_p|)$ with $E_p \equiv \sqrt{\p^2 + M^2}.$ In this way one obtains from
(\ref{SigBubble1})
\bea
\nonumber
\Sigma_\textrm{\tiny B1}(i\nu_q, \q ={\bf 0})&=&
\frac{e^4}{2\pi^4}\gamma_0\lim_{M^2\to 0}
\frac{\partial}{\partial M^2} \int_0^\infty d k 
\int_0^\infty d p\int_{-1}^1 d x\, k^2 p^2\, n_F(k) \sum_{r,s=\pm 1}
\Bigg\{\\
&&\frac{n_B(E_p)}{E_p}
\frac{2(i\nu_q (k+r s E_p)-k(s E_p-r p x)-M^2)-p(r p-s E_p x)}
{[2 k(p x-r s E_p)-M^2][2i\nu_q s E_p-(i\nu_q)^2-M^2]}
\Bigg\} \,+  \dots\, ,
\label{Eq:b_trees_regularized}
\eea
where the ellipsis denotes terms with fewer than two statistical factors, 
which are subleading in the high temperature limit. 
 
To exhibit the singular behavior of (\ref{Eq:b_trees_regularized}) as the regulator
$M^2$ is removed, one may take the derivative with respect 
to $M^2$ inside the integral, and naively take the limit $M^2\to 0$. After some 
straightforward but tedious algebra one obtains in this way,
\bea
\nonumber
&&-\frac{e^4}{\pi^4}\gamma_0\int_0^\infty d k\, k\, n_F(k)
\int_{0}^\infty d p\int_{-1}^{1} d x \,\frac{1}{[(i\nu_q)^2-4 p^2]}
\bigg\{ 
\left(\frac{p^2}{i\nu_q}+\frac{i\nu_q}{x^2-1}\right)\frac{d n_B(p)}{d p}\\
&&+n_B(p)\left[
\frac{2}{x^2-1}\left(\frac{p}{i\nu_q}+\frac{i\nu_q}{p}
\left(1+\frac{2}{x^2-1}\right)\right)+
\frac{i\nu_q p}{(i\nu_q)^2-4p^2}\left(1+\frac{4}{x^2-1}\right)
\right]\bigg\}\,,
\label{Eq:coll_x_IR}
\eea which factorizes into a hard $k$ integral and a soft $p$
integral.  This factorization property will persist even when the
$M^2$ regulator is retained, for the leading order terms in the high
temperature expansion. However, in several of the terms in
(\ref{Eq:coll_x_IR}) the soft $p$ integration is infrared divergent at
$p=0$, and is multiplied by an additional collinear divergence from
the $x= \cos\theta$ integral at $x= \pm 1$. Thus the most severe type
of divergence in (\ref{Eq:coll_x_IR}) is the product of an infrared
and a collinear divergence, each of which is linear. Hence one may
expect generically a quadratic $M^{-2}$ divergence from this singular
behavior of the bubble diagram as $M \rightarrow 0$.  This is the
reason for introducing a photon mass into the calculation to regulate
the double pole singularity of the bubble diagram. Since no infrared
or collinear divergences are encountered in the one-loop self-energy,
the divergence in the two-loop bubble diagram is the clearest
indication that a systematic resummation of the perturbative series is
needed.  This HTL resummation will eventually dress the photon, giving
it a Debye mass of order $eT$, and alter the counting of powers of the
coupling constant.

Retaining the mass regulator in $\Sigma_{\textrm{\tiny B1}}$ and
performing the sum over $r, s = \pm 1$ in (\ref{Eq:b_trees_regularized})
 one finds 
\bea
\nonumber
\Sigma_\textrm{\tiny B1}(i\nu_q)&=&
\frac{e^4}{4\pi^4 i\nu_q}\gamma_0\frac{\partial}{\partial M^2}
\left\{\int_0^\infty d k\, k\, n_F(k) \int_{-1}^{1} d x \int_0^\infty d p
\, \frac{p^2}{E_p}\, n_B(E_p)\right.
\\
&&
\left. \times \frac{2(i\nu_q)^2E_p(p x-E_p)-2p E_p (p-E_p x)(p x-E_p)
+M^2(i\nu_q)^2+M^4}
{(E_p^2-\mu_+^2)[(p x -E_p)^2-M^4/(4k^2)]}\right]\Bigg|_{M^2=0},
\label{Eq:Bbb1}
\eea
where $\mu_+= [(i\nu_q)^2+M^2]/(2 i\nu_q)$. The appearance of $k\, n_F(k)$ in
the first integral shows clearly that the $k$ integration is hard, {\it i.e.} dominated
by $k \sim T$. Since $M \ll T$, the $M^4/k^2 \sim M^4/T^2 \ll M^2$ term 
in the denominator can be neglected. Then the hard $k$ integral factorizes
completely. Multiplying both the numerator and the denominator 
by $(p x +E_p)^2$, and observing that all terms odd in $x$ vanish
by symmetry, we may perform the $x$ integral exactly and obtain
\bea
\nonumber
\Sigma_\textrm{\tiny B1}(i\nu_q)&=&
\frac{e^4 T^2}{24\pi^2 i\nu_q}\gamma_0\frac{\partial}{\partial M^2}
\left\{\left[(i\nu_q)^2+M^2\right]\int_0^\infty d p \frac{p^2}{E_p}
\frac{n_B(E_p)}{E_p^2-\mu_+^2} 
\right.\\
&&\left.
+2\int_0^\infty d p\, p^2 E_p\frac{n_B(E_p)}{E_p^2-\mu_+^2} 
-2i\nu_q\mu_+\int_0^\infty d p\ p\frac{n_B(E_p)}{E_p^2-\mu_+^2} 
\ln\left[\frac{E_p+p}{E_p-p}\right] \right\}\Bigg|_{M^2=0}\,.
\label{Eq:bubble_tree1_th}
\eea
As we shall show explicitly in Sec. \ref{sec:HTL} this is exactly the form
of one term which will be reproduced by a single insertion of a hard HTL photon 
self-energy in the one-loop $\Sigma$.

We note that $M^2$ in the square bracket of the first
term in (\ref{Eq:bubble_tree1_th}) comes from the last $M^4$ term 
of the numerator in (\ref{Eq:Bbb1}), which should not be neglected
since it will contribute after differentiation with respect to $M^2$.
The reason for this is the collinear singularity in the $x$-integral 
which is regulated by $M^2$ to give
\be
\int_{-1}^1 d x \frac{1}{(p x -E_p)^2}=\frac{2}{M^2}\,.
\ee 
This changes the $M^4$ dependence of this term in
 (\ref{Eq:bubble_tree1_th}) to $M^2$.

The soft $p$ integrals remaining in each of the three terms in (\ref{Eq:bubble_tree1_th}) 
are analyzed in Appendix~\ref{App:mitras_integral}. For the first integral we obtain
from (\ref{Eq:erdekesbol}), in the high temperature limit, 
$T \gg M, T \gg \omega$,
\be
\Sigma_\textrm{\tiny B1,a}(\omega) \simeq
\frac{e^4 T^2}{24 \pi^2 \omega}\gamma_0 \left[
\frac{\pi T}{M} +\ln\left(\frac{M}{\omega}\right) - \frac{1}{2}\ln\left(\frac{T}{\omega}\right)
- 2\pi i\,\frac{T}{\omega} + \dots \right]\,,
\label{Eq:BU1a}
\ee
upon analytical continuation $i\nu_q \rightarrow \omega$ to real frequencies.
The ellipsis in (\ref{Eq:BU1a}) refers to terms which first vanish in the asymptotic
high temperature limit $T/M, T/\omega \rightarrow \infty$ (with $M/\omega$ fixed),
and also remain finite in the second limit in which the regulator is removed, 
$M/\omega \rightarrow 0$. In other words, terms which are divergent as $M \rightarrow 0$
but subleading in the high temperature expansion have been discarded. It is
this particular sequence of limits which is relevant to the comparison to
the physics of hard thermal loops, and the fermion self-energy on-shell,
in which $\omega \sim M$ are both of order of the soft scale $eT$, and $e \ll 1$.

The second integral of (\ref{Eq:bubble_tree1_th}) gives after differentiation 
with respect to $M^2$ a finite contribution as $M\to 0.$ 
Using (\ref{Eq:int_in_1-loop_xi}) one obtains only
the purely imaginary part of the self-energy
\be
\Sigma_\textrm{\tiny B1,b}(\omega) \simeq -i \gamma_0\,\frac{e^4 T^3}{48\pi\omega^2}
+ \dots \label{Eq:BU1b}
\ee
in the high temperature expansion. 

Lastly the asymptotic behavior of the third integral in (\ref{Eq:bubble_tree1_th}) 
at high $T$ is studied in Appendix~\ref{App:mitras_integral}, and (\ref{la_int}) yields
\bea
\Sigma_\textrm{\tiny B1,c}(\omega)&\simeq&
\frac{e^4 T^2}{24 \pi^2 \omega}\gamma_0\, \left[
-\frac{\omega^2}{2M^2}\left(\ln\frac{T}{\omega}+\ln(4\pi)-\gamma_E\right)
+i \frac{\pi \omega T}{M^2}  \right.
\nonumber
\\
&&\left.\qquad\qquad\ \ +\frac{1}{2}\ln\frac{T}{\omega}\left(\ln\frac{T}{\omega}-
2\ln\frac{M}{\omega}\right) + \bigg(\ln(4\pi)-\gamma_E\bigg) \ln\frac{T}{M} 
-\frac{1}{2}\ln\frac{T}{\omega} + \dots \right]\,,
\label{Eq:BU1c}
\eea
where $\gamma_E$ is Euler's constant.

As anticipated by the discussion following  (\ref{Eq:coll_x_IR}),
both the real and imaginary parts of the last contribution to 
$\Sigma_\textrm{\tiny B}$ diverge quadratically as $M\to 0$. 
This severe quadratic infrared/collinear divergence will turn out 
to cancel against other terms in $\Sigma_\textrm{\tiny B}$
to be evaluated below. The most interesting term is the linearly 
divergent $1/M$ term in (\ref{Eq:BU1a}), both because
of its non-analyticity in $M^2$ and the fact that it gives rise
to the dominant term in the real part of $\Sigma$, if we anticipate that
after HTL resummation $M$ will be of order $eT$. The existence of this
term implies that the NLO correction to the real part of the fermion
self-energy will finally be of order $e^2 T$, rather than $e^3T \ln (1/e)$, 
as one might have supposed from the real part of the NLO one loop
self-energy in (\ref{Eq:1-loop-RI}). In other words, the form of the 
divergent terms as $M \rightarrow 0$ in the high temperature expansion
of the two-loopbubble self-energy diagram, which are not encountered
at one loop, indicate not only the need for resummation of higher loop 
diagrams in a systematic way, but moreover that these resummed higher loop
diagrams will actually give the {\it dominant} contribution to the NLO behavior 
of the real part of fermion self-energy ($e^2T$), compared to the NLO one loop
$e^3T \ln (1/e)$ behavior.

Proceeding to the second term $\Sigma_\textrm{\tiny B2}$ of (\ref{SigBubble})
one evaluates $B(r k,\k; r k+p_0, \k + \p ;i\nu_q-p_0,\q-\p)$ for 
$\q=0$ with the definition in (\ref{Eq:Dirac_bubble}) and obtains
\bea
\Sigma_\textrm{\tiny B2}(i\nu_q)&=&-\frac{e^4}{2 \pi^4} 
\gamma_0 \frac{\partial}{\partial M^2}
\int_0^\infty d k \int_0^\infty d p \int_{-\infty}^\infty d p_0
\int_{-1}^1 d x\, k^2 p^2\,
\sum_{r=\pm 1} \Bigg\{
\nonumber
\\
&&
\hspace{-2.7cm} n_F(r k) n_F(p_0+r k)\epsilon(p_0+r k) 
\delta\big((p_0+rk)^2-(\k+\p)^2\big)
\frac{2 (r k+p_0) (i\nu_q-p_0)+p^2+ p x (r p_0+2k)}
{[E_p^2-p_0^2][p^2-(i\nu_q-p_0)^2]}
\Bigg\}\,.
\label{Eq:SigB2hs_start}
\eea 
In this form one can see that because there is no $k$ in the denominator, 
the $k$ integral is clearly hard, since even taking account of the remaining 
delta function it would be UV quadratically  divergent if not for the statistical factors.
Thus $k \sim T$ and the $k$ integral will give a factor of $T^2$.
On the other hand the $p,p_0$ integrals are soft due to the appearance of
$p, p_0$ in the denominators, and after differentiation with respect to $M^2$
are rendered UV finite even without making use of the statistical factors.  
Hence $p$ and $p_0$ will be of order $\nu_q$ or $M$ and all expressions 
can be expanded in $p_0/k, p/k$. In the limit $p_0,p\ll k$ the
arguments of the Dirac $\delta$ are $(p_0\pm k)^2-(\k+\p)^2\simeq \pm
2 k(p_0\mp p x)$ and after the change of variable $x\to -x$ in the term 
coming from $r=-1$ one obtains
\bea
\nonumber
\Sigma_\textrm{\tiny B2}(i\nu_q)&\simeq&
-\frac{e^4}{2 \pi^4} i\nu_q \gamma_0 \frac{\partial}{\partial M^2}
\int_{-\infty}^\infty d p_0 \int_0^\infty d p\, p^2 
\frac{1}{[E_p^2-p_0^2][p^2-(i\nu_q-p_0)^2]}\times
\\
&& \int_{-1}^1 d x \int_0^\infty d k\, k^2
\Big[
n_F(k) n_F(k+p_0)+(1-n_F(k)) (1-n_F(k-p_0))
\Big]\delta(p_0-p x)\bigg|_{M^2=0}.\ \ \ \ 
\label{Eq:Btree5hs}
\eea
Next one can make use of a relation for the product of two Fermi-Dirac 
distributions in terms of products of a Bose-Einstein and a Fermi-Dirac
distribution functions (see {\it e.g.} Eq.(3.20) of \cite{baier88}): 
\be
n_F(x)n_F(x+y)=(1+n_B(y)) n_F(x+y)-n_F(x) n_B(y).
\label{Eq:FF2FB}
\ee
Using this relation with $x=k, y=p_0$ and $x=k, y=-p_0$, respectively, 
one obtains
\be
n_F(k) n_F(k+p_0)+(1-n_F(k)) (1-n_F(k-p_0))
=-n_B(-p_0) \big[n_F(k+p_0)-n_F(k-p_0)\big] +1.
\label{Eq:meta}
\ee
Discarding the vacuum piece, expanding the two terms in the square
brackets above for $k\gg p_0$ and keeping the first non-vanishing term
one substitutes (\ref{Eq:meta}) and 
\be
\int_0^\infty d k\, k^2 \frac{d n_F(k)}{d k}=
-2\int_0^\infty d k\, k n_F(k) =-\frac{\pi^2 T^2}{6}, 
\label{Eq:k2dfk}
\ee
into (\ref{Eq:Btree5hs}) to obtain
\bea
\Sigma_\textrm{\tiny B2}(i\nu_q)&=&-
\frac{e^4 T^2}{6\pi^2} i\nu_q\gamma_0
\frac{\partial}{\partial M^2} \int_{-1}^1 d x
\int_0^\infty d p \ x p^3\int_{-\infty}^\infty d p_0
\frac{\delta(p_0-x p)n_B(-p_0)}{(E_p^2-p_0^2)(p^2-(i\nu_q-p_0)^2)}
\bigg|_{M^2=0}.
\label{Eq:SigB2_HTL}
\eea
As we shall show in Sec. \ref{sec:HTL} this form is also exactly 
reproduced by an HTL insertion of a hard HTL photon 
self-energy in the one-loop $\Sigma$.

If one performs the $p_0$ integral in (\ref{Eq:SigB2_HTL}), keeping only the 
thermal contribution which is separated with the help of 
(\ref{Eq:BE_FD_decomp}), and taking into account the sign function 
coming from (\ref{Eq:BE_FD_decomp}), one obtains
\be
\Sigma_\textrm{\tiny B2}(i\nu_q)=
\frac{e^4 T^2}{6\pi^2}\gamma_0 \frac{\partial}{\partial M^2}
\int_0^1 d x\int_0^\infty d p 
\frac{x p^3 n_B(x p)}{[p^2(x^2-1)-M^2]}
\sum_{r=\pm 1}\frac{i\nu_q}{[i\nu_q-r p(1+x)][i\nu_q+r p(1-x)]}
\Bigg|_{M^2=0}.
\label{SigPiA1}
\ee
Using partial fractioning of the denominator, changing the
integration variable $p$ to $p/x$ and then performing the $x$
integral and sum over $r=\pm 1$, one obtains
\bea
\Sigma_\textrm{\tiny B2}(i\nu_q)&=&
\frac{e^4 T^2}{12\pi^2}\gamma_0\frac{\partial}{\partial M^2}
\bigg\{i\frac{\pi}{2} \int_0^\infty d p\frac{p^2 n_B(p)}{p^2-\mu_+^2}
-\mu_+\ln\left(\frac{M^2}{\nu_q^2}\right)
\int_0^\infty d p\frac{p\, n_B(p)}{p^2-\mu_+^2}
\nonumber\\
&&+\frac{\mu_+}{2}\int_0^\infty d p \frac{p\, n_B(p)}{p^2-\mu_+^2}
\ln\left(\frac{4 p ^2}{\nu_q^2}+1\right)
-i\int_0^\infty d p\frac{p^2 n_B(p)}{p^2-\mu_+^2}
\textnormal{arccot}\left(\frac{2 p}{\nu_q}\right)
\bigg\}\bigg|_{M^2=0},
\label{Eq:Pi2_HTL}
\eea
where $\mu_\pm=((i\nu_q)^2\pm M^2)/(2 i\nu_q).$ The asymptotic analysis of 
the above integrals is given in Appendix~\ref{App:mitras_integral}. 
Using (\ref{Eq:arctan_int}), (\ref{Eq:MB2}) and the series expansion 
given in (\ref{Eq:arccot_series}) one obtains after analytical continuation
\bea
\Sigma_\textrm{\tiny B2}(\omega)&\simeq&
\frac{e^4 T^2}{24 \pi^2 \omega}\gamma_0 \left[
\frac{\omega^2}{2M^2}\left(\ln\frac{T}{\omega}
+\ln(4\pi)-\gamma_E\right) -\frac{i\pi \omega T}{M^2}
+\ln\frac{T}{\omega}\ln\frac{M}{\omega}
-\frac{1}{4}\ln^2\frac{T}{\omega}\right.
\nonumber\\
&&\left.\qquad\qquad\ \ +\frac{1}{2}\big(1+\gamma_E-\ln(4\pi)\big)
\left(\ln\frac{T}{\omega}-2\ln\frac{M}{\omega}\right)
\right],
\label{B2f}
\eea
were we have again retained only the terms in the high temperature expansion 
which contribute at the order of interest or are singular in the $M^2\to 0$ limit.

For the last term $\Sigma_\textrm{\tiny B3}$ of (\ref{SigBubble})
and (\ref{SigBubble3}) one evaluates $B(k_0,\k;i\nu_q+k_0-p_0,\k+\p;p_0,-\p)$ 
and $B(k_0+p_0-i\nu_q,\k;k_0,\k+\p;p_0,-\p)$ with the
definition (\ref{Eq:Dirac_bubble}), and obtains then
\bea
\nonumber
\Sigma_\textrm{\tiny B3}(i\nu_q)&=&
\frac{e^4}{2\pi^4}\gamma_0\frac{\partial }{\partial M^2}
\int_0^\infty d k \int_0^\infty d p \int_{-\infty}^\infty d p_0
\int_{-1}^1 d x\, k^2 p^2
\frac{\delta (p_0^2-p^2)\,\epsilon(p_0)}{[E_p^2-(p_0 - i\nu_q)^2]}
\sum_{r=\pm 1} \, r\Bigg\{
\\
\nonumber
&&\ \frac{p\,x\,(2r k+i\nu_q-p_0) - 2 k p_0 +r p^2}
{[2kr(p r x+ p_0-i\nu_q)+  i\nu_q(2 p_0-i\nu_q)]} \  n_F(r k)\, n_F(-p_0)\\
&&   \hspace{-1.7cm} -
\left[\frac{(r |\k+\p|+p_0-i\nu_q)(2r|\k+\p|p_0+k p x+p^2) + r k p x |\k+\p|}
{k^2-(r|\k+\p|+p_0-i\nu_q)^2} \right]
 \frac{n_F(r |\k+\p|)}{|\k+\p|}\ n_F(p_0)
\Bigg\}\Bigg|_{M^2=0}\hspace{-.7cm},
\label{Eq:SigB3hs_start}
\eea
where $|\k + \p| = \sqrt{k^2 + p^2 +2 k p x}$. Since the large $k$ behavior of
the integrand (without the statistical factors) in both terms is $k^2$, the 
$k$ integral is hard, and $k \sim T$, while $p_0$ and $p$ are soft, of order
$\nu_q$ or $M$. Expanding the expression in the curly bracket of 
(\ref{Eq:SigB3hs_start}) for $p_0,p\ll k$, redefining $x\to -x$ in the terms 
with $r=-1$ and doing the sum over $r$ yields
\bea
\nonumber
\Sigma_\textrm{\tiny B3}(i\nu_q)&=&
\frac{e^4}{2\pi^4}\gamma_0\frac{\partial }{\partial M^2}
\int_0^\infty d k \int_0^\infty d p \int_{-\infty}^\infty d p_0
\int_{-1}^1 d x\, k^2\,p^2\,
\frac{\delta (p_0^2-p^2)\epsilon(p_0)}{E_p^2-(i\nu_q-p_0)^2}
\Bigg\{
\\\nonumber
&&\frac{p_0+p x}{p_0+p x-i\nu_q}[n_F(k)+n_F(-k)][n_F(p_0)+n_F(-p_0)]
\\\nonumber
&&+i\nu_q\frac{p^2}{2 k}\frac{x^2-1}{(p_0+p x-i\nu_q)^2}
[n_F(k)-n_F(-k)][n_F(-p_0)+n_F(p_0)]
\\
&&
+2 \frac{d n_F(k)}{d k} 
\left(p x +\frac{i\nu_q p x}{p x +p_0-i\nu_q}\right)n_F(p_0)
\nonumber
\\
&&-\frac{p_0}{2 k}[n_F(-p_0)-n_F(p_0)][n_F(k)-n_F(-k)]
\Bigg\}\bigg|_{M^2=0}.
\label{Eq:SigB3hs}
\eea
Note, that due to the relations (\ref{BE_FD_identities})
the first term in (\ref{Eq:SigB3hs}) is a vacuum piece and the second 
term contains only one statistical factor and can not contribute at leading \
order in the high temperature expansion. Disregarding these terms 
and using (\ref{Eq:k2dfk}) one obtains
\bea
\Sigma_\textrm{\tiny B3}(i\nu_q)&\simeq&
\frac{e^4 T^2}{6\pi^2}\gamma_0
\frac{\partial}{\partial M^2}\left\{
\int_0^\infty d p\, p^2\int_{-\infty}^\infty d p_0 \varepsilon(p_0)
\frac{\delta(p_0^2-p^2)n_F(p_0)}{E_p^2-(i\nu_q-p_0)^2}
\right.
\nonumber\\
&&\left.\times \left[p_0
+i\nu_q p\int_{-1}^1 d x \frac{x}{i\nu_q-p_0-x p}
\right]\right\}\bigg|_{M^2=0}.
\label{Eq:SigB3_HTL}
\eea
Performing the $x$ and $p_0$ integrals in $\Sigma_{B3}$ results in
\bea
\Sigma_{_{B3}}(i\nu_q)&\simeq&
-\frac{e^4 T^2}{24\pi^2}\gamma_0\frac{\partial}{\partial M^2}
\bigg\{4\mu_-\int_0^\infty d p\frac{p\, n_F(p)}{p^2-\mu_-^2}
+\mu_+\int_0^\infty d p \frac{p\, n_F(p)}{p^2-\mu_-^2}
\ln\left(\frac{4 p ^2}{\nu_q^2}+1\right)\nonumber\\
&&
+2 i \mu_-\mu_+ \int_0^\infty d p\frac{n_F(p)}{p^2-\mu_-^2}
\arctan\left(\frac{2 p}{\nu_q}\right)
\bigg\}\bigg|_{M^2=0},
\label{Eq:Pi3_HTL}
\eea
where $\mu_\pm=((i\nu_q)^2\pm M^2)/(2 i\nu_q)$. This form of the
third contribution to $\Sigma_B$ is also obtained directly from the
the insertion of a hard HTL photon self-energy in the one-loop $\Sigma$,
as we shall show explicitly in Sec.~\ref{sec:HTL}.

The asymptotic analysis of the above integrals is given in Appendix~\ref{App:mitras_integral}. 
Using (\ref{Eq:arctan_int}) and the series expansion given in (\ref{Eq:arccot_series}) one 
obtains after analytical continuation to real frequencies,
\bea
\Sigma_{_{B3}}(\omega)&\simeq& 
\frac{e^4 T^2}{48\pi^2\omega}\gamma_0
\left[-\frac{1}{2}\ln^2\frac{T}{\omega}
+(2+\gamma_E-\ln\pi)\ln\frac{T}{\omega} \right]
+i\cdot {\cal O}\left(\frac{e^4 T^2}{\omega}\ln\frac{T}{\omega}\right).
\label{B3f}
\eea
Here we have again retained only the terms in the high temperature expansion 
which contribute at the order of interest. The imaginary
part neglected in (\ref{B3f}) is of order $e^3T \ln (1/e)$
and subleading to the NLO terms of magnitude $e^2T$ in which we are
interested.

The sum of the contributions (\ref{Eq:BU1a}), (\ref{Eq:BU1b}) and 
(\ref{Eq:BU1c}) from  $\Sigma_{B1}$, (\ref{B2f}) from $\Sigma_{B2}$,
and (\ref{B3f}) from $\Sigma_{B3}$ is
\be
\Sigma_\textrm{\tiny B}(\omega) = \Sigma_\textrm{\tiny B1}(\omega) + 
\Sigma_\textrm{\tiny B2}(\omega) 
+ \Sigma_\textrm{\tiny B3}(\omega) \simeq
\frac{e^4 T^2}{24 \pi^2 \omega}\,\gamma_0 \left[
\frac{\pi T}{M} + \left(\frac{1}{2}+\ln 2\right)\ln \frac{T}{\omega}
- \frac{5\pi i T}{2\omega}\right]\,.
\label{sumB}
\ee
Remarkably, all the $M^{-2}$ and $\ln M$ infrared/collinear divergences 
cancel in this leading order high temperature form, as well as $\gamma_E$
and $\ln \pi$, but the $M^{-1}$ divergence remains. This cancellation is in 
line with the findings of \cite{majumder01} which demonstrates the cancellation 
of all collinear and infrared divergences in the final result for the imaginary part of 
the vector-boson self-energy in the limit where the dilepton invariant mass 
is much bigger than the temperature. In our case, the presence of the photon 
double pole, which produces a more singular behavior that the fermionic 
one present in \cite{majumder01}, results in the survival of the $M^{-1}$ 
divergence in the real part, coming from (\ref{Eq:BU1a}) in our final result 
given in (\ref{Eq:B_collected}). The cancellation of the $M^{-2}$ and $\ln M$ 
terms can be understood by a simple argument, implicit already in the work 
of Ref. \cite{GatKap}, which we discuss in Section \ref{sec:sum}.

The most important implication of (\ref{sumB}) is that because of its
$M^{-1}$ dependence, the photon mass regulator cannot be consistently 
taken to zero in the two-loop fermion self-energy. Only the bubble self-energy 
diagram contains this divergence. The second important feature of the
above analysis is that all of the contributions to the two-loop bubble self-energy 
that are of the same magnitude as the NLO one-loopterms in (\ref{Eq:1-loop-RI}),
(or larger in the case of the $M^{-1}$ term) are obtained from one hard and one 
soft momentum loop. The higher order terms which have been neglected
are soft-soft. There are no hard-hard contributions to $\Sigma_B$ at all.

The remaining two-loop linear divergence $\propto M^{-1}$ as $M\rightarrow 0$,
not present in the one-loop self-energy indicates the need for a 
systematic resummation of higher loop contributions and a reorganization 
of the perturbative series. Moreover, if we anticipate that after resummation 
the photon will acquire a Debye mass of order $eT$, this $M^{-1}$ term 
indicates that the correction to the real part of the fermion dispersion
relation on-shell will be of order $e^2T$, and not $e^3T\ln(1/e)$, as
might have been expected from the NLO one-loop correction to the real
part (\ref{Eq:1-loop-RI}).  

\subsection{The Rainbow Diagram}

Taking $\q=0$ in expressions (\ref{SigR1f}) and (\ref{Eq:R_B_2stat}) and
taking for term $\Sigma_\textrm{\tiny R2}$ the derivative with respect to $m^2$ 
and the limit $m^2=0$ one obtains
\bea
\nonumber
\Sigma_\textrm{\tiny R1}(i\nu_q,\q ={\bf 0})&=&
\frac{e^4}{8\pi^4}\gamma_0
\int_0^\infty d p\int_0^\infty d k\int_{-1}^1 d x \sum_{r,s=\pm 1}
\Bigg[\frac{r\, k\, n_F(p) n_B(k)}
{2(x-r s)((i\nu_q)^2-2 k p(x-r s)-2i\nu_q(r k+s p))}\\
\nonumber
&&-n_F(p) n_B(k) \frac{k}{2i\nu_q}
\frac{r}{(x-r s)(i\nu_q-2s p)}
+n_F(p)n_F(k) \frac{1}{2i\nu_q}\frac{s p-r k}{(x-r s)
(i\nu_q-2 s p)}\\ 
\nonumber
&&+n_B(p) n_B(k)\frac{k^2 p}{i\nu_q}
\frac{r}{(i\nu_q-2s p)((i\nu_q)^2-2 k p(x-r s)-2i\nu_q(r k+s p))}\\
&&+n_B(p)n_F(k) \frac{k p}{i\nu_q}
\frac{k r -(i\nu_q-s p)}
{(i\nu_q-2s p)((i\nu_q)^2-2 k p(x-r s)-2i\nu_q(r k+s p))}
\Bigg],
\label{Eq:rainbow_term_A_real}\\
\Sigma_\textrm{\tiny R2}(i\nu_q, \q = {\bf 0})&=&
\frac{e^4 T^2}{4 \pi^2}\gamma_0 \int_0^\infty d p\left[
\frac{p\ i\nu_q}{((i\nu_q)^2-4p^2)^2}
(n_B(p)-n_F(p))-
\frac{p^2}{i\nu_q ((i\nu_q)^2-4 p^2)}\frac{dn_F(p)}{d p}
\right].
\label{Eq:rainbow_term_B_real}
\eea
The first term and the very last term, involving $(i\nu_q - s p)$ in 
(\ref{Eq:rainbow_term_A_real}) can be handled in the following way.
If $k$ and $p$ are hard momenta in these terms, then the sums 
over $r,s$  either give integrands which vanish identically, or are odd 
under $x\rightarrow -x$, so vanish upon integration over $x$. Hence
these terms have no hard-hard contribution and must contain
at least one additional power of $\omega$ in the numerator,
relative to all terms in (\ref{Eq:1-loop-RI}). Therefore they are subleading.

Performing the sum over $r$ and $s$ one finds that
the sum of the second and third terms of (\ref{Eq:rainbow_term_A_real}) is
\be
-\frac{e^4}{8\pi^4i\nu_q}\gamma_0
\int_0^\infty d k\, k \big[n_B(k) + n_F(k)\big] \int_0^\infty d p\, n_F(p)
\frac{p} {[p^2 - (i\nu_q/2)^2]}\int_{-1}^1 \frac{d x}{1-x^2}\,.
\label{Eq:rainbow_term_A_real_2p3}
\ee
The imaginary part of this contribution is subleading, {\it cf.} (\ref{Eq:MitraF}),
while the real part follows the hard-soft pattern, and does contribute
at the same order as the NLO terms in (\ref{Eq:1-loop-RI}). Moreover 
because of the $x$ integral factor, it is collinearly divergent. However, taking 
into account the remaining terms in (\ref{Eq:rainbow_term_A_real}), {\it i.e.}
the fourth term and remaining part of the fifth term involving $k r$, 
we observe that in both of these $k$ must be treated as hard, while $p$
is soft. After partial fractioning and using interchange of $x \rightarrow -x$
freely, these remaining terms may be written in the form [{\it c.f.} in
Appendix~\ref{App:mitras_integral} the derivation of (\ref{Eq:L_pm_2}) from 
(\ref{Eq:L_pm_1})]
\be
\frac{e^4}{8\pi^4i\nu_q} \gamma_0\int_0^\infty d k\, k 
\big[n_B(k) + n_F(k)\big]
\int_0^\infty d p\, n_B(p) \int_{-1}^1 \frac{d x }{1-x^2} 
\left[ \frac{p}{p^2 - (i\nu_q/2)^2} - \frac{2p}{p^2 - (i\nu_q/(1+x))^2}\right]\,.
\label{Eq:rainbow_term_A_real_4p5a}
\ee
In view of (\ref{Eq:L_pm_2}) and (\ref{Eq:L_pm_result}) the real part of 
this expression cancels at leading order in the high temperature expansion 
the real part of (\ref{Eq:rainbow_term_A_real_2p3}), while its imaginary 
part vanishes at this order. Hence combining all terms of 
(\ref{Eq:rainbow_term_A_real}) gives no collinear divergence for either 
the real or imaginary  parts of the self-energy, and no contribution 
at the same NLO as (\ref{Eq:1-loop-RI}).

For the terms in (\ref{Eq:rainbow_term_B_real}), the factorization into a 
hard and soft momentum integral is again obvious. Performing an integration 
by parts in the last term of (\ref{Eq:rainbow_term_B_real}), one can
rewrite (\ref{Eq:rainbow_term_B_real}) in the form
\be
\Sigma_\textrm{\tiny R2}(i\nu_q)=
\frac{e^4 T^2 i\nu_q }{4\pi^2}\gamma_0 \int_0^\infty d p
\frac{p}{((i\nu_q)^2-4p^2)^2}\big[n_B(p)+n_F(p)\big].
\label{Eq:r_term_B_anal}
\ee
Differentiating the integrals $I_\pm(i a)$ defined in Appendix
\ref{App:mitras_integral} gives for this contribution
\be
\Sigma_\textrm{\tiny R}(i\nu_q) = \Sigma_\textrm{\tiny R2}(i\nu_q) = 
-i\frac{e^4 T^3}{32\pi (i\nu_q)^2}\gamma_0\,.
\ee
After analytic continuation $i\nu_q\to \omega+i 0^+$ the R2 term becomes purely 
imaginary,  while $\Sigma_\textrm{\tiny R1}$ does not contribute at all at this order comparable 
to the NLO one-loop self-energy. Hence the contribution to the fermion self-energy 
of the rainbow diagram is simply
\be
\Sigma_\textrm{\tiny R}(\omega)  \simeq 
\Sigma_\textrm{\tiny R2}(\omega) \simeq 
-i\frac{e^4 T^3}{32\pi\omega^2}\gamma_0.
\label{Eq:R_I_final}
\ee
at the same order as the NLO imaginary part of (\ref{Eq:1-loop-RI}). This 
leading order contribution of the two-loop rainbow diagram arises explicitly
from one hard and one soft loop integral. There are no hard-hard contributions
to the rainbow diagram.

\subsection{The Crossed Photon Diagram}

Evaluating the crossed photon diagram self-energy (\ref{SigCf}) at
$\q = 0$, by using the definition (\ref{Cdef}) one obtains
\bea
\nonumber
\Sigma_\textrm{\tiny C}(i\nu_q, \q = {\bf 0})&=&
\frac{e^4}{8\pi^4}\gamma_0
\int_0^\infty d p\int_0^\infty d k\int_{-1}^1 d x
\sum_{r,s=\pm 1}\Bigg[
2 n_B(p) n_F(k) \frac{(r k-s p)(i\nu_q r k+k p(x-r s))}
{(i\nu_q)^2(i\nu_q-2 r k)(i\nu_q-2 s p)(x-r s)}\\
\nonumber
&&\hspace{-1cm} -k p\, n_F(p) n_F(k) 
\frac{r k+s p -i\nu_q}
{(i\nu_q)^2 (i\nu_q-2 r k) (i\nu_q -2 s p)}\,
\frac{-2k p (x-r s)}{
((i\nu_q)^2-2 k p(x-r s)-2i\nu_q(r k+s p)) }\\
\nonumber
&&\hspace{-1cm} +k p\, n_B(p) n_B(k) 
\frac{r k+s p-i\nu_q}{(i\nu_q)^2(i\nu_q-2 r k)(i\nu_q-2 s p)}
\,
\frac{(i\nu_q)^2-2 k p(x-r s)-2i\nu_q(r k+s p)+(i\nu_q)^2}
{(i\nu_q)^2 -2 k p(x-r s)-2i\nu_q(r k+s p)} \\
\nonumber
&&\hspace{-1cm} +n_B(p)n_F(k) \frac{r k}{i\nu_q(i\nu_q-2 s p)(x-r s)}\,
\frac{(i\nu_q)^2-2 k p(x-r s)-2i\nu_q(r k+s p)-(i\nu_q)^2}
{(i\nu_q)^2 -2 k p(x-r s)-2i\nu_q(r k+s p)}\\
&&\hspace{-1cm} -2 n_F(p)n_F(k)
\frac{r k (k p(x-r s)+i\nu_q s p)}{i\nu_q(i\nu_q-2 s p)
((i\nu_q)^2 -2 k p(x-r s)-2i\nu_q(r k+s p))(x-r s)}
\Bigg].
\label{Eq:CR_2stat_q=0}
\eea
In the last term we have changed $x\to -x$ and $r\to -r$.
Performing partial fractioning in each term one observes that in some 
cases the $x$-dependence partially simplifies between the numerator and
the denominator. With simple algebraic manipulations one can separate 
in the integrand $x$-independent contributions and contributions which 
are explicitly collinearly divergent, obtaining  
\bea
\Sigma_\textrm{\tiny C}(i\nu_q)&=&\frac{e^4}{8\pi^4}\gamma_0\ 
\int_0^\infty d k \int_0^\infty d p \int_{-1}^1 d x \Bigg\{
\nonumber\\
& &n_B(p)n_F(k)
\left[
\frac{4k p}{i\nu_q}\left(
\frac{1}{(i\nu_q)^2-4k^2}-\frac{1}{(i\nu_q)^2-4p^2} \right)
-\frac{8k p}{i\nu_q((i\nu_q)^2-4p^2)}\frac{1}{x^2-1}\right]
\nonumber\\
&+&n_F(p)n_F(k)
\bigg[\frac{4k p}{i\nu_q((i\nu_q)^2-4p^2)}
-\frac{k p}{i\nu_q}\sum_{r,s=\pm 1}
\frac{1}{(i\nu_q)^2-2k p(x-r s)-2i\nu_q(r k+s p)}\nonumber\\
&&\qquad\qquad
+\, 2\, \frac{k^2p}{i\nu_q}\sum_{r,s=\pm 1}
\frac{1}{i\nu_q-2s p}
\frac{r}{(i\nu_q)^2-2k p(x-r s) - 2i\nu_q(r k+s p)}
\bigg]
\nonumber\\
&+&n_B(p) n_B(k)
\bigg[
-\frac{4k p}{i\nu_q((i\nu_q)^2-4p^2)}
-k p\sum_{r,s=\pm 1}
\frac{1}{i\nu_q-2s p}
\frac{1}{(i\nu_q)^2-2k p(x-r s)-2i\nu_q(k r+s p)}
\bigg]
\nonumber\\
&+&n_B(p)n_F(k)
\Bigg[\left(
\frac{i\nu_q}{(i\nu_q)^2-4(k-p)^2}-\frac{i\nu_q}{(i\nu_q)^2-4(k+p)^2}
+\frac{8k p}{i\nu_q((i\nu_q)^2-4p^2)}
\right)\frac{1}{x^2-1}\nonumber \\
&&\ \ 
-2k^2p \sum_{r,s=\pm 1}
\frac{1}{i\nu_q-2s p}
\frac{1}{i\nu_q-2(r k+s p)}
\frac{r}{(i\nu_q)^2-2k p(x-r s)-2i\nu_q(r k+s p)}
\Bigg]
\nonumber\\
&+&n_F(p)n_F(k)
\bigg[
\frac{i\nu_q}{2}\left(
\frac{1}{(i\nu_q)^2-4(k-p)^2}-\frac{1}{(i\nu_q)^2-4(k+p)^2}
\right)\frac{1}{x^2-1}\nonumber \\
&&\qquad\qquad -\,2\,\frac{k^2p}{i\nu_q}\sum_{r,s=\pm 1}
\frac{1}{i\nu_q-2s p}
\frac{r}{(i\nu_q)^2-2k p(x-r s)-2i\nu_q(r k+s p)}\nonumber\\
&&
\ \ -\,4\, \frac{k^2 p^2}{i\nu_q}\sum_{r,s=\pm 1}
\frac{1}{i\nu_q-2 s p}\frac{1}{i\nu_q -2(r k+s p)}
\frac{r s}{(i\nu_q)^2-2k p(x-r s)-2i\nu_q(r k+s p)}
\bigg]\Bigg\}.
\label{Eq:CR_2stat_q=0_complete}
\eea
Note that whenever possible we have exploited the interchanging of
$p\leftrightarrow k$ and $x\to -x.$ Note, also, that we have catalogued 
all the terms as they arise in order from the previous expression
(\ref{Eq:CR_2stat_q=0}), although actually the last term in the
fourth line with two Fermi-Dirac statistical factors exactly cancels 
the term in the ninth line of (\ref{Eq:CR_2stat_q=0_complete}).
Also noteworthy is that the terms with $(x^2 -1)^{-1}$ factors,
which are collinearly logarithmically divergent cancel 
whenever they multiply spatial momentum integrals which
factorize into separate integrals over $k$ and $p$, namely
the last terms of the second line and the sixth line respectively.

We shall show that the leading order terms in the high temperature
expansion of (\ref{Eq:CR_2stat_q=0_complete}) come from precisely
the remaining $x$-independent factorizable terms. Performing the 
interchange $k\leftrightarrow p$ in the first term inside the large round
parentheses of the second line of (\ref{Eq:CR_2stat_q=0_complete}) 
one can combine these terms to obtain
\bea
\Sigma_\textrm{\tiny C}(i\nu_q) &=&
\frac{e^4}{4\pi^4 i\nu_q}\gamma_0\int_0^\infty d k\, k \big[n_B(k)+n_F(k)\big]
\int_0^\infty d p \frac{p}{p^2-(i\nu_q)^2/4}\big[n_B(p)-n_F(p)\big] \nonumber\\
&= &\frac{e^4T^2}{16\pi^2 i\nu_q}\gamma_0 
\int_0^\infty d p \frac{p}{p^2-(i\nu_q)^2/4}\big[n_B(p)-n_F(p)\big].
\label{Eq:CR_final}
\eea

These leading factorizable terms contain no remaining collinear divergences.
Note, that we obtained the product of the same integrals which already have
appeared in the LO and NLO expressions of the HTL approximation to the
1-loop fermion self-energy (\ref{Eq:sigma-1loop_final}). The first integral 
over $k$ in (\ref{Eq:CR_final}) is clearly dominated by hard $k \sim T$, 
and given by (\ref{Eq:hard_integral}), while the second integral 
\rt{also encountered in the NLO correction of the one-loopself-energy} 
is given in
Appendix~\ref{App:mitras_integral} and is soft. 
Through the analytical continuation $i\nu_q\to\omega+i0^+$, using
(\ref{Eq:Mitra_corollary_R}) and (\ref{Eq:Mitra_corollary_I}), 
one obtains from these factorizable hard/soft terms
\be
\Sigma_\textrm{\tiny C}(\omega)=
\frac{e^4 T^2}{16\pi^2\omega}\gamma_0\left[
-\ln\frac{T}{\omega}+i\pi\frac{T}{\omega}
\right],
\label{CR_NLO}
\ee
which contains both real and imaginary parts.

It remains to show that all the remaining non-factorizable terms in
(\ref{Eq:CR_2stat_q=0_complete}) are subleading to (\ref{CR_NLO}).
We show first that these remaining terms do not contain any
contributions at all when both $k$ and $p$ are hard. 
The remaining non-factorizable terms are of two kinds, {\it viz.} those
involving the integrand
\be
\frac{i\nu_q}{(i\nu_q)^2-4(k-p)^2}-\frac{i\nu_q}{(i\nu_q)^2-4(k+p)^2}
\label{subCR1}
\ee
multiplied by the collinearly divergent integral $\int_{-1}^1 d x/(1-x^2)$
and either $n_B(p)n_F(k)$ or $n_F(p) n_F(k)$, and
the following four terms, involving sums over $r$ and $s$,
\bea
&&\sum_{r,s=\pm 1}\left\{n_F(p)n_F(k) \bigg[-\frac{k p}{i\nu_q}
\frac{1}{(i\nu_q)^2-2k p(x-r s)-2i\nu_q(r k+s p)}\right.\nonumber\\
&&
-4\frac{k^2 p^2}{i\nu_q} \frac{1}{i\nu_q-2 s p}\frac{1}{i\nu_q -2(r k+s p)}
\frac{r s}{(i\nu_q)^2-2k p(x-r s)-2i\nu_q(r k+s p)} \bigg]\nonumber\\
&&+ n_B(p) n_B(k)
\bigg[-k p \frac{1}{i\nu_q-2s p} \frac{1}{(i\nu_q)^2-2k p(x-r s)-2i\nu_q(k r+s p)}
\bigg]
\nonumber\\
&&\left.+ n_B(p)n_F(k) \left[ -2k^2p  \frac{1}{i\nu_q-2s p}
\frac{1}{i\nu_q-2(r k+s p)} \frac{r}{(i\nu_q)^2-2k p(x-r s)-2i\nu_q(r k+s p)}
\right]\right\}.
\label{subCR2}
\eea
For the first kind of terms in (\ref{Eq:CR_2stat_q=0_complete}), namely those
involving a difference like (\ref{subCR1}), it is clear that there is no hard-hard 
contribution, since the integrand (\ref{subCR1}) vanishes for $k\gg \omega$ 
or $p\gg\omega$. In fact, the integral over $k$ for these terms is dominated
by soft $k$ for which we can replace the Fermi-Dirac distribution $n_F(k)$ 
by $1/2$. Then the integral can be performed with the result
\be
\int_0^{\infty} d k\,n_F(k) \left[\frac{i\nu_q}{(i\nu_q)^2-4(k-p)^2}-
\frac{i\nu_q}{(i\nu_q)^2-4(k+p)^2}\right] \rightarrow \frac{1}{4} \ln \left[
\frac{i \nu_q + 2 p}{i\nu_q - 2 p}\right]\,.
\ee
The real part of this integrand vanishes for $i\nu_q \rightarrow \omega \rightarrow 0$, but
is proportional to $1/p$ for large $p$, the final integration over $p$ with a 
second $n_F(p)$ gives a contribution to $\Sigma_\textrm{\tiny C}$ proportional
to $e^4\omega \ln (T/\omega)$, which is subleading. In the cases where the 
final integration over $p$ involves the Bose-Einstein distribution $n_B(p)$,
there is an enhancement of $T/\omega$ and the contribution of the integral
is proportional to $e^4T$, which is still subleading. Therefore the contributions 
of the terms involving the difference (\ref{subCR1}) are subleading compared to
(\ref{Eq:1-loop-RI}) and hence NNLO. 

For the second kind of terms, that is those of (\ref{subCR2}), we observe that 
they again do not contain any hard-hard contributions, for when both $k$ and $p$ 
are hard, then one can ignore $i\nu_q \rightarrow \omega$ relative to $k$ and $p$ 
in factors of the  denominators in (\ref{subCR2}). In that limit, the third term vanishes 
upon summing over $r, s$, while the first and fourth terms of  (\ref{subCR2}) are explicitly 
odd under $x \rightarrow -x$, and vanish under integration over $x$. By symmetrizing
the remaining second term of (\ref{subCR2}) under $x \leftrightarrow -x$, we
find that it becomes proportional to
\be
\int_0^{\infty}d k\, n_F(k)\int_0^{\infty} dp\,n_F(p)\,\left[\frac{k p}
{k^2 - p^2}\right] = 0
\ee
by antisymmetry under interchange of $k$ and $p$. 

All terms in (\ref{subCR2}) vanish in the hard-hard limit, 
{\it i.e.} when $\omega$ is neglected in the denominators. The integral
over these terms must be proportional to at least one factor of
$\omega/T$, relative to (\ref{CR_NLO}) with one of the momenta soft.
Thus all the leading contributions from the high temperature expansion of the 
two-loop crossed photon self-energy diagram are obtained from the explicitly
factorizable terms in (\ref{Eq:CR_2stat_q=0_complete}), and given
completely by (\ref{CR_NLO}), which exhibit the pattern of 
one loop integration momentum hard, and the other soft, just as
in the bubble and rainbow diagrams considered previously.

%-------@#$--------- HARD THERMAL LOOP -----------------

\section{The Hard-Soft Pattern and Effective HTL Insertions
\label{sec:HTL}}
By explicit analysis of the two-loop self-energy we have found that the 
leading contributions in the high temperature limit which should be compared 
with the NLO one-loop contributions arise in all two-loop diagrams where one 
loop momentum is hard ($\sim T$) while the other is soft ($\sim \omega \sim eT$
for imaginary parts or over the range $0\le k < T$ for real parts giving rise
to logarithms). On the other hand, the basic one-loop HTL self-energies and 
vertices are derived precisely by assuming the internal loop momentum is hard, 
relative to the external momenta flowing into them which are assumed soft. 
Hence by replacing one of the loops (the hard one) of each two-loop diagram
by an HTL self-energy or vertex part, and evaluating the resulting effective 
{\it one-loop} soft integrals, we must obtain exactly the same result as our 
direct analysis of the leading high temperature behavior of the two-loop 
diagram. This reduction of two-loop diagrams with one loop hard and the other
soft is illustrated graphically in Figs.~\ref{Fig:B_HTL},~\ref{Fig:R_HTL} and
\ref{Fig:CR_HTL}. In this section we shall {\it prove} that the initially rather 
different looking expressions obtained by the insertion of one HTL
self-energy or vertex in the one-soft-loop diagrams of Figs.~\ref{Fig:B_HTL},
~\ref{Fig:R_HTL}, \ref{Fig:CR_HTL} yields precisely the same results
as that obtained in Section~\ref{sec:2loop_anal} for each diagram. This
furnishes a non-trivial check by a completely different method
on the detailed analysis of the two-loop fermion self-energy in
Section~\ref{sec:2loop_anal}.

The non-vanishing three- and four-point functions in the LO HTL approximation
are given in Minkowski space by \cite{braaten90b,lebellac96}
\bea
\label{Eq:2E_HTL_vertex}
\Sigma^\textnormal{HTL}(P)&=&m_f^2\int\frac{d \Omega}{4\pi} 
\frac{\slV}{P\cdot V+i\epsilon},\\
\label{Eq:2P_HTL_vertex}
\Pi_{\mu\nu}^\textnormal{HTL}(Q)&=&m_D^2\left[
-\delta_{0\mu}\delta_{0\nu}+Q_0
\int\frac{d \Omega}{4\pi}\frac{V_\mu V_\nu}{V\cdot Q+i\epsilon}
\right],
\\
\label{Eq:2E1P_HTL_vertex}
\Gamma_\mu^\textnormal{HTL}(P_1,P_2)&=&m_f^2 \int\frac{d \Omega}{4\pi} 
\frac{V_\mu \slV}{ (P_1\cdot V+i\epsilon)(P_2\cdot V+i\epsilon)},\\
\label{Eq:2E2P_HTL_vertex}
\Gamma_{\mu\nu}^\textnormal{HTL}(P_1,P_2,Q)&=&m_f^2\int\frac{d\Omega}{4\pi} 
\frac{V_\mu V_\nu\slV}
{((P_1+Q)\cdot V+i\epsilon)((P_2-Q)\cdot V+i\epsilon)}
\left(
\frac{1}{P_1\cdot V+i\epsilon}+\frac{1}{P_2\cdot V+i\epsilon}
\right),\ \ 
\eea
where $m_f=e T/\sqrt{8}$ is the fermion thermal mass, $m_D=e T/\sqrt{3}$ is
the Debye screening mass for photons, $d \Omega=d(\cos\theta) d\phi$ is
the measure for angular integration and $V=(1,{\bf v})$ with ${\bf v}^2=1$.

The HTL two-electron proper vertices obey the simple Ward identities 
\cite{lebellac96}
\bea
\label{HTLWard1}
(P_1 - P_2)^{\mu}\Gamma_{\mu}^\textnormal{HTL}(P_1,P_2) &=&
\Sigma^\textnormal{HTL}(P_2)-\Sigma^\textnormal{HTL}(P_1)\,,\\
\label{HTLWard2}
Q^{\nu}\Gamma_{\mu\nu}^\textnormal{HTL}(P,P,Q)&=&
\Gamma_{\mu}^\textnormal{HTL}(P,P-Q)-
\Gamma_{\mu}^\textnormal{HTL}(P+Q, P)\,,\\
\label{HTLWard3}
Q^{\mu}Q^{\nu}\Gamma_{\mu\nu}^\textnormal{HTL}(P,P,Q)&=&
\Sigma^\textnormal{HTL}(P+Q)-2\Sigma^\textnormal{HTL}(P)
+\Sigma^\textnormal{HTL}(P-Q)\,,
\eea
while the HTL two-photon polarization function is transverse:
\be
Q^{\nu}\Pi_{\mu\nu}^\textnormal{HTL}(Q) = 0\,.
\ee

\subsection{The HTL Reduced Bubble Diagram 
\label{ss:HTL_bubble}}

\begin{figure}[!t]
\includegraphics[keepaspectratio,width=0.75\textwidth,angle=0]{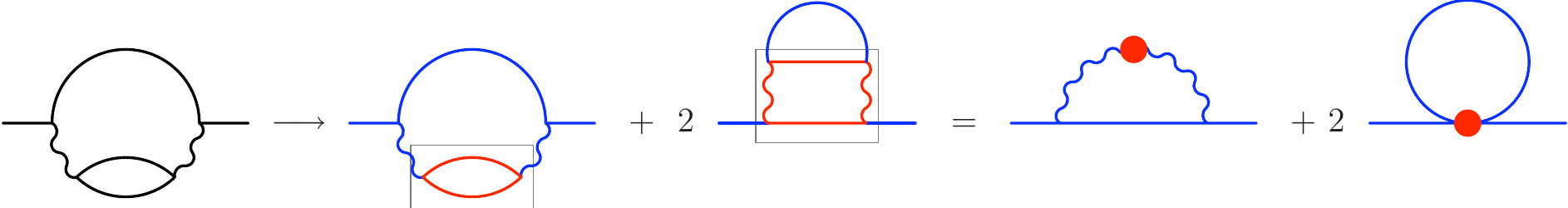}
\caption{Reduction of the two-loop bubble diagram to effective one-loop diagrams 
with HTL insertions. The framed part of the diagram represent 
the hard loop which are replaced by an effective HTL self-energy 
or vertex part.}
\label{Fig:B_HTL}
\end{figure}

Fig.~\ref{Fig:B_HTL} shows the diagrams arising when one loop is hard
and another is soft in the original two-loop bubble diagram and the hard
loop is replaced by the corresponding effective HTL two and four-point
function. The last tadpole diagram does not contribute because there is no
4-electron effective HTL vertex. The contribution of the one-loop
diagram with the insertion of the LO HTL photon polarization $\Pi$
(\ref{Eq:2P_HTL_vertex}) is
\be
\Sigma_\Pi(Q)=-e^2\int_P\gamma_\alpha G(Q-P) \gamma_\beta D^{\alpha\mu}(P)
\Pi^\textnormal{HTL}_{\mu\nu}(P) D^{\nu\beta}(P). 
\label{Eq:B_HTL}
\ee 

Working out the Dirac structure in Feynman gauge, we obtain two terms 
corresponding to the two structures of the HTL photon polarization in 
(\ref{Eq:2P_HTL_vertex})
\bea
\nonumber
\gamma^\alpha(\slQ-\slP)\gamma^\beta \delta_{0\alpha}\delta_{0\beta}&=&
(Q-P)_\rho \gamma^0\gamma^\rho\gamma^0=(Q-P)_0\gamma^0-
({\bf q}-{\bf p})\cdot {\bm \gamma}, \\
\nonumber
(Q-P)_\mu \gamma^\alpha\gamma^\mu\gamma^\beta V_\alpha V_\beta&=&
2(Q-P)^\alpha \gamma^\beta V_\alpha V_\beta-(\slQ-\slP) 
\slV \cdot \slV = 2 (Q-P)\cdot V \slV.
\eea

\begin{figure}[!t]
\includegraphics[keepaspectratio,width=0.7\textwidth,angle=0]
{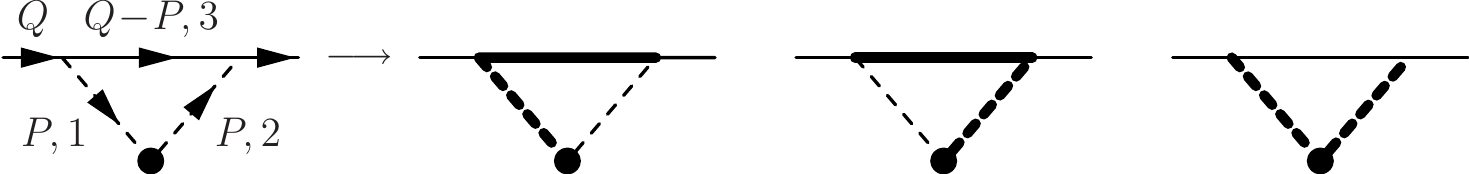}
\vspace*{-0.5cm}
\caption{The diagram corresponding to the first momentum integral of
(\ref{Eq:B_HTL_2pieces}) when it is decomposed into Gaudin tree graphs. 
The dashed (solid) line means bosonic (fermionic) propagator. 
See Fig.~\ref{Fig:b_tree} for other conventions on the lines.}
\label{Fig:B_HTL_tree}
\end{figure}

Using the relations above and setting $\q=0$, one can write 
\[ P_0\left(\frac{Q\cdot V}{V\cdot P+i\epsilon}-1\right)=
(Q-P)_0+\v\cdot\p\ Q_0 \frac{1}{V\cdot P+i\epsilon},\]
and hence,
\bea
\nonumber
\Sigma_\Pi(Q_0, \q ={\bf 0})&=&
-\frac{2}{3}e^4 T^2 Q_0 \gamma_0 \frac{\partial}{\partial M^2}
\int\frac{d \Omega}{4\pi}\int_P \v\cdot\p\, D_{M^2}(P) 
S(P)D(Q-P)\bigg|_{M^2=0}\\
&&
-i\frac{e^4 T^2}{3} \gamma_0
\frac{\partial}{\partial M^2} \int_P (Q_0 - P_0)\tilde D(Q-P) 
D_{M^2}(P)\bigg|_{M^2=0}\,,
\label{Eq:B_HTL_2pieces}
\eea
where $S(P)$ appearing here is defined by
\be
S(P) \equiv \frac{i}{P\cdot V+i\epsilon}\,,
\ee
and $D_{M^2}$ is defined by (\ref{DM2}). In imaginary time 
$S(i\omega_n,\p)=-i\Delta_S(i\omega_n,\p),$ ($\omega_n$ is bosonic) 
and the spectral reprezentation of $\Delta_S(i\omega_n,\p)$ is
\be
\Delta_S(i\omega_n,\p)=\frac{1}{\v\cdot\p-i\omega_n}=
\int_{-\infty}^\infty\frac{d p_0}{2\pi}
\frac{\rho_S(p_0,\p)}{p_0-i\omega_n},\qquad
\rho_S(p_0,\p)=2\pi \delta(p_0-\v\cdot\p)\,.
\label{Eq:D0_spectral}
\ee 
In imaginary time, since $\omega_n$ is bosonic and $\nu_q-\omega_n$ is 
fermionic, the Matsubara sum for the second term in (\ref{Eq:B_HTL_2pieces})
is given by (\ref{Eq:Matsubara-1loopFSE}). For the first term in 
(\ref{Eq:B_HTL_2pieces}) we need the Matsubara sum
\be
T\sum_n \prod_{i=1}^{3}\frac{1}{p_i^0-i\omega_i} =
-\frac{1}{p_1^0-p_2^0}\left[\frac{n_B(-p_2^0)}{p_3^0+p_2^0-i\nu_q}
-\frac{n_B(-p_1^0)}{p_3^0+p_1^0-i\nu_q}\right]
-\frac{n_F(p_3^0)}{(p_1^0+p_3^0-i\nu_q) (p_2^0+p_3^0-i\nu_q)}\,,
\label{Eq:BBF}
\ee
which was evaluated knowing that $\omega_1,\omega_2$ are 
bosonic while $\omega_3$ is fermionic. The corresponding Gaudin 
tree graphs are shown in Fig.~\ref{Fig:B_HTL_tree}. 

Using in (\ref{Eq:B_HTL_2pieces}) the spectral representation of the 
propagators and the result of the Matsubara sums given in
(\ref{Eq:Matsubara-1loopFSE}) and (\ref{Eq:BBF}), one obtains
\be
\Sigma_\Pi(i\nu_q)=\Sigma_{\Pi1}(i\nu_q)+
\Sigma_{\Pi2}(i\nu_q)+\Sigma_{\Pi3}(i\nu_q)\,,
\label{Eq:Pi_HTL}
\ee
with
\begin{subequations}
\bea
\Sigma_{\Pi1}(i\nu_q)&=&\frac{e^4 T^2}{6\pi^2}\gamma_0
\frac{\partial}{\partial M^2}\left\{
\int_0^\infty d p\, p^2\int_{-\infty}^\infty d p_0 \varepsilon(p_0)
\frac{\delta(p_0^2-E_p^2)n_B(-p_0)}{p^2-(i\nu_q-p_0)^2}\right.
\nonumber \\
&&\left.\times \left[i\nu_q-p_0 +i\nu_q p\int_{-1}^1 d x
\frac{x}{p_0-x p} \right]\right\}\bigg|_{M^2=0},
\label{Eq:Pi1a_HTL}
\\
\Sigma_{\Pi2}(i\nu_q)&=&-
\frac{e^4 T^2}{6\pi^2} i\nu_q\gamma_0
\frac{\partial}{\partial M^2} \int_{-1}^1 d x
\int_0^\infty d p \ x p^3\int_{-\infty}^\infty d p_0
\frac{\delta(p_0-x p)n_B(-p_0)}{(E_p^2-p_0^2)(p^2-(i\nu_q-p_0)^2)}
\bigg|_{M^2=0},
\label{Eq:Pi2a_HTL}
\\
\Sigma_{\Pi3}(i\nu_q)&=&
\frac{e^4 T^2}{6\pi^2}\gamma_0
\frac{\partial}{\partial M^2}\left\{
\int_0^\infty d p\, p^2\int_{-\infty}^\infty d p_0 \varepsilon(p_0)
\frac{\delta(p_0^2-p^2)n_F(p_0)}{E_p^2-(i\nu_q-p_0)^2}
\right.
\nonumber\\
&&\left.\times \left[p_0
+i\nu_q p\int_{-1}^1 d x \frac{x}{i\nu_q-p_0-x p}
\right]\right\}\bigg|_{M^2=0},
\label{Eq:Pi3a_HTL}
\eea
\end{subequations}
where $x=\cos\theta$ and the three $x$-dependent terms come from the 
first term of (\ref{Eq:B_HTL_2pieces}), while the two terms independent 
of $x$ come from the second term of (\ref{Eq:B_HTL_2pieces}).

For $\Sigma_{\Pi1}$ the comparison between the HTL expression
(\ref{Eq:Pi1a_HTL}) and the corresponding first contribution to the 
two-loop bubble diagram $\Sigma_{B1}$ is immediate, since by performing the
$x$ and $p_0$ integrals in (\ref{Eq:Pi1a_HTL}) one obtains exactly
(\ref{Eq:bubble_tree1_th}). For the other two contributions, $\Sigma_{\Pi2}$ 
and $\Sigma_{\Pi3}$, the expressions (\ref{Eq:Pi2a_HTL}) and (\ref{Eq:Pi3a_HTL}) 
derived by one HTL insertion on a soft loop coincide exactly with the leading order 
contributions (\ref{Eq:SigB2_HTL}) and (\ref{Eq:SigB3_HTL}) of the corresponding  
terms in the two-loop bubble self-energy, $\Sigma_{B2}$ and $\Sigma_{B3}$ 
respectively. The hard $k$ integral in all terms  of $\Sigma_B$ has converted
into an HTL photon self-energy insertion, with the final $p$ integral remaining
soft, thus proving explicitly the reduction of the two-loop bubble to
the one soft loop illustrated in Fig.~\ref{Fig:B_HTL}.

As a technical aside, we remark that there is a graphical correspondence between 
the Gaudin tree graphs of the two-loop bubble diagram and those of the corresponding 
HTL reduced diagrams, namely the first four diagrams of Fig.~\ref{Fig:b_tree} 
correspond to the first diagram of Fig.~\ref{Fig:B_HTL_tree}, the fifth diagram of Fig.~\ref{Fig:b_tree} corresponds to the second diagram of Fig.~\ref{Fig:B_HTL_tree} and the
last two diagrams of Fig.~\ref{Fig:b_tree} correspond to the third diagram of Fig.~\ref{Fig:B_HTL_tree}. The second correspondence between $\Sigma_{\Pi2}$ and  
$\Sigma_{B2}$ is not obvious graphically because the fifth diagram of Fig.~\ref{Fig:b_tree}
gives a product of two Fermi-Dirac statistical factors coming from the independent
frequencies corresponding to the thin lines of the fermionic bubble, whereas
the second diagram of Fig.~\ref{Fig:B_HTL_tree} contains just one Bose-Einstein
statistical factor coming from the only independent frequency in this one-loop
diagram represented by its one thin, dashed bosonic line. The detailed
correspondence of these terms is revealed only by the analysis described in 
Sec.~\ref{sec:2loop_anal} between  (\ref{Eq:SigB2hs_start}) and (\ref{Eq:SigB2_HTL}),
and relies upon the interesting relation (\ref{Eq:meta}) between a product of two Fermi-Dirac statistical factors and a mixed product of one Fermi-Dirac and one Bose-Einstein factor.
The Fermi-Dirac factor contributes to the hard-loop HTL insertion, leaving behind
just the soft-loop Bose-Einstein factor required to match the expression in (\ref{Eq:Pi2a_HTL}).

In this way the equivalence of the high temperature asymptotic
contributions of the two-loop bubble diagram with the corresponding 
HTL reduced diagram in Fig.~\ref{Fig:B_HTL} is proven.

\subsection{The HTL Reduced Rainbow Diagram}

\begin{figure}[!t]
\includegraphics[keepaspectratio,width=1.0\textwidth,angle=0]{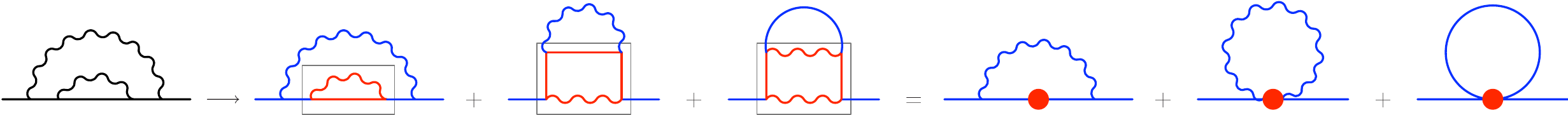}
\caption{Reduction of the two-loop rainbow diagram to effective one-loop diagrams 
with HTL insertions. The framed part of the diagram represent 
the hard loop which are replaced by an effective HTL self-energy 
or vertex part.}
\label{Fig:R_HTL}
\end{figure}

Fig.~\ref{Fig:R_HTL} shows the three diagrams obtained by considering a hard
and a soft loop in the original two-loop rainbow diagram. The last
tadpole diagram does not give contribution to the fermion self-energy
at the NLO HTL order because there is no 4-electron HTL effective
vertex.  The tadpole diagram containing the two-electron--two-photon
4-point function $\Gamma_{\mu\nu}^\textnormal{HTL}$ vanishes in Feynman 
gauge, since cf. (\ref{Eq:2E2P_HTL_vertex}) one has the contraction,
$g^{\mu\nu} V_\mu V_\nu=V_\mu V^\mu=1-{\bf v}\cdot{\bf v}=0$.

The contribution of the remaining effective one-loop diagram with the insertion of
the LO HTL fermion self-energy given in (\ref{Eq:2E_HTL_vertex}) reads
\be
\Sigma_\Sigma(Q)=-e^2\int_P\gamma_\alpha G(P) \Sigma^\textnormal{HTL} G(P)\gamma_\beta 
D^{\mu\alpha}(Q-P). 
\label{Eq:R_HTL}
\ee 
In Feynman gauge the Dirac algebra gives 
\[\gamma_\alpha\slP\slV\slP\gamma^\alpha=
-P^2\gamma_\alpha\slV\gamma^\alpha+2P\cdot V\gamma_\alpha\slP\gamma^\alpha
=2 P^2\slV-4 P\cdot V \slP.\]
Using the above expression in (\ref{Eq:R_HTL}), the integral reduces to
the sum of two terms:
\be
\Sigma_\Sigma(Q)=
\frac{e^4 T^2}{4}\int\frac{d\Omega}{4\pi}\int_P \slV S(P) 
\tilde D(P) D(Q-P)+
i\frac{e^4T^2}{2}\frac{\partial}{\partial m^2}\int_P 
G_{m^2}(P) D(Q-P)\bigg|_{m^2=0}.
\label{Eq:R_HTL_2pieces}
\ee
The propagators $S(P)$ and $G_{m^2}(P)$ were defined in
(\ref{Eq:D0_spectral}) and (\ref{Eq:handle_dp_in_rainbow}). In 
imaginary time $S(i\omega_n,\p)=-i\tilde\Delta_S(i\omega_n,\p),$ 
($\omega_n$ is fermionic) and $\tilde\Delta_S(i\omega_n,\p)$ has the same 
spectral representation as $\Delta_S(i\omega_n,\p)$ in (\ref{Eq:D0_spectral}).

The second term in (\ref{Eq:R_HTL_2pieces}) is exactly the term $\Sigma_\textrm{\tiny R2}$
encountered in the calculation of the two-loop rainbow given by
(\ref{Eq:rainbow_term_B_rt}). This can be seen immediately by applying
(\ref{Eq:diff_of_tads}) to the first integral in (\ref{Eq:rainbow_term_B_rt}). 

\begin{figure}[!t]
\includegraphics[keepaspectratio,width=0.7\textwidth,angle=0]
{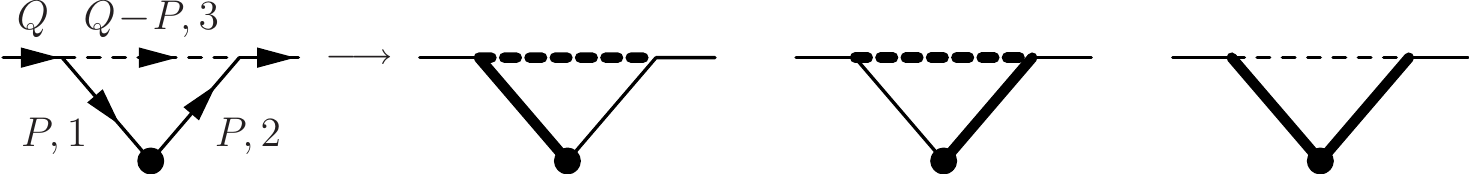}
\vspace*{-0.5cm}
\caption{The diagram corresponding to the first momentum integral of 
(\ref{Eq:R_HTL_2pieces}) when it is decomposed in Gaudin tree graphs. 
The dashed (solid) line means bosonic (fermionic) propagator. 
See Fig.~\ref{Fig:b_tree} for other conventions  on the lines.}
\label{Fig:R_HTL_tree}
\end{figure}

Since with the direct two-loop calculation we have seen that at leading 
order in the high temperature expansion the non-vanishing contribution 
comes only from $\Sigma_\textrm{\tiny R2}$, it remains to show that the first term of
(\ref{Eq:R_HTL_2pieces}) does not contribute at this order.
In imaginary time this term is
\bea
\Sigma_{\Sigma 1}(i\nu_q,\q)=-\frac{e^4 T^2}{4}\int\frac{d \Omega}{4\pi}
\int_\p 
\left(\prod_{i=1}^{3}\int\frac{d p_i^0}{2\pi}\right)\!
\rho(p_1^0,\p_1)  \rho_S(p_2^0,\p_2) \rho(p_3^0,\p_3)
T\sum_n\prod_{i=1}^{3} \frac{1}{p_i^0-i\omega_i},
\label{Eq:R_HTL_termA}
\eea
where $\p_1=\p_2=\p,$ $\p_3=\q-\p,$ $\omega_1=\omega_2=\omega_n,$ and
$\omega_3=\nu_q-\omega_n$. Since $\omega_1$ and $\omega_2$ are 
fermionic Matsubara frequencies, the result of the sum is 
\bea
T\sum_n \prod_{i=1}^{3}\frac{1}{p_i^0-i\omega_i}=
\frac{1}{p_1^0-p_2^0}\left[\frac{n_F(-p_2^0)}{p_3^0+p_2^0-i\nu_q}
-\frac{n_F(-p_1^0)}{p_3^0+p_1^0-i\nu_q}\right]
+\frac{n_B(p_3^0)}{(p_1^0+p_3^0-i\nu_q) (p_2^0+p_3^0-i\nu_q)},
\label{Eq:ffb_sum}
\eea
where the three terms correspond to the Gaudin tree graphs of
Fig.~\ref{Fig:R_HTL_tree}. 

Substituting (\ref{Eq:ffb_sum}) into (\ref{Eq:R_HTL_termA}), performing
the frequency integrals, and taking $\q = 0$, one obtains the three terms
\be
\Sigma_{\Sigma 1}(i\nu_q, \q = {\bf 0})=
\Sigma_{\Sigma 1a}(i\nu_q)+\Sigma_{\Sigma 1b}(i\nu_q)
+\Sigma_{\Sigma 1c}(i\nu_q)\,,
\ee
where
\begin{subequations}
\bea
\Sigma_{\Sigma1a}(i\nu_q)&=&
\frac{e^4 T^2}{16\pi^2}\gamma_0 
\int_{-1}^1 d x \frac{1}{x^2-1}\int_0^\infty d p
\frac{n_F(-p x)} {p^2-(i\nu_q-p x)^2},
\label{Eq:R_HTL_tree1}\\
\Sigma_{\Sigma1b}(i\nu_q)
&=&-\frac{e^4 T^2}{32\pi^2 i\nu_q}\gamma_0\int_{-1}^1 d x \frac{1}{1-x^2}
\int_0^\infty d p\, n_F(p) \frac{p}{p^2-(i\nu_q/2)^2},
\label{Eq:R_HTL_tree2}\\
\Sigma_{\Sigma1c}(i\nu_q)
&=&\frac{e^4 T^2}{32 \pi^2 i\nu_q}\gamma_0
\int_{-1}^1 d x \frac{1}{1-x^2}
\int_0^\infty d p\left[ 
\frac{p\, n_B(p)}{p^2-(i\nu_q/2)^2}-2 \frac{p\, n_B(p)}{p^2-(i\nu_q/(1+x))^2}
\right]. 
\label{Eq:R_HTL_tree3}
\eea
\end{subequations}
\vspace{-6mm}

\noindent
For $\Sigma_{\Sigma 1a}$ one observes that in the UV the integrand is
well-behaved allowing the use of $n_F(-p x)\simeq 1/2$. 
Then the momentum integral gives
\be
\frac{1}{2}\int_0^\infty d p \frac{1}{p^2-(i\nu_q-p x)^2}=
\frac{\pi}{4i\nu_q}+\frac{i}{4 i\nu_q} \ln\frac{1-x}{1+x}.
\ee
After analytical continuation one sees that the
imaginary part contains an odd function in $x$ whose integral vanishes. The
real part is collinearly divergent, but nevertheless it is subleading because
it does not contain the factor $\ln(T/\omega)$.

Using (\ref{Eq:Mitra_corollary_R}) and (\ref{Eq:L_pm_result}), the real part 
of  $\Sigma_{\Sigma 1c}$ cancels at leading order in the high 
temperature expansion against the real part of $\Sigma_{\Sigma 1b}$.
As for the imaginary part, as shown in Appendix~\ref{App:mitras_integral}
neither  $\Sigma_{\Sigma 1b}$ nor $\Sigma_{\Sigma 1c}$ acquires a
contribution at leading order in the high temperature
expansion [see (\ref{Eq:MitraF}) and (\ref{Eq:L_pm_result})].
This completes the proof that the first term in (\ref{Eq:R_HTL_2pieces}) 
does not contribute at leading order in the high temperature expansion.
Hence the only contribution of the reduced HTL rainbow diagram in 
Fig.~\ref{Fig:R_HTL} at leading order in the high temperature expansion
is given by the second term of  (\ref{Eq:R_HTL_2pieces}),
which agrees exactly with the corresponding rainbow diagram contribution
of (\ref{Eq:R_I_final}).

\subsection{The HTL Reduced Crossed Photon Diagram}

\begin{figure}[t]
\includegraphics[keepaspectratio,width=1.0\textwidth,angle=0]{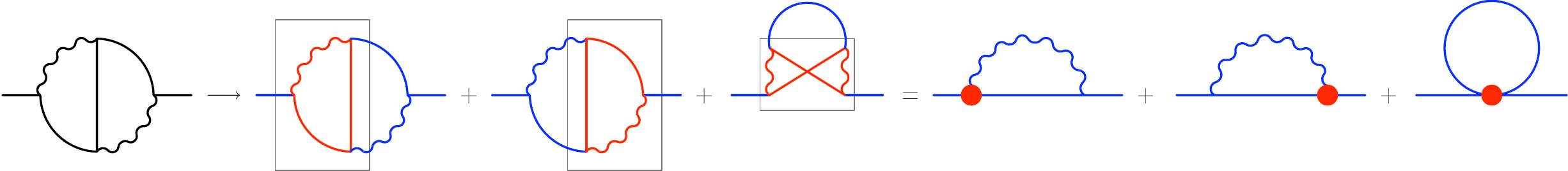}
\caption{Reduction of the two-loop crossed photon diagram to effective 
one-loop diagrams with HTL insertions. The framed part of the diagram represent 
the hard loop which are replaced by an effective HTL vertex part.}
\label{Fig:CR_HTL}
\end{figure}

The reduction of the two-loop crossed photon diagram into diagrams
containing one HTL insertion and one soft loop is shown in Fig.~\ref{Fig:CR_HTL}. 
The last diagram in Fig.~\ref{Fig:CR_HTL} does not contribute to the fermion
self-energy at NLO HTL order because there is no 4-electron HTL effective
vertex. The contribution of the two remaining  diagrams in Fig.~\ref{Fig:CR_HTL} 
in which one of the original loop integral is replaced by the effective HTL
vertex given by (\ref{Eq:2E1P_HTL_vertex}) are equal and give
\be
\Sigma_\Gamma(Q)=-2i e^2\int_P \gamma_\alpha G(Q-P) 
\Gamma_\nu^\textnormal{HTL}(Q-P,Q) D^{\nu\alpha}(P)\,.
\label{Eq:HTL_CR_a}
\ee 
In Feynman gauge ($\xi=1$) one has 
$\gamma_\alpha (\slQ-\slP) V^\alpha \slV=2(Q-P)\cdot V \slV$ and the 
angular and momentum integrals factorize:
\be
\Sigma_\Gamma(Q)=i\frac{e^4 T^2}{2} \int\frac{d\Omega}{4\pi} 
\frac{\slV}{Q\cdot V}\int_P \tilde D(Q-P) D(P).
\label{Eq:HTL_CR_b}
\ee
In imaginary time, using the spectral representation of the propagators 
and performing the Matsubara sum coming from the momentum integral of 
(\ref{Eq:HTL_CR_b}), one obtains
\be
\Sigma_\Gamma(i\nu_q,\q)=-\frac{e^4 T^2}{2}\gamma_0
\int\frac{d\Omega}{4\pi}
\frac{\slV}{Q\cdot V}
\int_\p
\int_{-\infty}^\infty\frac{d p_0}{2\pi}
\int_{-\infty}^\infty\frac{d r_0}{2\pi}
\rho(p_0,\p) \rho(r_0,\q-\p)
\frac{n_F(r_0)+n_B(-p_0)}{p_0+r_0-i\nu_q}.
\label{Eq:HTL_CR}
\ee
For $\q=0$ one has $Q\cdot V=i\nu_q$ and one frequency integral
can be performed by undoing the spectral representation of the propagator, 
while the other frequency integral can be performed by the use the Dirac 
$\delta$-function in the spectral function. Retaining only the terms with 
two statistical factors, one obtains
\be
\Sigma_\Gamma (i\nu_q)=
\frac{e^4 T^2}{16\pi^2 i\nu_q}\gamma_0
\int_0^\infty d p \frac{p}{p^2-(i\nu_q)^2/4}\big[n_B(p)-n_F(p)\big]\,,
\label{Eq:HTL_CR_final}
\ee
which is exactly the result of (\ref{Eq:CR_final}). 

Thus, the full leading order high temperature contributions of the perturbative 
two-loop fermion self-energy coincide with those of the corresponding one-loop
reduced HTL diagrams, according to Figs.~\ref{Fig:B_HTL},~\ref{Fig:R_HTL}, and 
\ref{Fig:CR_HTL}, just as required by the hard-soft pattern exhibited in the
two-loop self-energy.

%------@#$---------- GAUGE DEPENDENCE-----------------

\section{Gauge Parameter Dependence at Two Loops
\label{sec:xi}}

Having computed the full two-loop electron self-energy in the Feynman
gauge $\xi =1$ we consider the dependence of the self-energy upon
the gauge parameter $\xi$ at two-loops. For $\xi=1$ we have shown that 
all the leading order terms in the high temperature expansion of the two-loop 
self-energy exhibit a factorized hard-soft pattern. This pattern should
continue to hold for $\xi \neq 1$, because in Euclidean time the additional 
factor of $(1-\xi) K^{\mu}K^{\nu}/K^2$ in the photon propagator is homogeneous
of order zero under rescaling $K^{\mu} \rightarrow \lambda K^{\mu}$. Hence
it neither enhances nor suppresses the loop momentum Matsubara sum
and integration with respect to the $\xi = 1$ case, and we expect the
same hard-soft pattern to hold. Assuming this to be the case, we can 
analyze the $\xi$ dependence of the remaining soft loop integration
in the reduced HTL description of the last section. 

One immediate effect of the factorization into one hard, one soft 
loop is that all $\xi$ dependence drops out of the internal hard thermal
loop, so that it is impossible to find any quadratic dependence on 
$\xi$ in the leading high temperature terms of the two-loop self-energy,
despite the fact that the original diagrams contain two photon propagators.
The remaining $\xi$ dependence can at most be linear. In this section 
we analyze the linear $\xi$ dependent terms in the two-loop self-energy, 
assuming the hard-soft pattern and HTL reduction to effective one-loop 
diagrams.\footnote{Some preliminary work on the gauge parameter dependence 
was reported in \cite{CarMot}, with however a differing result for the real part, 
which we calculate fully here.} We note also that owing  to the transversality of the 
HTL photon polarization $\Pi^\textnormal{HTL}_{\mu\nu}$, the HTL reduced
bubble diagram has no $\xi$ parameter dependence at all. 

For the rainbow diagram the first framed self-energy correction 
diagram of Fig.~\ref{Fig:R_HTL} gives
\be
^\xi\Sigma_\Sigma(Q)=(1-\xi) i e^2 \int_K \left[\gamma_{\mu} 
\frac{1}{\slQ -\slK} \Sigma^\textnormal{HTL}(Q-K) 
\frac{1}{\slQ -\slK} \gamma_{\nu}\right] \frac{K^{\mu}K^{\nu}} {(K^2)^2}, 
\ee
while the second framed diagram of Fig.~\ref{Fig:R_HTL} gives
the tadpole contribution
\be
^\xi\Sigma_T(Q)=
(1-\xi)i e^2\int_{K} \Gamma_{\mu\nu}^{HTL}(Q, Q, K) 
\frac{K^{\mu}K^{\nu}} {(K^2)^2} \,.
\ee 
In these expressions $\Gamma_{\mu}^{HTL}(P_1, P_2)$ and
$\Gamma_{\mu\nu}^{HTL}(Q, Q, K)$ are the HTL vertices
defined by (\ref{Eq:2E1P_HTL_vertex}) and (\ref{Eq:2E2P_HTL_vertex}).

For the crossed photon diagram the reduction of one of the loops 
into a HTL vertex yields the $\xi$ dependent terms
\be
^\xi\Sigma_\Gamma(Q)=(1-\xi) i e^2 
\int_{K} \left[\gamma_{\mu} \frac{1}{\slQ -\slK}\Gamma^{HTL}_{\nu}(Q-K, Q)+ 
\Gamma_{\mu}^{HTL}(Q, Q-K) \frac{1}{\slQ -\slK} 
\gamma_{\nu}\right] \frac{K^{\mu}K^{\nu}}{(K^2)^2} 
\ee
from the first two framed diagrams of Fig.~\ref{Fig:CR_HTL}, 
while the third framed diagram of Fig.~\ref{Fig:CR_HTL} gives no contribution,
since the corresponding LO HTL 4-fermion vertex vanishes.

Since the HTL vertex functions obey the Ward Identities (\ref{HTLWard1}), 
(\ref{HTLWard2}), and (\ref{HTLWard3}) , we may contract the $K^{\mu}K^{\nu}$ 
vectors into these vertices  to express all the $\xi$ dependent terms in terms 
of the HTL self-energy $\Sigma^\textnormal{HTL}$ in the form
\bea
\nonumber
^\xi\Sigma_2(Q)&\equiv& {^\xi\Sigma}_\Gamma(Q)+
{^\xi\Sigma}_\Sigma(Q)+{^\xi\Sigma}_T(Q)\\
&=&(1-\xi)i e^2 \int_K\ \frac{1}{(K^2)^2} \left\{
\left[ \slQ \frac{1}{\slQ -\slK} - 1\right] 
\left[  \Sigma^\textnormal{HTL}(Q-K)-
\Sigma^\textnormal{HTL}(Q)\right]\right.\nonumber\\
&& \left.+ \left[\Sigma^\textnormal{HTL}(Q-K) -  
\Sigma^\textnormal{HTL}(Q)\right]
\left[ \frac{1}{\slQ -\slK} \slQ - 1\right] \right\}\nonumber\\
&&
\nonumber
+(1-\xi)i e^2 \int_K\ \frac{1}{(K^2)^2}
\left[\slQ \frac{1}{\slQ -\slK} - 1\right] 
\Sigma^\textnormal{HTL}(Q-K) 
\left[\frac{1}{\slQ -\slK} \slQ - 1\right]
\\
&& + (1-\xi)i e^2 \int_K \frac{1}{(K^2)^2} 
\left\{\Sigma^\textnormal{HTL}(Q + K)- 2 \Sigma^\textnormal{HTL}(Q) 
+ \Sigma^\textnormal{HTL}(Q-K)\right\}\nonumber\\
&=& (1-\xi)i e^2 \int_K \frac{1}{(K^2)^2} 
\left\{- \slQ \frac{1}{\slQ -\slK}  \Sigma^\textnormal{HTL}(Q)
- \Sigma^\textnormal{HTL}(Q) \frac{1}{\slQ -\slK}\slQ\right. \nonumber\\
&&\left.+ \slQ \frac{1}{\slQ -\slK}  \Sigma^\textnormal{HTL}(Q-K) 
\frac{1}{\slQ -\slK}\slQ +\Sigma^\textnormal{HTL}(Q+K)\right\}\,,
\eea
for the sum of these two-loop $\xi$ dependent contributions. In order
to find the magnitude of this two-loop gauge dependent self-energy,
we substitute the LO HTL condition, $\slQ = \Sigma^\textnormal{HTL}(Q)$ 
and add the result to the one-loop $\xi$ dependent self-energy (\ref{oneloopxi}), 
to obtain 
\bea
\nonumber
^\xi\Sigma_\textnormal{on-shell}(Q)&\equiv&
[{^\xi\Sigma}_1(Q)+{^\xi\Sigma}_2(Q)]_\textnormal{on-shell}
= (1-\xi)i e^2 \int_K \frac{1}{(K^2)^2} 
\left\{-\slQ \frac{1}{\slQ -\slK}\slQ - (\slQ + \slK) \right.\\
&&\left.
+ \slQ \frac{1}{\slQ -\slK} \Sigma^\textnormal{HTL}(Q-K) 
\frac{1}{\slQ -\slK}\slQ + \Sigma^\textnormal{HTL}(Q+K)
\right\}_\textnormal{on-shell}\,
\label{Eq:gauge1}
\eea
for the sum. 

We note that the relative sign of the first two terms in (\ref{Eq:gauge1})
is opposite that for one-loop $\xi$ dependent self-energy (\ref{oneloopxi}).
Following the familiar methods these two terms can 
be written in the forms
\begin{subequations}
\begin{eqnarray}
&& -(1-\xi)i e^2 \int_K \frac{1}{(K^2)^2} \slQ \frac{1}{\slQ -\slK}\slQ
= - \frac{(1-\xi) e^2}{2 \pi^2}\, (i \nu_q)^2 \gamma_0\, F_1\,,\\
&& -(1-\xi)i e^2 \int_K \frac{1}{(K^2)^2} (\slQ + \slK) =
-\frac{(1-\xi) e^2}{2 \pi^2}\,(i \nu_q) \gamma_0\, F_2\,,
\end{eqnarray}
\label{xiIandxiII}
\end{subequations}
\vspace{-6mm}

\noindent
for $\bf q = 0$, where
\begin{subequations}
\begin{eqnarray}
&&F_1 \equiv \frac{\partial}{\partial M^2} \int_0^{\infty} k^2 dk\,
\int_{-\infty}^{\infty} \frac{dk_0}{2 \pi}\, \rho_{M^2} (k_0, {\bf k}) 
\int_{-\infty}^{\infty}\frac{dp_0}{2 \pi}\, p_0\ \rho (p_0, -{\bf k})\,
\left[\frac{n_B(-k_0)+n_F(p_0)}{k_0+p_0-i\nu_q}\right]\,,\\
&&F_2 \equiv \frac{\partial}{\partial M^2} \int_0^{\infty} k^2 dk\,
\int_{-\infty}^{\infty} \frac{dk_0}{2 \pi}\, \rho_{M^2} (k_0, {\bf k}) 
\, n_B(k_0)\,,
\end{eqnarray}
\label{F1andF2def}
\end{subequations}
\vspace{-6mm}

\noindent
with the limit $M \rightarrow 0$ (if it exists) to be taken after integration 
is understood. After performing the integrations over all variables which 
are {\it not} arguments of the statistical distributions, and an integration
by parts in the fermionic term, we find that the thermal part of
$F_1$ may be expressed in the form
\begin{eqnarray}
&&F_1 = - \frac{1}{8i\nu_q} \int_0^{\infty} \frac{dE}{E^2 - (i\nu_q/2)^2}\,
\frac {d}{dE} [E^2 n_F(E)] \nonumber\\
&& \qquad - \frac{1}{8(i\nu_q)^2} \int_0^{\infty} dE\,E\,n_B(E)
\left[\frac{E+i\nu_q}{(E+ i\nu_q/2)^2} - \frac{E-i\nu_q}{(E - i\nu_q/2)^2}\right]\nonumber\\
&& \qquad - \frac{1}{8i\nu_q} \int_M^{\infty} \frac{dE\, n_B(E)}{\sqrt{E^2-M^2}}
\left[\frac{E+i\nu_q}{E+ i\nu_q/2} + \frac{E-i\nu_q}{E - i\nu_q/2}\right]\,,
\label{F1}
\end{eqnarray}
where the finite $M$ regulator must be retained only in the last integral.
Likewise for $F_2$,
\be
F_2 = -\frac{1}{2} \int_M^{\infty} \frac{dE\, n_B(E)}{\sqrt{E^2-M^2}} \,.
\label{F2}
\ee
The high temperature asymptotic forms of each of the integrals in $F_1$ 
and $F_2$ may be obtained by the methods of Appendix \ref{App:mitras_integral},
 with the results
\begin{subequations}
\begin{eqnarray}
&&F_1 = -\frac{1}{4\omega}\left[\frac{\pi T}{M} - \ln\frac{\omega}{M}
- \frac{3i\pi T}{2\omega}\right] + \cdots\\
&&F_2 = -\frac{1}{4}\left[\frac{\pi T}{M} - \ln\frac{T}{M}\right] + \cdots
\end{eqnarray}
\label{F12eval}
\end{subequations}
\vspace{-6mm}

\noindent
after continuation to real time, $i\nu_q \rightarrow \omega + i 0^+$.
Thus each of the first two terms in (\ref{Eq:gauge1}) contain both
linear and logarithmic divergences as the photon regulator mass
$M \rightarrow 0$. These add in (\ref{Eq:gauge1}) but cancel when combined 
with the opposite relative sign in (\ref{oneloopxi}), to yield the finite 
result (\ref{Eq:1loop_xi_RI}) for the high $T$ asymptotic form of the gauge 
dependent part of the one-loop self-energy.

For the third and fourth terms in (\ref{Eq:gauge1}) one uses the definition 
of $\Sigma^\textnormal{HTL}$ given in (\ref{Eq:2E_HTL_vertex}), performs 
the Dirac algebra, using 
\be
\slQ (\slQ - \slK) \slV (\slQ -\slK) \slQ
= 2 (Q - K) \cdot V [ (Q-K)^2 \slQ - K^2 \slQ + Q^2\slK]
+ (Q-K)^2 [Q^2 \slV - 2 Q \cdot V \slQ]
\ee
and the Matsubara sums, to obtain for $\bf q = 0$, 
\be
(1-\xi)i e^2 \int_K \frac{1}{(K^2)^2} \slQ \frac{1}{\slQ -\slK} 
\Sigma^\textnormal{HTL}(Q-K) \frac{1}{\slQ -\slK}\slQ = 
\frac{(1 - \xi) e^2}{\pi^2} m_f^2 (i \nu_q)\gamma_0 
\left(G_1 - G_2 + i\nu_q\,G_3 - \frac{i \nu_q}{4}\, G_4\right)\,,
\label{xiIII}
\ee
and
\be
(1-\xi)i e^2 \int_K \frac{1}{(K^2)^2}\Sigma^\textnormal{HTL}(Q+K) = 
-\frac{(1 - \xi) e^2}{4\pi^2} m_f^2 \gamma_0\,H\,,
\label{xiIV}
\ee
where
\begin{subequations}
\begin{eqnarray}
&&G_1 \equiv \frac{\partial}{\partial M^2} \int_0^{\infty} k^2 dk\,
\int_{-\infty}^{\infty} \frac{dk_0}{2 \pi}\, \rho_{M^2} (k_0, {\bf k}) 
\int_{-\infty}^{\infty}\frac{dp_0}{2 \pi}\, \rho (p_0, -{\bf k})\,
\left[\frac{n_B(-k_0)+n_F(p_0)}{k_0+p_0-i\nu_q}\right]\,,\\
&&G_2 \equiv \frac{\partial}{\partial m^2} \int_0^{\infty} k^2 dk\,
\int_{-\infty}^{\infty} \frac{dk_0}{2 \pi}\, \rho (k_0, {\bf k}) 
\int_{-\infty}^{\infty}\frac{dp_0}{2 \pi}\, \rho_{m^2} (p_0, -{\bf k})\,
\left[\frac{n_B(-k_0)+n_F(p_0)}{k_0+p_0-i\nu_q}\right]\,,\\
&&G_3 \equiv \frac{\partial}{\partial M^2}\frac{\partial}{\partial m^2} 
\int_0^{\infty} k^2 dk\, \int_{-\infty}^{\infty} \frac{dk_0}{2 \pi}\,k_0\,
\rho (k_0, {\bf k}) \int_{-\infty}^{\infty}\frac{dp_0}{2 \pi}\, 
\rho_{m^2} (p_0, -{\bf k})\, \left[\frac{n_B(-k_0)+n_F(p_0)}{k_0+p_0-i\nu_q}\right],\\
&&G_4 \equiv \frac{\partial}{\partial M^2} \int_0^{\infty} k^2 dk\,
\int_{-\infty}^{\infty} \frac{dk_0}{2 \pi}\, \rho_{M^2} (k_0, {\bf k}) 
\int_{-\infty}^{\infty}\frac{dp_0}{2 \pi}\, \rho (p_0, -{\bf k})
\int_{-1}^1 dx \int_{-\infty}^{\infty} dr_0\, \delta(r_0 + kx) \times\nonumber\\
&& \qquad \left\{\frac{n_B(k_0)}{(k_0+p_0-i\nu_q)(k_0+r_0-i\nu_q)}
+ \frac{1}{p_0 - r_0}\left[\frac{n_F(p_0)}{k_0+p_0-i\nu_q} 
-\frac{n_F(r_0)}{k_0+r_0-i\nu_q}\right]\right\}\,,
\label{threetree}
\end{eqnarray}
\label{G1234defs}
\end{subequations}
\vspace{-6mm}

\noindent
and
\be
H \equiv \frac{\partial}{\partial M^2} \int_0^{\infty} k^2 dk\,
\int_{-\infty}^{\infty} \frac{dk_0}{2 \pi}\, \rho_{M^2} (k_0, {\bf k}) 
\int_{-1}^{1} dx\int_{-\infty}^{\infty}dr_0\,\delta(r_0 + kx) \,
\left[\frac{n_B(-k_0)+n_F(r_0)}{k_0+r_0+i\nu_q}\right]\,,
\label{Hdef}
\ee
The Matsubara sum needed to obtain (\ref{threetree}) is given by 
a formula analogous to (\ref{Eq:ffb_sum}). corresponding to the Gaudin tree graphs of Fig.~\ref{Fig:R_HTL_tree}. 
In order to deal with the double poles in both the bosonic and fermionic
propagators we have made use of the massive regulator (\ref{DM2}) as before,
with $M$ the bosonic mass and $m$ the fermionic one, the limits
$M,\, m \rightarrow 0$ (if they exist) to be taken at the end.

The integrals in (\ref{G1234defs}) and (\ref{Hdef}) are similar to those we 
have encountered above and in the previous Feynman gauge analysis. 
The most efficient method of evaluating them is to write them as a sum of
terms and in each term use the delta functions to evaluate all the integrals 
which do not involve the Bose-Einstein or Fermi-Dirac statistical functions,
leaving this single frequency integral for last, using relations 
(\ref{BE_FD_identities}) to retain only the thermal parts as integrals
over positive range. In this way we obtain
\begin{subequations}
\begin{eqnarray}
&&G_1 = -\frac{1}{8\,(i\nu_q)^2} \int_0^{\infty} dE\,E\,[n_F(E) + n_B(E)]
\left[\frac{1}{(E+ i\nu_q/2)^2} + \frac{1}{(E - i\nu_q/2)^2}\right]\nonumber\\
&&\qquad\qquad + \frac{1}{8}\int_M^{\infty} \frac{dE}{\sqrt{E^2-M^2}}
\,\frac{n_B(E)}{E^2 - (i\nu_q/2)^2} \,,\\
&&G_2 =  \frac{1}{8\,(i\nu_q)^2} \int_0^{\infty} dE\,E\,[n_F(E) + n_B(E)]
\left[\frac{1}{(E+ i\nu_q/2)^2} + \frac{1}{(E - i\nu_q/2)^2}\right]\nonumber\\
&&\qquad\qquad - \frac{1}{8}\int_m^{\infty} \frac{dE}{\sqrt{E^2-m^2}}
\,\frac{n_F(E)}{E^2 - (i\nu_q/2)^2} \,,\\
&&G_3 = \frac{1}{16\,(i\nu_q)^2}\int_0^{\infty} dE\,n_B(E)
\left[\frac{1}{(E+ i\nu_q/2)^2} - \frac{1}{(E - i\nu_q/2)^2}\right]\nonumber\\
&&\qquad\qquad + \frac{1}{8\,(i\nu_q)^3} \int_0^{\infty} dE\,E^2\, n_B(E)
\left[\frac{1}{(E+ i\nu_q/2)^3} + \frac{1}{(E - i\nu_q/2)^3}\right]\nonumber\\
&& \quad - \frac{1}{8\,(i\nu_q)^3} \int_0^{\infty} dE\,E\, n_F(E) \left[
\frac{E+i\nu_q}{(E+ i\nu_q/2)^3} + \frac{E-i\nu_q}{(E - i\nu_q/2)^3}\right]\nonumber\\ 
&& \qquad\qquad + \frac{1}{16\,(i\nu_q)^2}\int_m^{\infty} \frac{dE\, n_F(E)}{\sqrt{E^2-m^2}}
\left[\frac{E + i\nu_q}{(E+ i\nu_q/2)^2} - \frac{E-i\nu_q}{(E - i\nu_q/2)^2}\right]\,,\\
&&G_4 = - \frac{1}{8\,(i\nu_q)^2} \int_0^{\infty} dE\, n_B(E) 
\left\{ \frac{1}{(E+i\nu_q/2)^2} \ln\left(\frac{2E
+ i\nu_q}{i\nu_q} \right) -  \frac{1}{(E-i\nu_q/2)^2}
\ln\left(\frac{2E - i\nu_q}{-i\nu_q}\right)\right\}\nonumber\\
&&\qquad\qquad - \frac{1}{8\,(i\nu_q)^2} \int_M^{\infty} \frac{dE\,n_B(E)}{\sqrt{E^2-M^2}}\left[
\frac{E+i\nu_q}{(E+ i\nu_q/2)^2} - \frac{E-i\nu_q}{(E - i\nu_q/2)^2}\right] \nonumber\\
&& +\frac{1}{8\,(i\nu_q)^2} \int_0^{\infty} dE\, n_F(E) 
\left\{ \frac{1}{(E+i\nu_q/2)^2} \ln\left[\frac{4E^2}{-2iE\nu_q 
- (i\nu_q)^2}\right] -  \frac{1}{(E-i\nu_q/2)^2} 
\ln\left[\frac{4E^2}{2iE\nu_q - (i\nu_q)^2}\right]\right\}  \nonumber\\
&&\qquad\qquad + \frac{1}{8\,(i\nu_q)^2} \int_0^{\infty} dE\, n_F(E) 
\left[ \frac{1}{(E+i\nu_q/2)^2} -  \frac{1}{(E-i\nu_q/2)^2}\right]\,,
\end{eqnarray}
\label{G1234int}
\end{subequations}
\vspace{-6mm}

\noindent
and
\be
H = \frac{1}{4\,i\nu_q} \int_M^{\infty} \frac{dE\,n_B(E)}{\sqrt{E^2-M^2}}
\left[\frac{E+i\nu_q}{E+ i\nu_q/2} + \frac{E-i\nu_q}{E - i\nu_q/2}\right]
+ \frac{1}{2\,i\nu_q}  \int_0^{\infty} \frac{dE\,E\,n_F(E)}{E^2 - (i\nu_q/2)^2} \,,
\label{Hint}
\ee
where we have set $M$ and $m$ to zero wherever the integrals are finite after differentiation,
retaining the regulator masses only when they are needed.

The high temperature asymptotic forms of the integrals (\ref{G1234int}) and (\ref{Hint}), 
evaluated by the methods of Appendix \ref{App:mitras_integral} are
\begin{subequations}
\begin{eqnarray}
&&G_1 = \frac{1}{4\omega^2}\left[ -\frac{\pi T}{M} + \ln\frac{\omega}{M} + \frac{2\pi i T}{\omega}\right] + \dots\\
&&G_2 = \frac{1}{4\omega^2}\ln\frac{\omega}{m} + \dots\\
&&G_3 = \frac{1}{4\omega^3}\left[ \ln\frac{\omega}{m} + \frac{i\pi T}{\omega}\right] + \cdots\\
&&G_4 = -\frac{1}{2\omega^3}\left[\frac{\pi T}{M} - \ln\frac{\omega}{M} - \frac{3\pi i T}{\omega}\right] + \dots
\end{eqnarray}
\label{G1234eval}
\end{subequations}
\vspace{-6mm}

\noindent
and
\be
H = \frac{1}{2\omega}\left[\frac{\pi T}{M} - \ln\frac{\omega}{M} - \frac{i\pi T}{\omega}\right] + \dots
\label{Heval}
\ee
upon analytic continuation to real time. 

Assembling the results for the asymptotic forms (\ref{F12eval}), (\ref{G1234eval}) and (\ref{Heval})
in (\ref{Eq:gauge1}), (\ref{xiIandxiII}), (\ref{xiIII}) and (\ref{xiIV}), with $\omega = m_f = eT/\sqrt{8}$,
we find finally
\be
^\xi\Sigma_\textnormal{on-shell}(Q) = (1-\xi)  \frac{e^2}{8\pi^2}\,\gamma_0 \left[
- m_f \ln\frac{T}{m_f} + \frac{5 i \pi T}{2} \right] + \dots
\label{xiresult}
\ee
where we have everywhere neglected terms of order $e^2 m_f \sim e^3 T$ subdominant to the 
logarithm in the real part and $e^2 T$ in the imaginary part. Although all dependence on the 
regulator masses $M$ and $m$ have dropped out of the sum of terms in (\ref{Eq:gauge1}), neither 
the real nor imaginary parts of the gauge dependent self-energy vanish, even when the on-shell 
LO HTL condition is used. In addition to the infrared divergence encountered in the two-loop 
bubble diagram, this residual gauge parameter dependence in a physical quantity
such as the fermion lifetime in the plasma indicates that a resummation of 
the perturbative series in $e^2$ is necessary in high temperature QED.
Such a resummation was performed for QED in \cite{meg07,meg08} within the HTL 
resummation program. After HTL resummation both the electron damping rate and 
the electron mass were calculated at NLO and shown to be independent of the gauge 
fixing parameter $\xi.$ 

%------@#$---------- SUMMARY OF RESULTS  -----------------

\section{Summary and Discussion}
\label{sec:sum}

We have presented a thorough analysis of the leading order 
contributions to the high temperature limit of the
two-loop electron self-energy at rest with respect to the plasma
($\q=0$). The three two-loop self-energy diagrams
illustrated in Fig.~\ref{Fig:diagrams} yield (\ref{sumB}), 
(\ref{Eq:R_I_final}), and (\ref{CR_NLO}) at leading order 
for $T \gg \omega$ in Feynman gauge, {\it i.e.},
\begin{subequations}
\bea
\Sigma_\textrm{\tiny B}(\omega)\Big\vert_{\xi =1} &\simeq&
\frac{e^4 T^2}{24 \pi^2 \omega}\,\gamma_0 \left[
\frac{\pi T}{M} + \left(\frac{1}{2}+\ln 2\right)\ln \frac{T}{\omega}
- \frac{5i\pi}{2}\,\frac{T}{\omega}\right]\,,
\label{Eq:B_collected}\\
\Sigma_\textrm{\tiny R}(\omega) \Big\vert_{\xi =1} &\simeq& 
-i\frac{e^4 T^3}{32\pi\omega^2}\,\gamma_0\,,
\label{Eq:R_collected}
\\
\Sigma_\textrm{\tiny C}(\omega)\Big\vert_{\xi =1} &\simeq&
\frac{e^4 T^2}{16\pi^2\omega}\,\gamma_0\left[
-\ln\frac{T}{\omega}+ i \pi\,\frac{T}{\omega}
\right]\,.
\label{Eq:C_collected}
\eea 
\label{Sig2Collected}
\end{subequations}
\vspace{-.6cm}

\noindent
As explained in the Introduction, the photon mass $M$ was introduced to 
regulate the infrared and/or collinear divergences in the self-energy
bubble diagram. All terms which are either independent of temperature
or vanish in the high temperature limit $T/\omega,\ T/M \rightarrow \infty$
have been omitted from (\ref{Sig2Collected}). Terms which remain finite
in the subsequent limit of removing the regulator, $M/\omega \rightarrow 0$ 
have also been omitted.

The first important observation about the leading high temperature behavior 
of the two-loop self-energy given in (\ref{Sig2Collected}) is that it is more 
infrared singular, both as $\omega \rightarrow 0$ and $M \rightarrow 0$ 
than the corresponding one-loop LO and NLO self-energy, 
\be
\Sigma_\textnormal{1-loop} (\omega,{\bf 0}) = 
\frac{e^2 T^2}{8\omega}\,\gamma_0
+\xi\frac{e^2 \omega}{8\pi^2}\,\gamma_0\,
\ln\frac{T}{\omega}
+ i (1-3\xi)\frac{e^2 T}{16\pi}\,\gamma_0,  
\label{Eq:collecting1}
\ee
calculated in (\ref{Eq:1-loop-RI}) and (\ref{Eq:uj_on-shell}) and Refs.
\cite{mitra00,wang04} in covariant $\xi$ gauges.
Since to LO, $\omega = m_f = eT/\sqrt{8}$, the NLO one-loop self-energy
gives a correction to its real part of order $e^3 \ln (1/e)$, and
an imaginary part of order $e^2T$. Using the lowest order value
for $\omega = m_f$ for an electron on-shell and at rest in the plasma, we observe
that the leading order terms in the high temperature expansion at two loops, 
given by (\ref{Eq:R_collected}) and (\ref{Eq:C_collected}) are comparable 
to the NLO terms from one-loop, again $e^3 \ln (1/e)$ in its real part, 
and $e^2T$ in its imaginary part for this value of $\omega \sim eT$.
The higher power of $e^2$ in the numerator of the two-loop self-energy 
is compensated by the more singular behavior of the denominators 
(\ref{Sig2Collected}) when $\omega \sim eT$. The $1/M$ term in the 
real part of the bubble contribution (\ref{Eq:B_collected}) violates the 
rule of finite NLO contribution to the real part of $\Sigma$ of order
$e^3 \ln (1/e)$, and gives a true, uncanceled infrared divergence 
as the photon mass regulator $M \rightarrow 0$ in bare perturbation theory.
The singular infrared behavior of (\ref{Sig2Collected}), and the uncanceled 
linear divergence as $M \rightarrow 0$ in (\ref{Eq:B_collected}) clearly 
demonstrate the breakdown of the usual perturbative loop expansion, 
even for very weak coupling $e \ll 1$.

The gauge dependent terms of the self-energy up to two-loop order on-shell
are given by (\ref{xiresult}), {\it viz.}
\be
^\xi\Sigma_\textnormal{on-shell}(Q) = 
[{^\xi\Sigma}_1(Q)+{^\xi\Sigma}_2(Q)]_\textnormal{on-shell} =
(1-\xi) \frac{e^2}{8\pi^2}\,\gamma_0 \left[
- m_f \ln\frac{T}{m_f} + \frac{5 i \pi T}{2} \right]\,,
\label{gaugedep12}
\ee
and are also of the same order as the NLO one-loop result (\ref{Eq:collecting1}).
This non-cancellation of the gauge parameter dependence even on-shell
is a second indication of the breakdown of weak coupling perturbation theory at high temperatures, the need for resummation and eventual restructuring of the usual 
(bare) perturbative loop expansion.

In our direct evaluation of the two-loop self-energy in bare perturbation theory, 
the photon mass $M$ is a regulator, introduced to define the double pole terms
in (\ref{Mreg}), and should be taken to zero. However, anticipating the result of
HTL resummation of the NLO terms, we can estimate instead that after resummation
$M$ will be replaced by the photon Debye mass, so that $M \sim eT$ finally. With this estimate 
we observe that the $1/M$ term in (\ref{Eq:B_collected}) gives a contribution to the
real part of the self-energy on-shell which is of order $e^2T$, parametrically {\it larger} 
than the $e^3 \ln(1/e)$ correction of both the NLO one-loop terms and the other leading 
order two-loop terms. Thus, the electron effective thermal mass at next-to-leading order 
in the HTL approximation is of order ${\cal O}(e^2 T)$, {\it not} of order $e^3 T \ln(1/e)$ 
as one might expect from the NLO one-loop result, and {\it both} the real and imaginary 
parts of the full resummed self-energy contain order $e^2T$ corrections. In fact, 
the resummation within the HTL effective theory has been calculated in both its 
imaginary and real parts, and a gauge parameter independent result of order 
${\cal O}(e^2 T)$ for both has been obtained \cite{meg07,meg08}. 

The origin of this result can be understood intuitively also in the following way.
From (\ref{Eq:sigma-1loop_final}) in the LO HTL one-loop self-energy 
there is a contribution proportional to the Bose-Einstein statistical 
factor of the photon of the form
\be
\frac{e^2}{2\pi^2 \omega} \gamma_0\int_0^\infty d k \frac{k}{e^{k/T} - 1}\,.
\label{1loopSE}
\ee
Because the Bose-Einstein distribution behaves like $T/k$ at small $k$,
this hard integral converges linearly as $k\rightarrow 0$, and as we have noted
is dominated by $k \sim T$. However at two loops the bubble diagram has an 
additional photon propagator, and photon self-energy mass insertion proportional
to $\Pi^\textnormal{HTL}/k^2 \sim e^2T^2/6k^2$ relative to the one-loop contribution
(\ref{1loopSE}). Thus  we should expect an integral with an additional power of
$e^2T^2/6k^2$ in the integrand, and the two-loop bubble to make 
a contribution to $\Sigma$ of
\be
\frac{e^4T^2}{12\pi^2 \omega}\gamma_0 \int_0^\infty \frac{d k}{k} 
\frac{1}{e^{k/T} - 1} \sim\frac{e^4T^3}{12\pi^2 \omega}
\int_0^\infty \frac{d k}{k^2} \rightarrow \infty \,. \nonumber
\ee
which is now dominated by {\it soft} momenta and in fact diverges linearly at
small $k$. When this linear IR divergence is cut off by the introduction
of a photon mass regulator $M$, so that $k^2$ is replaced by $k^2 + M^2$,
we therefore expect a non-analytic contribution of order of
\be
\frac{e^4T^3}{12\pi^2 \omega} \gamma_0 \int_0^\infty \frac{d k}{k^2 + M^2} =
\frac{e^4T^3}{24\pi \omega M}\gamma_0\,,
\label{B1M}
\ee
which is indeed what we find in (\ref{Eq:B_collected}). Notice that
this estimate does not allow for a $M^{-2}$ or $\ln M$ term, and
is consistent with earlier approximations to the bubble diagram, {\it c.f.}
\cite{GatKap}. There is no divergence, logarithmic or otherwise, from the 
corresponding HTL insertion of a fermion self-energy and double fermion 
propagators in the rainbow diagram. Hence there is no need for
any fermion mass regulator in any two-loop elf-energy diagram.
From (\ref{Eq:B_collected}) or (\ref{B1M}) the order $e^2 T$ contribution
to the real part of the self-energy in the HTL resummed evaluation of 
\cite{meg08} has its origin in the appearance of a non-analytic linear 
$1/M$ dependence on the photon Debye mass $M \sim m_f \sim e T$ 
of the two-loop (and higher) bubble self-energy diagram(s). 

Since the HTL resummation program requires a hard-soft factorization 
at $\ell$ loops, in which exactly one of the loops is soft and the rest are
hard, the hard-soft pattern we have verified explicitly at $\ell=2$ 
is a necessary condition for HTL resummation method to work,
and provides a non-trivial check on its consistency. At $\ell >2$ loops 
the HTL resummation contains a geometric series of multiple 
hard photon and electron self-energy insertions on the basic one-loop 
topology of Fig. \ref{Fig:Sigma1}, but in which the bare vertex is corrected 
only to one hard loop order. Thus for $\ell > 2$ HTL resummation when 
expanded in bare perturbative diagrams contains only a subset of all 
self-energy diagrams with $\ell -1$ hard loops and one soft loop. To 
demonstrate that the HTL resummation captures all contributions of 
comparable order for $\omega \sim eT$, one would have to show that 
for $\ell > 2$ all the contributions to $\Sigma$ at order $e^2T$ arise from
\begin{itemize}
\item factorized $\ell$ loop diagrams with exactly $\ell -1$ hard loops
imbedded in one remaining soft loop; but
\item only the subset of diagrams of this kind that can be represented 
as multiple insertions of either the HTL photon or electron self-energy,
with no more than the one-loop HTL vertex insertion.
\end{itemize}
Thus the diagrams of different topology, such as a double rainbow
two-loop vertex correction on the basic one-loop topology should give 
a subleading correction to the self-energy at $\ell =3$ and be negligible 
compared to $e^2T$ for an on-shell fermion at rest. The HTL resummation 
program essentially assumes that this holds.

Let us emphasize again that there are {\it no} hard-hard contributions at
order $e^2 T$ in the two-loop self-energy as enumeration of the generic power 
counting rules given in \cite{braaten90b,lebellac96} might lead one to
expect, and moreover that this absence of hard-hard contributions at two-loop order is
in fact a {\it necessary} condition for HTL resummation. The soft-soft contributions 
to the two-loop self-energy are non-vanishing but subdominant to the NLO 
one-loop terms in $\Sigma$, and will enter at higher orders in the consistent
expansion in $e$.

Our method has been based on isolating the leading order singularities 
as $\omega \rightarrow 0$ of the asymptotics of the high temperature 
expansion, at each loop order separately, and then comparing the 
magnitude of these contributions when the fermion is on-shell
at $\omega \simeq eT/\sqrt{8}$. This suggests that collecting all 
the contributions which are subdominant to the NLO ones we have
been considering, {\it i.e.} the NNLO order contributions, would require 
{\it two} soft loops and $\ell -2$ strictly hard ones of order $T$, which 
would require either at least a three-loop computation, or an inductive 
general argument to verify explicitly. If this is the case, then collecting infrared 
singularities order by order in the high temperature expansion at each $\ell$ 
becomes possible, and a reordering of the usual loop expansion can be carried 
out in a completely systematic and gauge invariant way. At NLO this reordering 
is just one-loop HTL resummation, while at higher orders it may require 
additional resummation of both self-energy and vertex parts. 

The HTL resummation procedure works when all energies and momenta in the
remaining loops are soft $\sim eT$. This one-soft-loop HTL resummed 
contribution should always give the order $\omega/T \sim e$ corrections 
to the leading order HTL self-energy. On the other hand, uncritical use of HTL 
resummation has no justification {\it a priori} when extended to higher orders 
in $\omega/T$ in the high temperature expansion, or in loop integrals containing 
hard momenta, and may lead to incomplete or incorrect results when misapplied
in this way. This seems to be the case in the use of HTL resummed propagators 
in the photon polarization function for the calculation of the photon damping rate 
in QED, or photon and dilepton production in QCD \cite{AGKP,zaraket98,AGZ}. 
In the photon self-energy (unlike the fermion self-energy considered in this paper) 
the NLO result {\it vanishes}, because of the appearance of the statistical factor 
$1 - 2 n_F(\omega/2) \simeq \omega/4T$ which produces a further suppression \cite{zaraket98}. 
Because of this suppression the first non-zero contribution to the photon damping 
rate in QED and the complete calculation of the photon and dilepton production rate 
in QCD both require additional diagrams and depend upon additional physical processes
(bremsstrahlung) than those captured in the one-loop HTL resummed photon 
polarization tensor $^*\Pi_{\mu\nu}$. Based on the results of this paper, we would 
conjecture that a completely analogous calculation of the two-loop perturbative
photon polarization in QED would give a vanishing contribution from both hard-hard 
and hard-soft loops, and yield a non-zero contribution only when both loops are soft, 
which is down by $(\omega/T)^2 \sim e^2$ and is NNLO compared to the LO 
one-loopresult $e^2T^2$. Hence it cannot be captured by the usual HTL resummation.

Once HTL resummation has been carried out, one obtains a new HTL improved
self-energy which contains the correct physics to NLO, at the soft scale
$eT$.  Thus we would expect that the main results of our detailed two-loop
study for fermions at rest ${\bf q} = {\bf 0}$ should continue to hold for
fermions which are slow moving, $|{\bf q}| \lesssim eT$.  However, there is
no necessity for the expansion in $\q^2/\omega^2$ to be analytic at $\q^2 =
0$.  Non-analytic behavior has been reported in scalar QED
\cite{ThoTrax,AbaBou}, but the spinor QED or QCD cases have not been fully
investigated to our knowledge.  It is known that even the HTL resummed
fermion self-energy is inadequate to describe the damping rate or
energy loss of a fast moving fermion with $|{\bf q}| \gg eT$
\cite{altherr93,Pis93}, and a further or different sort of resummation is
required \cite{blaizot97}.  The appearance of additional logarithmic terms
of the form $\ln (\q^2/\omega^2)$ for $|{\bf q}| \gg eT$ does not mean that
HTL resummation is inconsistent, but only that it must be properly applied. 
This example of the damping rate of a fast moving fermion, as those
mentioned above of the photon damping rate in QED and both the photon
production and dilepton production rates in a hot QCD plasma simply lie
outside of the range of strict applicability of HTL resummation, which is
valid for soft external four-momenta only.

To investigate the breakdown of HTL resummation, the systematic method 
which we have used in this paper for the two-loop perturbative self-energy
could be applied to the HTL one-loop resummed self-energy (photon or 
fermion) as well. Thus, if this HTL resummed self-energy is given
{\it off-shell} for all $\omega$ and ${\bf q}$, one could examine its 
high temperature limit, or equivalently, its infrared singularities
at ${\bf q} = {\bf 0}$ and $\omega \rightarrow 0$, as well as its
behavior at large $|\q| \gg eT$. Then, depending on the result one 
could examine terms at two and higher loops in the effective HTL theory 
to determine when they become of the same magnitude as the LO
HTL results. It is known that the breakdown of the HTL resummation 
program is associated with the appearance of collinear 
divergences \cite{Reb} and requires an additional resummation of
ladder-like diagrams corresponding to multi-loop vertex corrections
not contained in the usual HTL resummed self-energy \cite{AGMZ}.
Such ladder sums are also required to compute transport coefficients
in hot gauge theories \cite{Jeon,GagJeon}. Quantitative control of the breakdown 
of the HTL resummation program by the extension of the methods
utilized in this paper to determine the most infrared sensitive contributions
at each HTL resummed loop order can provide detailed guidance to its 
proper generalization to more complete resummation algorithms
in hot gauge theories.

If one finds that higher order terms beyond NLO can be resummed
in a consistent gauge invariant way, this would provide the
theoretical justification for a systematic extension of the HTL program to
higher orders, including also resummed vertex parts and ladder diagrams, 
which might enable the extension of resummation methods to lower energy 
and momentum scales and the hydrodynamic limit, where weak coupling 
methods are usually thought to fail entirely, particularly in QCD.  

Physically, taking account of infrared and collinear singularities by a 
reordering of the  loop expansion and resummation of self-energies and 
proper vertices relate the dressed quasiparticles and many-body interactions 
in the medium to the single particle excitations of the theory in the absence of
a medium, {\it i.e.} in the vacuum at zero temperature and density. The 
dispersion relations of the dressed quasiparticles for low frequencies 
and soft momenta are required to describe the collective modes of the 
plasma in the hydrodynamic limit, and to determine transport coefficients 
such as electrical conductivity, and shear and bulk viscosity. A thorough 
understanding of the quasiparticle degrees of freedom at high temperatures
or densities and the interactions they undergo in thermodynamic
equilibrium or under slight departures therefrom is thus an entry point 
into the systematic study of non-equilibrium processes in the plasma
and self-consistent methods for studying them systematically as well.

\section*{Acknowledgments}
E. M. would like to thank Margaret E. Carrington for invaluable discussions,
and in particular for sharing with us her detailed calculations of the two-loop 
self-energy in the real time formalism, which indicated the non-cancellation of
the $\xi$ dependence in the imaginary part of $\Sigma$ up to two-loop order, 
prior to the initiation of this work. Zs. Sz. would like to thank Los Alamos 
National Laboratory for its hospitality and visitor support under LDRD Grant 
20060049DR. E. M. would like to thank A. Patk{\'o}s for his hospitality and 
visitor support from the E{\"o}tv{\"o}s Lor\'and University, Budapest, Hungary. 
Both authors thank the KITP at UCSB, and organizers of the KITP Program, 
Nonequilibrium Dynamics in Particle Physics and Cosmology, Jan.-Mar. 2008,
during which a significant segment of this research was carried out, supported 
in part by the National Science Foundation under Grant No. NSF PHY05-51164. 
Zs. Sz. was supported by the Hungarian Scientific Research Fund (OTKA) under 
Postdoctoral Grant No. PD 050015 and also T068108 and K77534.

%------@#$---------- APPENDIX -----------------

\appendix

\section{Gaudin's Method for Matsubara sums
\label{App:gaudins_method}}

The Gaudin method \cite{gaudin65} relies on a systematic decomposition 
of a product of $N$ fractions, each linearly depending upon Matsubara
frequencies and corresponding to a finite temperature propagator, into 
sums of products of fractions, in each of which the only Matsubara frequencies 
which appear are the {\it independent} ones to be summed.  Since
the number of such independent Matsubara frequencies in a Feynman
diagram with $\ell$ loops is $\ell$, no matter how many propagators ($N$) 
it may contain, the independent Matsubara sums are more easily 
performed in this method when $\ell < N$.  

The necessary systematic decomposition of the original product of denominators
containing the Matsubara frequencies in a given Feynman diagram
can itself be represented graphically.  For simplicity and definiteness we 
consider only the case of one external Euclidean frequency $\nu$ as in 
the case of the multi-loop Feynman self-energy diagrams for $\Sigma$ 
treated in the text.

To each Feynman multi-loop diagram one associates a set of Gaudin 
graphs $\cal G$. The set is composed of tree diagrams  specified by
all the possible ways of connecting every vertex of the original Feynman 
diagram without forming any closed loop. In each of these Gaudin 
graphs the propagator lines which belong to the tree can be drawn 
with thick lines and form a set denoted by $\cal T_{\cal G}$. The 
complement of this set, namely the propagator lines of
the original Feynman diagram not contained in the given Gaudin tree
graph is denoted by $\bar{\cal T}_{\cal G}$. If the total number of propagator 
lines in the original graph is $N$, we must have
\be
N_{{\cal T}_g} + N_{\bar{\cal T}_g}=N\,,
\ee
where $N_{{\cal T}_g}$ is the number of lines in the Gaudin tree 
graph $g\in {\cal G}$ , and $N_{\bar{\cal T}_g}$ the number of lines in 
its complement. For an original Feynman diagram with $\ell$ loops, 
in fact $N_{{\bar{\cal T}}_g} = \ell$.
The propagator lines in ${\bar{\cal T}_g}$ will be treated as the lines with
the independent Matsubara frequencies to be summed $\omega_k$, 
$k = 1, \dots, N_{\bar{\cal T}_g}$, while the frequencies in ${\cal T}_g$, 
denoted $\Omega_j$, $j = 1, \dots,  N_{{\cal T}_g}$ are treated as dependent. 
The frequency $\Omega_j (i\nu, p_1^0, \dots, p_\ell^0)$ of each line of ${\cal T}_g$ is 
expressed in terms of the independent real frequencies $p_k^0$, $k = 1,\dots,\ell$
by simply replacing each $i\omega_k$ with $p_k^0$ and the external 
imaginary frequency $i\nu$, using energy conservation at each vertex.

The general Gaudin decomposition formula then reads
\be
\prod_{i=i}^{N}\frac{e^{i\omega_i\tau_i}}{p_i^0-i\omega_i}
=\sum_{g \in {\cal G}}
\left[
e^{i\nu T_e}
\prod_{j\in {\cal T}_g}
\frac{1}{p_j^0-i\Omega_j(i\nu,p_k^0)}
\prod_{k\in \bar {\cal T}_g}
\frac{e^{i \omega_k T_k}}{p_k^0-i\omega_k}
\right]\,.
\label{app:Gaudin_decomp}
\ee
The time arguments in the exponentials are introduced as regulators
so that the sums over the independent Matsubara frequencies
in Eqs. (\ref{ap_Eq:sumbose}) and (\ref{ap_Eq:sumfermi}) below converge. 
These time arguments are somewhat arbitrary and will be taken to
zero at the end. They are needed only to guarantee that all sums
converge, and results of Matsubara sums which are absolutely
convergent will not depend on how the $\tau_i$ are assigned.
For definiteness one may choose the following rules to fix the $\tau_i$
\begin{itemize}
\item
Specify the orientation of each line of a diagram, label it with
integer number ($k$) and associate to each line a positive number
$\tau_k$. A possible choice is $\tau_k=k\epsilon,$ with $\epsilon$
positive.
\item
$T_k$ for a given line is calculated by choosing a loop
containing that line and adding the $\tau_i$ of all the lines 
of the loop, formed with the line $k\in\bar{\cal T}_g$ and lines 
of the tree ${\cal T}_g$, with $\pm$ sign if the orientation of the line
agrees with/is opposite to the orientation of 
the line $k\in\bar{\cal T}_g$.
\item
 $T_e$ is determined by the $\tau$'s of the lines of
${\cal T}_g$ which connects the two external lines. In this case the
reference orientation is the orientation of the external line.
\end{itemize}

To perform the sums over the independent Matsubara frequencies
set $\tau_k=k\epsilon,$ let $\epsilon \to 0^+$ and apply the summation 
formulas
\begin{alignat}{3}
T\sum_n\frac{e^{i\omega_n\tau}}{p^0-i\omega_n}&=\varepsilon(\tau)
n_B(\varepsilon(\tau)p^0)e^{p^0\tau}, \quad
&\textnormal{for}\quad &\omega_n=2\pi n T, 
\label{ap_Eq:sumbose}\\
T\sum_n\frac{e^{i\omega_n\tau}}{p^0-i\omega_n}&=-\varepsilon(\tau)
n_F(\varepsilon(\tau)p^0)e^{p_0\tau}, \quad
&\textnormal{for}\quad &\omega_n=(2n+1)\pi T \,,
\label{ap_Eq:sumfermi}
\end{alignat}
valid for $-\beta<\tau<\beta$. This associates a statistical factor 
to each line of $\bar{\cal T}_g$. The $e^{i\omega_n \tau}$ factors are
necessary as regulators so that the Matsubara sums are well defined
and convergent. After the sums have been performed, the limit
$\epsilon \rightarrow 0, \tau \rightarrow 0$ may be taken. 

The simplification of the Gaudin method over other methods for
performing the sums results from the fact that the frequencies of the
last product in (\ref{app:Gaudin_decomp}) are disentangled and
independent of each other. For each of them the Matsubara sum may be 
performed separately, using only the fundamental formulas (\ref{ap_Eq:sumbose})
and (\ref{ap_Eq:sumfermi}), each independent sum yielding its own 
statistical factor depending on whether that line is bosonic or fermionic.
With all Matsubara sums performed, the finite temperature amplitude
is expressed in terms of $N$ timelike momentum integrals over 
$p^0_i$, $i = 1, \dots, N_{{\cal T}_g}$ and $p^0_j$, $j = 1, \dots, N_{\bar{\cal T}_g}$
and $\ell$ spatial momentum integrals. The first set of $N- \ell$ timelike
integrations can be performed immediately by undoing the spectral
representation of the propagators. For subtleties and a
more extensive discussion of the Gaudin method the reader is referred
to Ref.~\cite{reinosa06}.

\section{Asymptotic analysis of integrals \label{App:mitras_integral}}

We present here the high temperature expansion of various integrals
which were encountered in the text. We start with the evaluation of the 
integral 
\bea
I_\pm(i a)&\equiv&\int_0^\infty d x g_\pm(x)\,  \frac{1}{x^2 - (i a)^2} \nonumber\\
&=& \frac{1}{2i a}\int_0^\infty d x g_\pm(x)\left[ \frac{1}{x-i a} -
\frac{1}{x+i a}\right],
\label{Ipm}
\eea
where $g_\pm(x) \equiv x/(\exp(x)\pm 1)$. This integral appears in the calculation
of the one-loop self-energy at NLO in the HTL approximation, and is typical
of the various integrals over soft momenta that are encountered in the
two-loop self-energy. Its real part emerging after analytical continuation 
to real frequencies
\be
i a = \frac{i\nu_q}{2T} \rightarrow \frac{\omega}{2T} + i 0^+
\label{analcont}
\ee
was evaluated only numerically in \cite{mitra00}.

We note first that although $\omega \ll T$, we cannot set $a=0$ in the
integrand, since then $g_-(x)/x^2\sim x^{-2}$ as $x\to 0$ and
the integral would be linearly divergent. As we will see, the linear
divergence survives in the imaginary part of $I_-$ after analytic
continuation but not in the real part, giving a logarithmic dependence
on $T/\omega$ instead. Indeed integrating (\ref{Ipm}) by parts twice,
we obtain
\bea
I_\pm(i a) &=& \frac{1}{2i a}\,g_\pm (0) \, \left[\ln (i a) - \ln (-i a)\right]
-\frac{1}{2} g_\pm'(0) \left[ \ln (i a) + \ln (-i a)\right] \nonumber\\
&&- \frac{1}{2i a} \int_0^\infty d x g_\pm''(x)
\left[ (x+i a)\ln (x+i a) - (x-i a) \ln (x-i a)\right]\,.
\label{Ipm2}
\eea
Under the analytic continuation (\ref{analcont}),
\bea
\ln(i a) \rightarrow \ln \left(\frac{\omega}{2T}\right),\qquad
 \ln (-i a) \rightarrow \ln \left(\frac{\omega}{2T}\right) - i \pi\,,
\eea
so that
\be
I_\pm \left(\frac{\omega}{2T} + i 0^+\right) = \frac{i\pi T}{\omega} \, g_\pm (0)
+ g_\pm'(0) \left[\ln \left(\frac{T}{\omega}\right) + \frac{i\pi}{2}\right]
+ \dots\ ,
\label{Integval}
\ee
where the ellipsis denotes (real) constant terms, and terms that vanish
as $\omega/T \rightarrow 0$.

Using then $g_+(0) = 0,$ $g_-(0) =1,$ and $g_\pm'(0) =\pm 1/2,$
we find that to leading order in the high temperature expansion
\bea
\label{Eq:Mitra_corollary_R}
-\textnormal{Re} \int_0^{\infty} d p\frac{p\, n_B(p)}{p^2 - (\omega+i0^+)^2/4}
&=&\textnormal{Re} \int_0^{\infty} d p\frac{p\, n_F(p)}{p^2 - (\omega+i0^+)^2}
=\frac{1}{2}\ln\frac{T}{\omega},\\
\textnormal{Im} \int_0^{\infty} d p \frac{p\, n_B(p)}{p^2 - (\omega + i0^+)^2/4} &=&
\frac{\pi T}{\omega}.
\label{Eq:Mitra_corollary_I}
\eea
Note that for $T \gg \omega$ only the integral with the Bose-Einstein
statistical factor has an imaginary part. These results to leading logarithmic
order may also be obtained more rapidly by simply substituting the first
two terms of the Taylor expansion of $g_{\pm}(x)$ into (\ref{Ipm}),
and cutting off the logarithmic UV divergence by a cutoff of order of $T$.

An analytic calculation which captures the large temperature behavior of 
(\ref{Ipm}) beyond the leading order given in (\ref{Eq:Mitra_corollary_R}) 
and (\ref{Eq:Mitra_corollary_I}) can be performed with the standard 
technique presented in detail in Appendix~C of \cite{dolan74}. 
Using the identities
\be
\frac{1}{\exp(x)\pm 1}=\pm \frac{1}{2}\mp 
\sum_{n=-\infty}^\infty \frac{x}{x^2+\omega_\pm^2(n)},
\label{Eq:sum_rep}
\ee
where $\omega_+(n)=(2n+1)\pi$ and $\omega_-(n)=2n\pi,$ one separates 
the integral given in the first line of (\ref{Ipm}) into two integrals 
which are regularized by inserting $x^\epsilon$ in the integrands. 
Interchanging the order of the summation and integration one uses partial 
fractioning and exploits the properties of the sums to write 
\bea
\nonumber
I_\pm(i a)&=&\lim_{\epsilon\to 0}\bigg\{\pm\frac{1}{2}\int_0^\infty d x
\frac{x^{1+\epsilon}}{x^2+a^2}
+l_\pm \int_0^\infty d x\frac{x^\epsilon}{x^2+a^2} \\
&&
\pm 2 a^2 \int_0^\infty d x  \frac{x^\epsilon}{x^2+a^2} 
\sum_{n=l_\pm}^\infty \frac{1}{\omega_\pm^2(n)-a^2} 
\mp 2 \sum_{n=l_\pm}^\infty \frac{\omega_\pm^2(n)}{\omega_\pm^2(n)-a^2}
\int_0^\infty d x  \frac{x^\epsilon}{x^2+\omega_\pm^2(n)}\bigg\},
\label{Eq:MitraDJ}
\eea
where $l_\pm=-(-1\pm 1)/2$. The second term is the contribution of 
the static Matsubara mode $n=0$ in $I_-(a).$ In the second and third 
terms the limit $\epsilon\to 0$ can be taken. In the last term one does 
the integral and then one expands the summand around $a=0$ in order to be able 
to do the sum over $n.$ Note, that for bosons this expansion can be safely 
done once we separated the static mode. The limit $\epsilon\to 0$ is taken at 
the very end which results in the cancellation of the $1/\epsilon$ 
type singularity present in the first and last terms of (\ref{Eq:MitraDJ}). 
With these sequence of steps one obtains
\begin{subequations}
\bea
I_+(i a)&=&-\frac{1}{2}\left(\ln \frac{2 a}{\pi}+\gamma_E\right)+
\frac{\pi a}{8} - \frac{7 \zeta(3)}{8\pi^2} a^2+ {\cal O}(a^3),
\label{Eq:MitraF}\\
I_-(i a)&=&\frac{\pi}{2a}+\frac{1}{2}\left(\ln \frac{a}{2\pi}+
\gamma_E\right) -\frac{\pi a}{24}+ \frac{\zeta(3)}{8\pi^2}a^2+{\cal O}(a^3),
\label{Eq:MitraB}
\eea
\label{Eq:uj_Mitra_BF}
\end{subequations}
where $a=\nu_q/(2T)$ and $\gamma_E$ is the first Stieltjes constant 
(Euler's constant).

We can use the result (\ref{Integval}) or (\ref{Eq:uj_Mitra_BF})
for the integral (\ref{Ipm}) to evaluate the integral appearing in
(\ref{Eq:rainbow_term_A_real_4p5a}) and (\ref{Eq:R_HTL_tree3})
at leading order in the high temperature expansion.
One defines the integral
\be
L_\pm(i\nu_q)\equiv\int_{-1}^1 d x \int_0^\infty d p\, k\, p f_\pm(p) 
\sum_{r,s=\pm 1} \frac{r}
{(i\nu_q-2s p)((i\nu_q)^2-2k p(x-r s)-2i\nu_q(r k+s p))},
\label{Eq:L_pm_1}
\ee
where $f_-(x)=n_B(x)$ and $f_+(x)=n_F(x)$. In the limit $k\gg p,\nu_q$
one approximates 
$-2k p(x-r s)-2i\nu_q r(k+ r s k p -r i\nu_q/2)\simeq 
-2k[p(x-r s)+r i\nu_q]$, then one uses partial fractioning and the change 
$x\to-x$ in some terms to rewrite the integral in the following form:
\bea
L_\pm(i\nu_q)
&\simeq& \int_{-1}^1 d x \frac{1}{1-x^2} \int_0^\infty d p \left[
\frac{p f_\pm(p)}{p^2-(i\nu_q/2)^2}
-2\frac{p f_\pm(p)}{p^2-(i\nu_q/(1+x))^2} \right].
\label{Eq:L_pm_2}
\eea
Then one uses (\ref{Eq:Mitra_corollary_R}) 
and (\ref{Eq:Mitra_corollary_I}) to write
\be
L_\pm(\omega+i0^+)\simeq \int_{-1}^1 d x \frac{1}{1-x^2}
\left[-l_\pm x\frac{i\pi T}{\omega} \pm\frac{1}{2}
\left(2\ln\frac{\omega}{T(1+x)}-\ln\frac{\omega}{2T}\right)
\right],
\label{Eq:L_pm_highT}
\ee
where $l_\pm=-(-1\pm 1)/2.$
The imaginary part vanishes since the integrand is odd in $x.$ Keeping in
the real part the terms which are leading order in the temperature one has
\be
L_\pm(\omega+i0^+)\simeq \mp\frac{1}{2}\ln\frac{T}{\omega}\int_{-1}^1 d x 
\frac{1}{1-x^2} +0\, i.
\label{Eq:L_pm_result}
\ee

With the same method one can perform the asymptotic analysis of the integrals
appearing in (\ref{Eq:Pi2_HTL}) and (\ref{Eq:Pi3_HTL}).
Introducing a new variable of integration $x=p/T,$ the notation 
$a=\nu_q/(2T),$ $b_\pm=(\nu_q^2\pm M^2)/(2\nu_q T)$ and taking  
the derivative with respect to $M^2$ of the various terms in (\ref{Eq:Pi2_HTL}) 
and (\ref{Eq:Pi3_HTL}) one obtains 
\begin{subequations}
\bea
\frac{\partial}{\partial M^2}\bigg\{\mu_+
\int_0^\infty d x g_\pm(x)\frac{\ln(1+\frac{x^2}{a^2})}{x^2+b_\pm^2}
\bigg\}\bigg|_{M=0}&=&
\frac{1}{2 i\nu_q}\bigg\{
\mp\left(2+a\frac{\partial}{\partial a}\right) I_\pm(i a) \\
&&+\left(1\mp a\frac{\partial}{\partial a}\right)
\int_0^\infty d x g_\pm(x)
\frac{\ln\left(1+\frac{x^2}{a^2}\right)}{x^2+a^2}\bigg\},
\label{Eq:log_int}\\
\frac{\partial}{\partial M^2}\bigg\{
\int_0^\infty\!\! d x x g_-(x)
\frac{\cot^{-1}(\frac{x}{a})}{x^2+b_-^2}
\bigg\}\bigg|_{M=0}
&=&\frac{a}{\nu_q^2}\left[\frac{1}{2 a}
\frac{\partial}{\partial a}\left(a^2 I_-(i a)\right)
-\frac{\partial}{\partial a} \int_0^\infty \!\! d x x g_-(x)
\frac{\cot^{-1}(\frac{x}{a})}{x^2+a^2}\right],\hspace{1.2cm}
\label{Eq:arccot_int}\\
\frac{\partial}{\partial M^2}\bigg\{
2i\frac{\mu_-\mu_+}{T} 
\int_0^\infty \frac{d x }{e^x+1}\frac{\tan^{-1}(\frac{x}{a})}{x^2+b_+^2} 
\bigg\}\bigg|_{M=0}&=&\frac{i}{4 T}\bigg\{
\frac{\partial}{\partial a} I_+(i a) 
-2 a\frac{\partial}{\partial a}
\int_0^\infty \frac{d x}{e^x+1}
\frac{\tan^{-1}(\frac{x}{a})}{x^2+a^2} 
\bigg\},\ \ 
\label{Eq:Uj_arctan_int}\\
\frac{\partial}{\partial M^2}\bigg\{
4 \mu_- I_+(i b_+)
\bigg\}\bigg|_{M=0}&=&-\frac{2}{i\nu_q}\left(
1+a\frac{\partial}{\partial a} \right)I_+(i a),\\
\frac{\partial}{\partial M^2}\bigg\{
\int_0^\infty dx g_-(x)\frac{x}{x^2+b_-^2}
\bigg\}\bigg|_{M=0}&=&
-\frac{a}{\nu_q^2} \frac{\partial}{\partial a}
\int_0^\infty dx g_-(x)\frac{x}{x^2+a^2},
\eea
\label{Eq:arctan_int}
\end{subequations}
where again $g_\pm(x)=x/(\exp(x)\pm 1).$ The only term where one cannot take 
the limit $M^2=0$ is
\be
\frac{\partial}{\partial M^2}\bigg\{
b_-\ln\frac{M^2}{\nu_q^2} I_-(ib_-)
\bigg\}=\left[
\frac{a}{M^2}-\frac{a}{\omega^2}\left(1+\ln\frac{M^2}{\nu_q^2}\right)
\left(1+a\frac{\partial}{\partial a}\right)
\right] I_-(ia)+{\cal O}(M^2).
\label{Eq:MB2}
\ee
Here we expanded in powers of $M.$ In the expressions above 
$I_\pm (i a)$ is given by (\ref{Ipm}). 

Then the method described between (\ref{Eq:sum_rep}) and (\ref{Eq:MitraF})
gives the following series:
\begin{subequations}
\bea
\int_0^\infty d x g_+(x)\frac{\ln\left(1+\frac{x^2}{a^2}\right)}{x^2+a^2}
&=&
\frac{1}{2}\ln^2 a+\left(\gamma_E-\ln\frac{\pi}{2}\right) \ln a + C_1
-\frac{a \pi}{4}(1-\ln 2)+{\cal O}(a^2),
\label{Eq:logF_series}
\\
\int_0^\infty d x g_-(x)\frac{\ln\left(1+\frac{x^2}{a^2}\right)}{x^2+a^2}
&=&
\frac{\pi}{a}\ln 2 - \frac{1}{2}\ln^2 a -(\gamma_E-\ln(2\pi))\ln a
+C_2+\frac{a\pi}{12} (1-\ln 2) +{\cal O}(a^2),\ \ \
\label{Eq:logB_series}
\\
\int_0^\infty \frac{d x}{e^x+1}
\frac{\tan^{-1}(\frac{x}{a})}{x^2+a^2}
&=&
\frac{\pi^2}{16a}+\frac{\pi}{8}\ln a+C_3+
\frac{a \pi}{8}(\ln 2 -\frac{7}{\pi^3}\zeta(3))+{\cal O}(a^2),
\label{Eq:arctan_series}
\\
\int_0^\infty d x x g_-(x)
\frac{\cot^{-1}(\frac{x}{a})}{x^2+a^2}
&=&
\frac{\pi}{2} \ln 2 +\frac{a}{2} \left(-1+\gamma_E+\frac{\pi^2}{8}
+\ln\frac{a}{2\pi}\right)
+{\cal O}(a^2),
\label{Eq:Uj_arccot_series}
\\
\int_0^\infty d x g_-(x)\frac{x}{x^2+a^2}
&=&
\frac{a\pi}{4}-\ln a+{\cal O}(a^2),
\eea
\label{Eq:arccot_series}
\end{subequations}
where the constants are given in terms of the first two Stieltjes constants
$\gamma_E$ and $\gamma_1$ and the Glaisher's constant $G$ as
$C_1=\frac{\pi^2}{8}-\gamma_E\ln\frac{\pi}{2}-\frac{1}{2} \ln^2 2-
\frac{1}{2}\ln\frac{4}{\pi} \ln \pi-\gamma_1,$
$C_2=-\frac{\pi^2}{8}-\frac{1}{2}\ln^2(2\pi)+\gamma_E\ln(2\pi)
+\gamma_1,$ and $C_3=\frac{\pi}{24}(3\gamma_E+4\ln 2 -36\ln G).$

In order to derive (\ref{Eq:BU1a}), (\ref{Eq:BU1b}), and (\ref{Eq:BU1c}) 
first we define two integrals,
\bea
J_-(\mu_-)&\equiv&\int_0^\infty \frac{d p}{E_p}\, \frac{n_B(E_p)}{p^2-\mu_-^2}
=\frac{\pi T}{2 \mu_+^2}\left[\frac{1}{(-\mu_-^2)^{1/2}}-\frac{1}{M}\right]
-\frac{1}{2\mu_-\mu_+}\left[
\ln\frac{M}{i\nu_q}+i\frac{\pi}{2} \right]
+{\cal O}\left(\frac{1}{Tm},\frac{1}{T\mu_-}\right),\ \ \ 
\label{Eq:J-def}\\
T_B^\beta(M)&\equiv&\frac{1}{2\pi^2}\int_0^\infty d p\, p^2 \frac{n_B(E_p)}{E_p}
=\frac{T^2}{12}-\frac{M T}{4\pi}-\frac{M^2}{8\pi^2}
\ln\left(c \frac{M}{T}\right)
+{\cal O}\left(\frac{M^4}{T^2}\right),
\label{Eq:BTad-HTE}
\eea
where $\mu_\pm=((i\nu_q)^2\pm M^2)/(2i\nu_q),$ $E_p^2=p^2+M^2,$ and 
$\ln(c)=\gamma_E-1/2-\ln(4\pi).$ The integrals were evaluated in the high 
temperature expansion with the method described between 
(\ref{Eq:sum_rep}) and (\ref{Eq:MitraF}).
The first term in the expression 
of $J_-(\mu_-)$ is the contribution of the static Matsubara mode and 
can be obtained also with the approximation $n_B(E_p)\simeq T/E_p.$

Then using that
\be
\int_0^\infty d p \frac{n_B(E_p)}{E_p}=-2 \int_0^\infty d p\, p^2 
\frac{\partial }{\partial p^2} \frac{n_B(E_p)}{E_p}=
-2\frac{\partial}{\partial M^2} \int_0^\infty d p\, p^2 \frac{n_B(E_p)}{E_p}
=-4\pi^2 \frac{\partial }{\partial M^2} T_B^\beta(M),
\label{Eq:atiras}
\ee
the first two integrals in (\ref{Eq:bubble_tree1_th}) can be written as
\bea
\int_0^\infty d p
\frac{p^2}{E_p} \frac{n_B(E_p)}{E_p^2-\mu_+^2}&= &
-4\pi^2\frac{\partial T_B^\beta(M)}{\partial M^2}
+\mu_-^2 J_-(\mu_-),
\label{Eq:erdekes_tad}\\
\int_0^\infty d p
\frac{p^2 E_p n_B(E_p)}{E_p^2-\mu_+^2}&=&
2\pi^2\left[1-2\mu_+^2\frac{\partial}{\partial M^2}\right]T_B^\beta(M)
+\mu_+^2 \mu_-^2 J_-(\mu_-).
\label{Eq:int_Tad_dTad}
\eea
Using the expressions given in (\ref{Eq:J-def}) and (\ref{Eq:BTad-HTE})
one obtains
\bea
\frac{\partial}{\partial M^2}\left[ \left((i\nu_q)^2+M^2\right)
\int_0^\infty d p \frac{p^2}{E_p} \frac{n_B(E_p)}{E_p^2-\mu_+^2} \right]
&\simeq& \pi T\left[\frac{1}{M}-\frac{2i}{i\nu_q}\right]
+\ln\frac{M}{i\nu_q}-\frac{1}{2}\ln\frac{T}{i\nu_q}+C,
\label{Eq:erdekesbol}\\
\frac{\partial}{\partial M^2}\int_0^\infty d p
\frac{p^2 E_p n_B(E_p)}{E_p^2-\mu_+^2}&\simeq&
\frac{1}{8}-i\frac{\pi T}{4i\nu_q},
\label{Eq:int_in_1-loop_xi}
\eea
where $C=(1+\gamma_E-\ln(4\pi))/2$ and the expressions on the right hand sides 
were expanded for small $M$. After analytical continuation 
($i\nu_q\to\omega+i0^+$) one obtains (\ref{Eq:BU1a}) and (\ref{Eq:BU1b}).

For the last term of (\ref{Eq:bubble_tree1_th}) one may
introduce $x=E_p/T$ as a new variable of integration and obtain
\be
I_B\equiv\frac{\partial}{\partial M^2}\left\{
(\nu_q^2-M^2)\int_\mu^\infty d x \frac{x}{e^x-1}\frac{1}{x^2+b^2}
\ln\left[\frac{x+\sqrt{x^2-\mu^2}}{x-\sqrt{x^2-\mu^2}}\right]
\right\},
\label{Eq:I_B}
\ee
where $\mu=M/T$ and $b=(\nu_q^2-M^2)/(2\nu_q T).$ In order to obtain
the contribution of (\ref{Eq:I_B}) for $M,|i\nu_q| \ll T$ at leading
order in the high-$T$ expansion one splits the integral into two
pieces by introducing the point $\bar x$ satisfying 
$0<\mu, b\ll\bar x\ll 1.$ For $x>\bar x$ one expands the logarithm for 
large $x$ and use the method described between 
(\ref{Eq:sum_rep}) and (\ref{Eq:MitraF}).
For $x<\bar x$ one expands $x/(e^x-1)\simeq 1-x/2$ and one scales the
variables with $\mu.$ Writing the logarithm as a difference of two
logarithms one introduces in the resulting two integrals the argument
of the respective logarithm as a new variable of integration. The
integrals can be done with the symbolic program Mathematica. Adding
the two parts the $\bar x$ dependence cancels and the result is
\bea 
I_B&=&-\frac{\pi \nu_q T}{M^2}
+\frac{\nu_q^2}{2 M^2}\left[ \ln\frac{T}{\nu_q}+\ln(4\pi)-\gamma_E \right]
+\frac{1}{2} \ln\left(\frac{T}{\nu_q}\right) 
\ln\left(\frac{\nu_q T}{M^2}\right) 
\nonumber
\\
&&+\big(\ln(4\pi)-\gamma_E\big)\ln\frac{T}{M} -\frac{1}{2}\ln\frac{T}{\nu_q}
+ C_B +{\cal O}\left(\frac{\nu_q}{T},\frac{M}{T}\right),
\label{la_int}
\eea
where $C_B=\frac{\gamma_E}{2}-\gamma_1+\frac{\pi^2}{12}+\frac{1}{2}\ln^2(4\pi)
-\left(\gamma_E+\frac{1}{2}\right)\ln(4\pi).$ After analytical continuation 
one obtains (\ref{Eq:BU1c}).

The high temperature asymptotics of integrals of the form
\be
F = \int_0^{\infty} \frac{dE}{\sqrt{E^2 - M^2}}\ n_B(E)\, f(E)\,,
\ee
with $f(E)$ a regular differentiable function of $E$ at $E=0$, encountered
in Sec. \ref{sec:xi} can be handled by similar methods. Introducing an 
intermediate scale $\bar E$, such that $M \ll \bar E \ll T$, one can write
\be
F = \int_M^{\bar E} \frac{dE}{\sqrt{E^2 - M^2}} \left(\frac{T}{E} - \frac{1}{2} + \dots\right)
\left(f(0) + E f'(0) + \dots\right) + \int_{\bar E}^{\infty}\frac{dE}{E}\,n_B(E)f(E) + \dots\,,
\ee
where the ellipsis gives terms of order $M/T$ or smaller. Performing the first integral
exactly and integrating the second by parts twice gives
\begin{eqnarray}
&&F = T f(0) \left[\frac{\pi}{2M} - \frac{1}{\bar E}\right] + \left[T f'(0)  - \frac{1}{2}\right]\ln\frac{\bar E}{M}
- n_B(E)f(E)\Big\vert_{\bar E}^{\infty} \nonumber\\
&& + \frac{d}{dE} \left[E n_B(E) f(E)\right] \ln\frac{E}{T}\Big\vert_{\bar E}^{\infty} + \int_{\bar E}^{\infty} dE \,\frac{d^2}{dE^2} \left[E n_B(E)f(E)\right] \ln\frac{E}{T}\nonumber\\
&& \qquad = \frac{\pi T}{2M} f(0) + T f'(0) \ln\frac{T}{M} - \frac{f(0)}{2} \ln\frac{T}{M} + T f'(0) + {\cal O}(1)
\label{asympF}
\end{eqnarray} 
in the high temperature limit. This gives the results (\ref{F12eval}) and is used
in obtaining (\ref{G1234eval}) and (\ref{Heval}). A result analogous to (\ref{asympF}) is
easily obtained for the Fermi-Dirac distribution $n_F$. Finally we note that the
same method of introducing an intermediate scale, $|i\nu_q| \ll \bar E \ll T$ may be
used to extract the asymptotic behavior of the finite integrals in Eqs. (\ref{G1234int})
as well.

\end{document}